%% file: talehybridCompositionPRD.tex
\documentclass[%
 reprint,
superscriptaddress, amsmath, amssymb, aps, reprint, makecell,
]{revtex4-2}

\usepackage{graphicx}
\usepackage{dcolumn}
\usepackage{bm}
\usepackage{comment}
\usepackage{xcolor} 
\colorlet{mdtBlue}{blue!50!black}
\usepackage{hyperref}
\hypersetup{
    colorlinks=true,
    linkcolor=mdtBlue,
    citecolor=mdtBlue,
    filecolor=black,
    urlcolor=mdtBlue
}
\usepackage{amsmath}
\usepackage{physics}
\usepackage{graphicx}
\usepackage{multirow}
\usepackage{tabularx}
\usepackage{adjustbox}
\usepackage{siunitx}
\usepackage{array}
\usepackage{url}
\usepackage{cleveref}
\usepackage{multirow}
\usepackage{makecell}
\usepackage{soul} 
\usepackage[small]{caption}
\usepackage[small]{ragged2e}
\usepackage[subrefformat=parens]{subcaption}
\usepackage{lineno}

\renewcommand{\thetable}{\arabic{table}}
\usepackage{CJKutf8}

\newcommand{\xmax}{X_\mathrm{max}}
\newcommand{\mxmax}{\langle X_\mathrm{max} \rangle}

\newcommand{\sxmax}{\sigma(\xmax)}

\newcommand{\mlna}{\langle \mathrm{ln}\,A \rangle}

\newcommand{\fl}{\mathrm{fluorescence}}
\newcommand{\Fl}{\mathrm{Fluorescence}}
\newcommand{\Ch}{\mathrm{Cherenkov}}

\newcommand{\add}[1]{\textcolor{black}{#1}}

\nolinenumbers
\begin{document}

\preprint{APS/123-QED}

\title{Cosmic ray mass composition measurement in the energy range from $10^{16.5}$ eV to $10^{18.5}$ eV observed with the TALE hybrid detector}

\input{TA-author-20250704-revtex}

\date{\today}

\begin{abstract}
\newpage
We report on the cosmic ray mass composition measured by the Telescope Array Low-energy Extension (TALE) hybrid detector. 
The TALE detector consists of a fluorescence detector (FD) station with 10 FD telescopes located at the Telescope Array (TA) Middle Drum FD Station (itself made up of 14 FD telescopes), and a surface detector (SD) array of scintillators. 
The array consists of 40 SDs with 400 m spacing and 40 SDs with 600 m spacing. 
In this paper, we present results on the measurement of the depth of shower maxima ($\xmax$) in the energy range from $10^{16.5}$ eV to $10^{18.5}$ eV collected over five years of the TALE hybrid detector.
The $\xmax$ distributions were analyzed and compared with Monte Carlo simulations of proton, helium, nitrogen, and iron primaries, using the QGSJet\,II-04 hadronic interaction model.
Our results indicate that the elongation rate of the mean $\xmax$, which is defined as the slope of $\mxmax$ versus cosmic ray energy, exhibits a break around $10^{17}$ eV. 
Up to this energy, the composition becomes increasingly heavy, characterized by a growing dominance of heavy nuclei and a steadily decreasing fraction of light primaries.
Beyond this energy, the proton fraction increases significantly with energy.
These findings suggest a transition from Galactic to extra-Galactic cosmic ray sources around the so-called second knee.
\end{abstract}

\maketitle

\section{Introduction} \label{sec:intro}
The origin and nature of cosmic rays have remained a mystery for over a century since their discovery.
Cosmic rays are observed across an extremely broad energy range, from $10^8$ eV to beyond $10^{20}$ eV, with a flux that steeply decreases with increasing energy, approximately following a power-law function of $E^{-\gamma}$.
One of the well-known features in the spectrum is the ``knee"~\cite{bib:knee1959} at around $10^{15.6}$ eV, where the spectrum steepens from a spectral index from $\gamma \sim 2.7$ to $\gamma \sim 3.0$.

Recent diffuse gamma-ray observations by the Tibet AS$\gamma$~\cite{bib:TibetASgamma} experiment and LHAASO~\cite{bib:LHAASO} have provided strong evidence for the existence of PeVatrons in the Milky Way, which are believed to be capable of accelerating particles to PeV energies.
The steepening of the spectrum at the knee is often interpreted as the maximum energy limit of Galactic cosmic ray accelerators or as the efficient escape of cosmic rays from the Galaxy.
Both scenarios are typically described in terms of particle rigidity, defined as $R = pc/(eZ)$, where $p$ is the momentum, $c$ is the speed of light, $e$ is the elementary charge, and $Z$ is the atomic number.

For heavier nuclei the maximum energy can be up to $Z$ times greater than that of protons, this rigidity-dependent scaling is known as the Peters cycle~\cite{bib:Peters1961PrimaryCR}.
Consequently, if the knee corresponds to the maximum energy of Galactic protons, the maximum energy for heavy nuclei should be expected at higher energies, around $10^{17}$\,eV, where indeed a knee like structure, known as the ``second knee," is observed and has been confirmed by several measurements~\cite{bib:KASCADESpectrum, bib:tunkaSpectrum, bib:YakutskSpectrum, bib:iceTopResults, bib:taleSpectrum}.
This hypothesis leads to an increasingly heavier average composition with increasing cosmic ray energy up to the second knee.

On the other hand, recent results from the Pierre Auger Observatory (Auger)~\cite{PierreAuger:2014gko, PierreAuger:2014sui, bib:AugerXmax} and the Telescope Array (TA)~\cite{bib:tahybrid}, based on measurements of the depth of shower maximum ($\xmax$) using hybrid detectors that combine fluorescence and surface detectors, indicate that the mean mass composition shifts back to lighter components, particularly protons, at energies around the ankle, approximately $5\times10^{18}$\,eV~\cite{ta_sd_spec2013, auger_prd2020}.
Also, Auger reported evidence of anisotropy in the arrival directions of cosmic rays above $8 \times 10^{18}$\,eV, strongly suggesting that these ultra-high-energy cosmic rays (UHECRs) originate from extragalactic sources~\cite{bib:AugerDipole}.
Thus, the energy range from $10^{17}$\,eV to $10^{18}$\,eV is particularly intriguing, as it is expected to be the transition region where the dominant contribution shifts from Galactic to extra-Galactic cosmic rays, and the mass composition transitions from heavy to light nuclei ~\cite{bib:Candia2002qd}.

Building upon the success of $\xmax$-based mass composition measurements with hybrid detectors at ultrahigh energies by Auger and TA, the Telescope Array Low-energy Extension (TALE) was developed to apply the same method at lower energies covering the transition region, to investigate the change in cosmic ray origin in greater detail.
This paper presents the results from the cosmic ray mass composition measurements in the energy range from $10^{16.5}$\,eV to $10^{18.5}$\,eV, collected over five years with the TALE hybrid detector.
We describe the detector and operations in Sec.\,\ref{sec:ta}, followed by event reconstruction methods in Sec.\,\ref{sec:reconstruction}.
In Sec.\,\ref{sec:simulation}, we describe the Monte Carlo simulations used in this work.
Using these simulations, we evaluate the event reconstruction accuracy and assess the detector performance through the observational data and Monte Carlo comparisons.
Section\,\ref{sec:systematic} examines the systematic uncertainties on the $\xmax$ measurement.
Finally, we discuss the results and their implications for understanding cosmic ray sources in Sec.\,\ref{sec:result}, and conclude with a summary in Sec.\,\ref{sec:conclusion}.

\section{Detector and dataset} \label{sec:ta}

\begin{figure}[t]
\begin{center}
\vspace{+5mm}
\includegraphics[width=1.\linewidth]{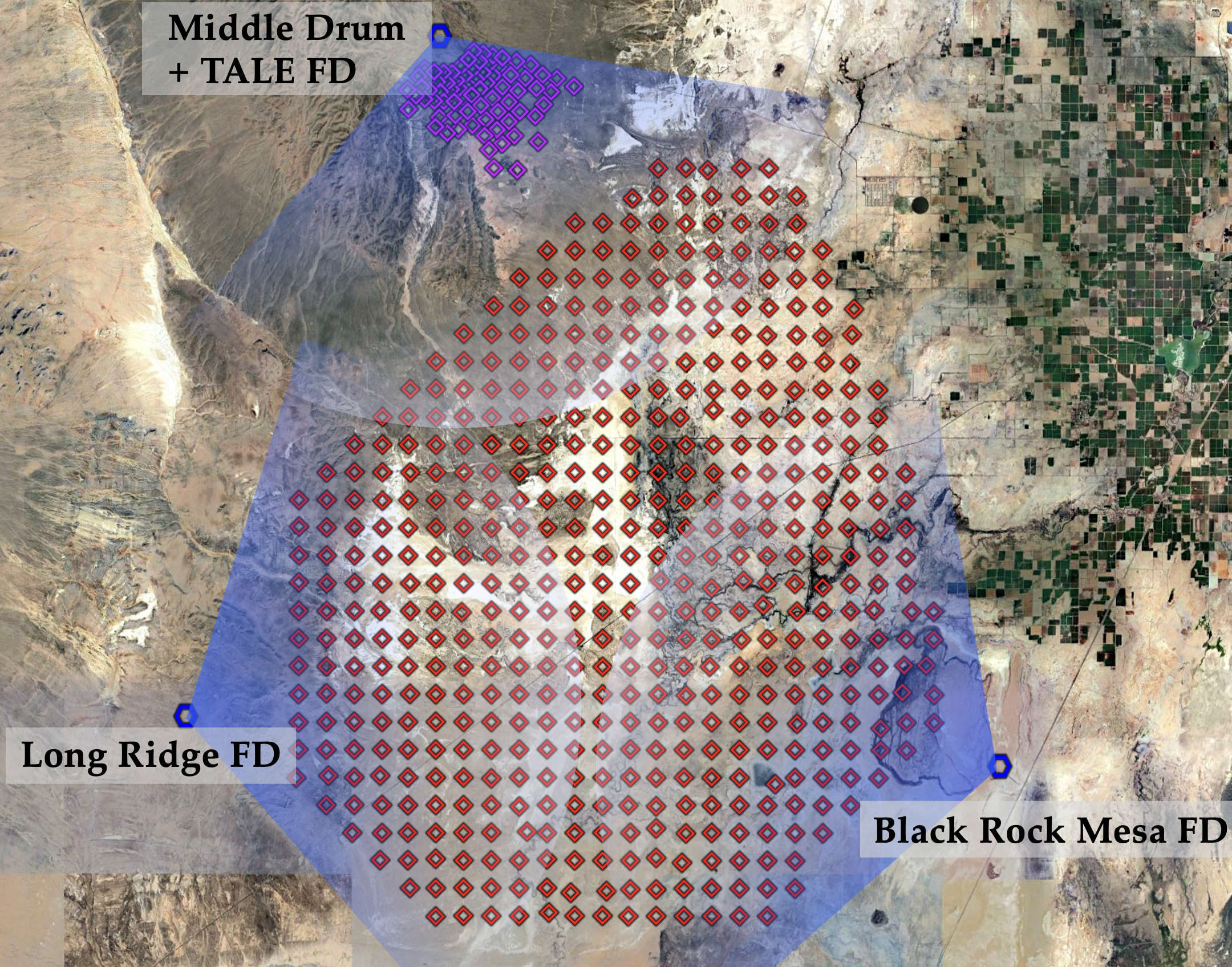}
\vspace{+5mm}
\caption{\justifying{Map of the Telescope Array detectors. Red diamonds represent the locations of each surface detector. Three FD stations are shown as blue hexagons. A transparent blue fan shape represents a field of view for each FD. The TALE SDs are deployed in the northwest part of the TA site, shown by magenta diamonds}.}
\label{fig:taMap}
\end{center}
\end{figure}

The TA is the largest cosmic ray observatory in the Northern Hemisphere, located at approximately $39.3^\circ$ N latitude and $112.9^\circ$ W longitude in Utah, USA.
The primary aim of the TA experiment is to study UHECRs by detecting extensive air showers (EAS) induced by these particles as they interact with Earth's atmosphere.
The TA experiment comprises a surface detector (SD) array and three fluorescence detector (FD) stations, which together provide hybrid detection capability\,(Fig.\,\ref{fig:taMap}).

The SD array consists of 507 scintillation detectors arranged in a square grid with 1200 m spacing, covering a total area of approximately 700 km$^2$.
The three TA FD stations are located at Black Rock Mesa (BRM), Long Ridge (LR), and Middle Drum (MD).
Each station contains telescopes that observe the atmosphere above the surface detector array, detecting faint ultraviolet fluorescence light emitted by atmospheric molecules when they are excited by charged particles in an EAS.
In addition, the telescopes also detect Cherenkov photons produced by shower particles, which contribute to the observed light signal.
Each telescope is equipped with a segmented spherical mirror and a camera composed of 256 hexagonal photomultiplier tubes (PMTs), providing a field of view of $1^\circ \times 1^\circ$ per PMT~\cite{bib:Tokuno2012, bib:HiResI}.
The stations have an elevation viewing range from 3$^\circ$ to 31$^\circ$.

\begin{figure}[t]
\begin{center}
\includegraphics[width=1.05\linewidth, trim=0 0 0 2cm, clip]{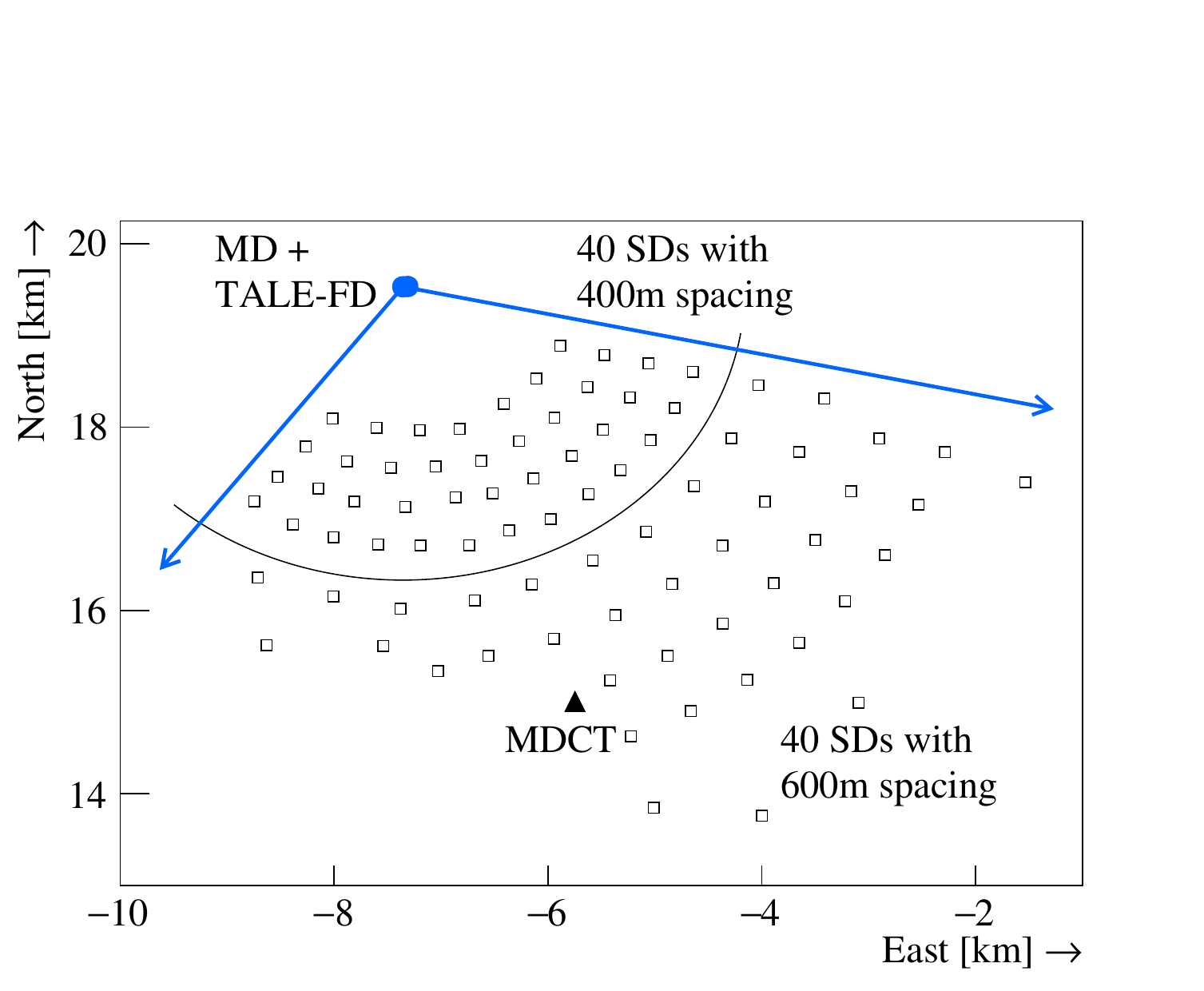}
\caption{\justifying{Layout of the TALE detector. Open square boxes represent the locations of the TALE SDs, and a small filled circle corresponds to the location of the MD / TALE FD station. The arrows represent the azimuthal viewing ranges of the FDs. The curved boundary indicates the separation between the 40 SDs with 400\,m spacing and those with 600\,m spacing. The central data acquisition tower, labeled MDCT in the figure, stands for the Middle Drum Control Tower, which collects trigger and waveform data from the SDs.}}
\label{fig:taleMap}
\end{center}
\end{figure}
\begin{figure}[h]
\begin{center}
\vspace{+5mm}
\begin{minipage}[l]{0.4025\hsize}
\centering
\begin{center}
\includegraphics[width=1.\linewidth]{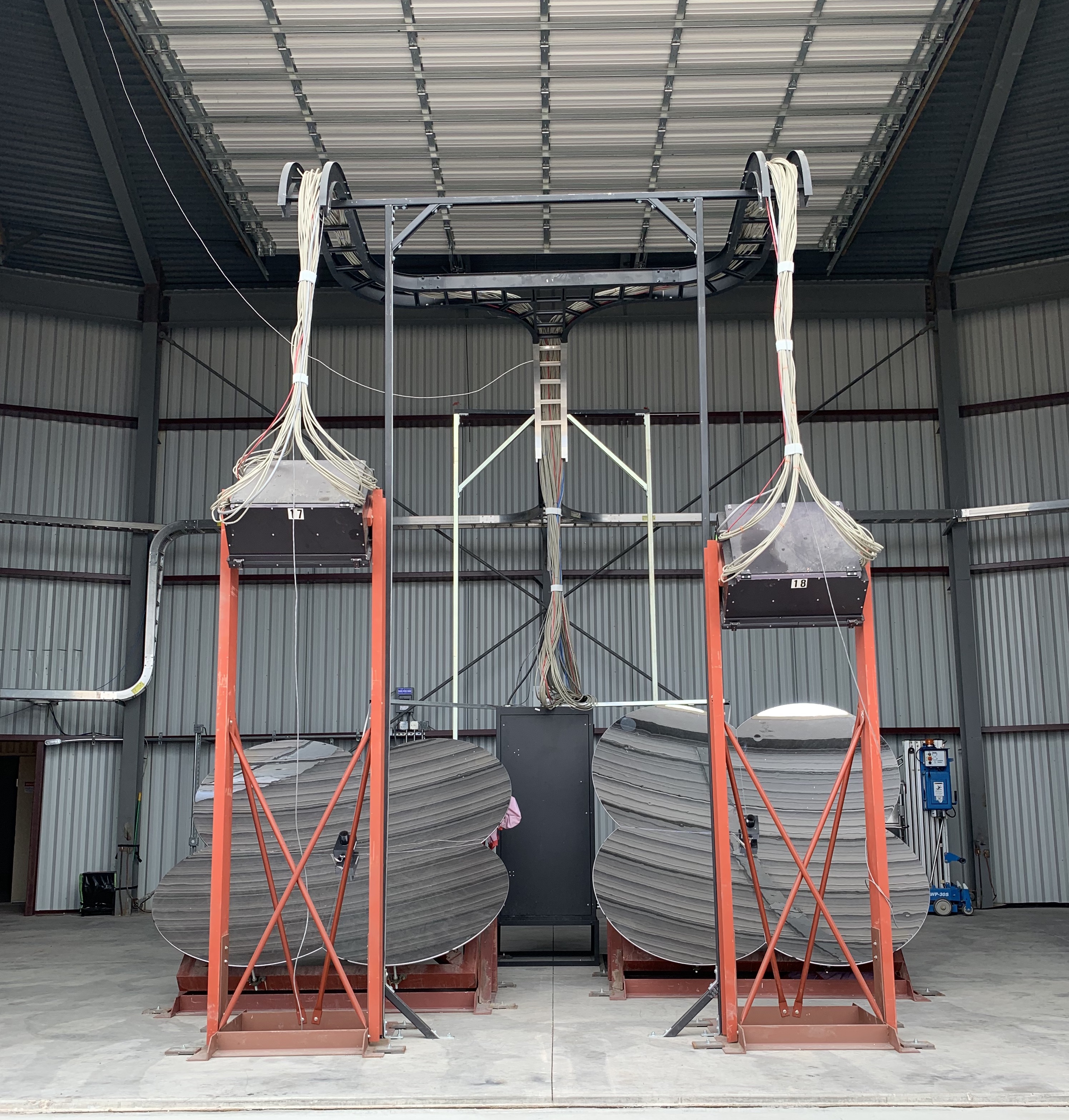}
\end{center}
\end{minipage}
\begin{minipage}[r]{0.55\hsize}
\centering
\includegraphics[trim=0cm 0cm 1cm 0cm,width=1.\linewidth]{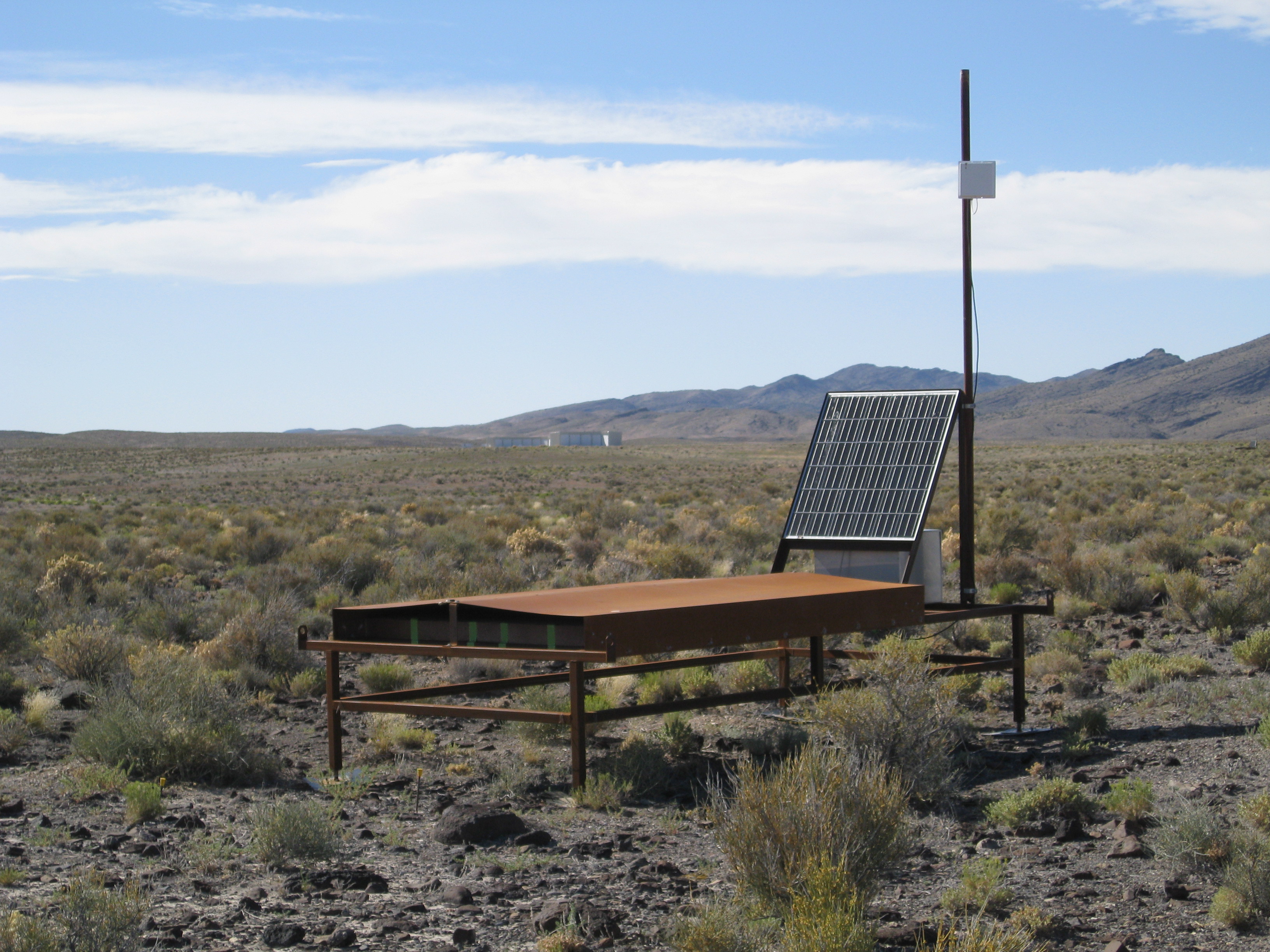}
\end{minipage}
\caption{\justifying{Left: Photograph of TALE telescopes at the FD station. Right: A deployed SD in the experimental site}.}
\label{fig:taleDetector}
\end{center}
\end{figure}

TALE was constructed to extend the TA's sensitivity to lower energy cosmic rays, down to approximately $10^{16}$ eV.
The TALE detector configuration is shown in Fig.\,\ref{fig:taleMap}.
TALE FD consists of an additional FD station with ten telescopes, installed at the same site as the MD FD station.
While both FD stations observe in the same azimuthal directions, they are arranged to cover different elevation ranges: TALE FD covers from $31^\circ$ to $59^\circ$, overlapping with the upper part of the MD-FD’s field of view.
This configuration enables the TALE detector to observe the full development of air showers across a broad range of elevations.

The ten telescopes used in TALE were refurbished from components previously employed in the HiRes experiment~\cite{bib:hires2}.
The mirror of a telescope is composed of four segmented clover-shaped spherical mirrors with a total area of 5.2 m$^{2}$, and the camera consists of 16 $\times$ 16 PMTs at the focal plane of the mirror, as shown in the left panel of Fig.\, \ref{fig:taleDetector}.
The UV filter is installed at the front of the camera box to reduce the night sky background noise.
Each PMT signal is digitized using a flash analog-to-digital converter (FADC) with 10\,MHz sampling and eight-bit resolution.
Summations of waveforms for each row and column, in a total of 32 channels, are also recorded and are used in trigger logic, which looks for a threefold coincidence in rows or columns.

In addition to the FD station, the TALE setup includes an SD array of 80 scintillators deployed with 400 m and 600 m spacing, covering an area of about 20 km$^2$.
All SDs are connected to a central data acquisition tower via wireless LAN communication operating at 2.4 GHz band.

Each SD shown in the right panel of Fig.\,\ref{fig:taleDetector} has two layers of plastic scintillators, each with an area of 3 m$^{2}$ and a thickness of 1.2\,cm.
The scintillation light produced by energy deposition from charged particles is guided to the PMTs, which are connected separately to each layer via wavelength-shifting fibers\,~\cite{bib:tasdNIM}.
The PMT signals are digitized by FADC modules with 12-bit resolution and 50 MHz sampling.

Whenever the SDs detect signals above a certain intensity, they autonomously issue two types of triggers, called Level-0 (Lv.\,0) and Level-1 (Lv.\,1) triggers.
A Lv.\,0 trigger is generated when the integrated digitized waveforms within an eight-slice time window (160\,ns total) for both the upper and lower layers exceed 15 FADC counts above the pedestal -- corresponding to approximately 0.3 minimum ionizing particles (MIPs). 
In this case, the SD stores the waveforms from both layers in a local buffer, covering a total of 2.56\,$\mu$s. 
All waveforms stored by the Lv.\,0 triggers are integrated over 2.56 $\mu$s, and the pedestal is subtracted.
If the resulting value exceeds 150 ADC counts (equivalent to 3 MIPs), a Lv.\,1 trigger is issued, and the SD records the timing of the Lv.\,1 trigger.
Each SD transmits a list of Lv.\,1 trigger timestamps to the DAQ PC at the central data acquisition tower (labeled MDCT in Fig.\,\ref{fig:taleMap}) once per second.
If four or more SDs have a Lv.\,1 trigger within a 32$\mu$s window, then the DAQ PC requests SDs that stored a Lv.\,0 trigger within $\pm$\,32 $\mu$s of the event to send the waveform data to the PC for storage.

In addition, an external trigger from the TALE FD to the TALE SD, a so-called hybrid trigger system, was installed in 2018 to trigger low-energy cosmic ray air showers, which cannot be detected by the SD trigger alone.
When the DAQ PC receives the hybrid trigger, the Lv.\,0 trigger waveforms within $\pm$\,32 $\mu$s from the hybrid trigger timing are collected.

An accurate measurement of $\xmax$, which is the most sensitive parameter to cosmic ray mass, requires the use of the FD.
Furthermore, if the SD array simultaneously detects the same events as the FD, the hit information at the ground can constrain the shower core, allowing for a more precise determination of the shower geometry, specifically the core location and arrival direction of the air shower.
For these reasons, hybrid events, which are observed simultaneously by both FD and SDs, enable high-quality $\xmax$ measurements.

Therefore, in this study, we utilize approximately five years (from November 2017 to March 2023, with a total observation time of $\sim$2500 hours) of TALE hybrid data, including the MD telescope data, which, due to their lower field of view, can detect showers with deeper $\xmax$ values.
This dataset comprises approximately 9200 events, which represent the number of events that passed through the reconstruction and event selection procedures described in subsequent Sections.
Before the implementation of the hybrid trigger, hybrid events were identified by time matching within a coincidence window of 500\,$\mu$s.
After the trigger system was introduced, events were combined using the time information recorded by the hybrid trigger.

\begin{figure*}
\centering
\includegraphics[width=1.\linewidth]{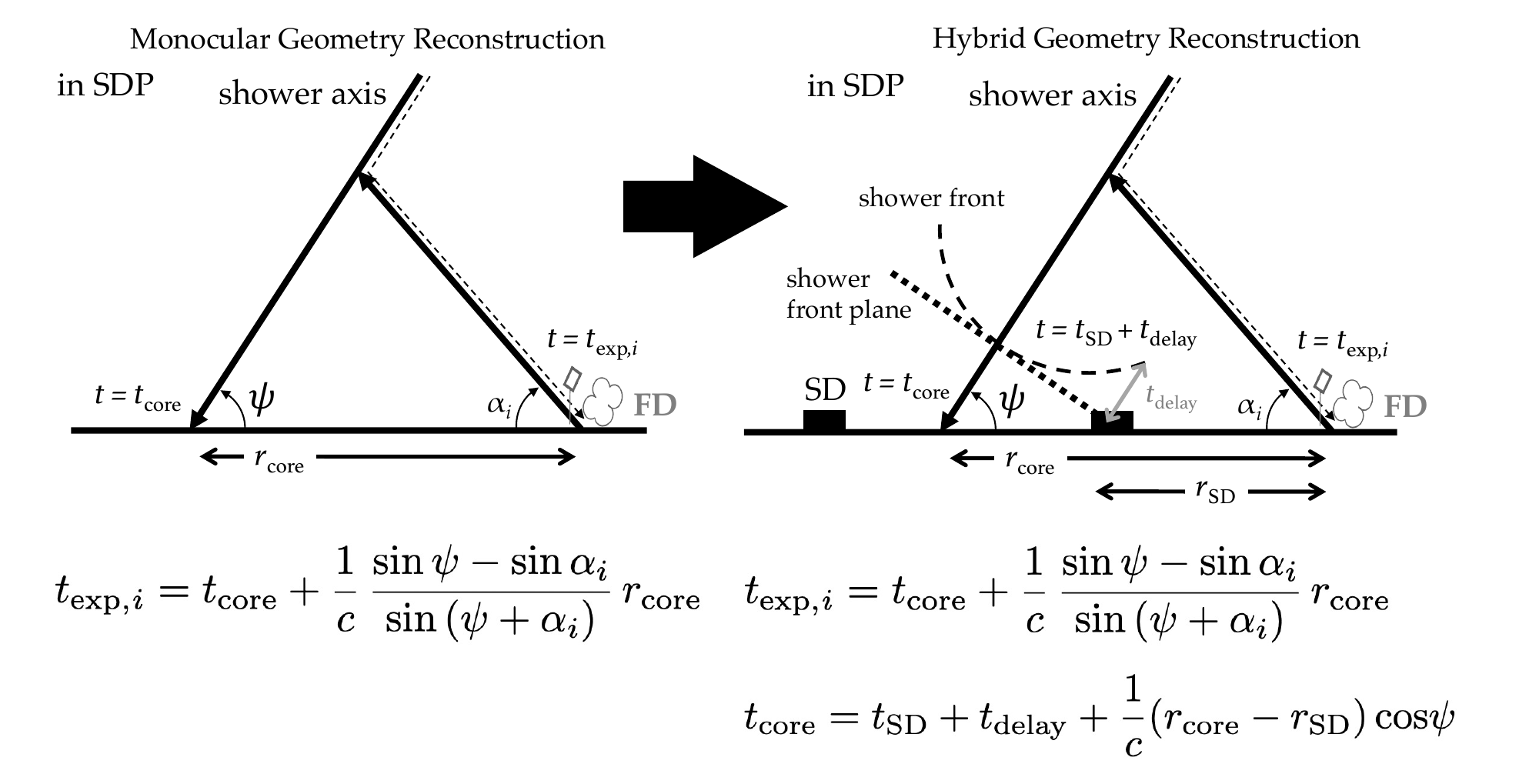}
\caption{\justifying{Schematic comparison of monocular and hybrid shower geometry reconstruction. The relations between the measured values, $t_{i}$ and $\alpha_{i}$, and the fitting parameters, which are $t_{\rm{core}}$, $r_{\rm{core}}$ and $\psi$, are shown. In the FD monocular measurement on the left side, the expected detection timing of each PMT is expressed using these three parameters. In the hybrid geometry reconstruction on the right side, the parameter $t_{\rm{core}}$ is removed by two observables measured by an SD, $t_{\rm{SD}}$ and $r_{\rm{SD}}$.}}
\label{fig:geometryReconstruction}
\end{figure*}

\section{Event Reconstruction} \label{sec:reconstruction}
The reconstruction processes for the hybrid events consist of the following three steps: 1) signal selection processes for both observed FD and SD signals, 2) determination of the shower detector plane (SDP), which is a plane including the shower axis and the FD location, and 3) reconstruction of the shower geometry and longitudinal profile.

At first, all triggered SD waveforms are scanned to calculate the arrival time of the air shower particles and the number of deposited particles in each SD.
For the FD, the PMTs to be used in the reconstruction are selected by rejecting spatially and temporally isolated PMTs from the shower image.

After the selected PMTs are defined, the SDP is calculated.
To determine the SDP, we minimize the $\chi^{2}$ function defined in Eq.\,(\ref{eq:chi2_sdp}), which takes into account the detected number of photoelectrons $N_{\rm{pe}}$ and the pointing directions of the PMTs: 
\begin{equation}
	\chi^2 = \sum_i \frac{(\vb*{n}_i \cdot \vb*{n}_{\mathrm{SDP}})^2 \cdot w_i}{\sigma_i^2}, 
 \label{eq:chi2_sdp}
\end{equation}
where $\vb*{n}_i$ is the pointing unit vector of the $i$th PMT, $\vb*{n}_{\mathrm{SDP}}$ is the unit normal vector of the SDP, $\sigma_i$ is the angular uncertainty related to the field of view of the PMT, which is to be set as a constant $\sigma_i = 1^\circ$ for all PMTs, and $w_i$ is the weight factor of the $i$th PMT, defined as $w_i = {N_{\mathrm{pe}, \, i}}/{\sum\limits_i N_{\mathrm{pe}, \, i}}$.
The use of photoelectron counts as weights is motivated by the fact that fluorescence and Cherenkov photons are emitted with a finite lateral spread and detected as a track with a characteristic width.
Since the detected photon intensity depends on the angular distance of each PMT from the shower axis, PMTs that record larger signals are expected to be located closer to the shower axis.

After we determine the SDP, the shower geometry can be calculated by fitting the arrival time of light at the telescope for each good PMT as a function of the viewing angle of the PMT in the SDP,
\begin{equation}
	t_{\rm{exp},\it{i}} =  t_{\mathrm{core}} + \frac{1}{c} \, \frac{\mathrm{sin}\,\psi - \mathrm{sin}\,\alpha_i}{\mathrm{sin}\,(\psi + \alpha_i)}\,r_{\mathrm{core}},
    \label{eq:2}
\end{equation}
where $\alpha_i$ is the viewing angle of the $\it{i}$th PMT on the SDP, $t_{\rm{core}}$ is the timing when the air shower reaches the ground, $r_{\rm{core}}$ is the distance from the FD station to the shower core, and $\psi$ is the shower inclination angle on the SDP (relations are shown in the left panel of Fig.\,\ref{fig:geometryReconstruction}).
This procedure describes the monocular reconstruction using only FD information.
In the case of hybrid reconstruction, where at least one SD near the shower core also records the shower signal, $t_{\rm{core}}$ is expressed by
\begin{equation}
    t_{\rm{core}} = t_{\rm{SD}} \,+\,t_{\rm{delay}} + \frac{1}{c}(r_{\rm{core}} - r_{\rm{SD}})\,\rm{cos}{\psi},
    \label{eq:3}
\end{equation}
where $t_{\rm{SD}}$ is the timing of the leading edge of the SD signal, and $r_{\rm{SD}}$ is the distance from the FD to the SD, as shown in the right panel of Fig.\,\ref{fig:geometryReconstruction}.
The term $t_{\rm delay}$ accounts for the curvature delay of the shower front; the shower front is not planar, and the curvature correction is essential for an accurate hybrid geometry reconstruction.
The curvature delay $t_{\rm delay}$ is parametrized using the modified Linsley shower-shape function \cite{Teshima:1986rq} and is applied to each event.

The parameter $t_{\rm{core}}$ can be removed by two observables measured by an SD, $t_{\rm{SD}}$ and $r_{\rm{SD}}$ if at least one SD simultaneously measures the air shower signal at the ground.
If multiple SDs detect signals on the ground, chi square fitting is performed on all candidates, and the geometry giving the least chi-square is selected.
This restriction in the position and hit timing of the shower core by the SD data leads to the improvement of the accuracy of the shower geometry determination.

\begin{figure}[t]
\begin{center}
\includegraphics[width=1.05\linewidth,trim=0.5cm 0cm 0cm 0.5cm]{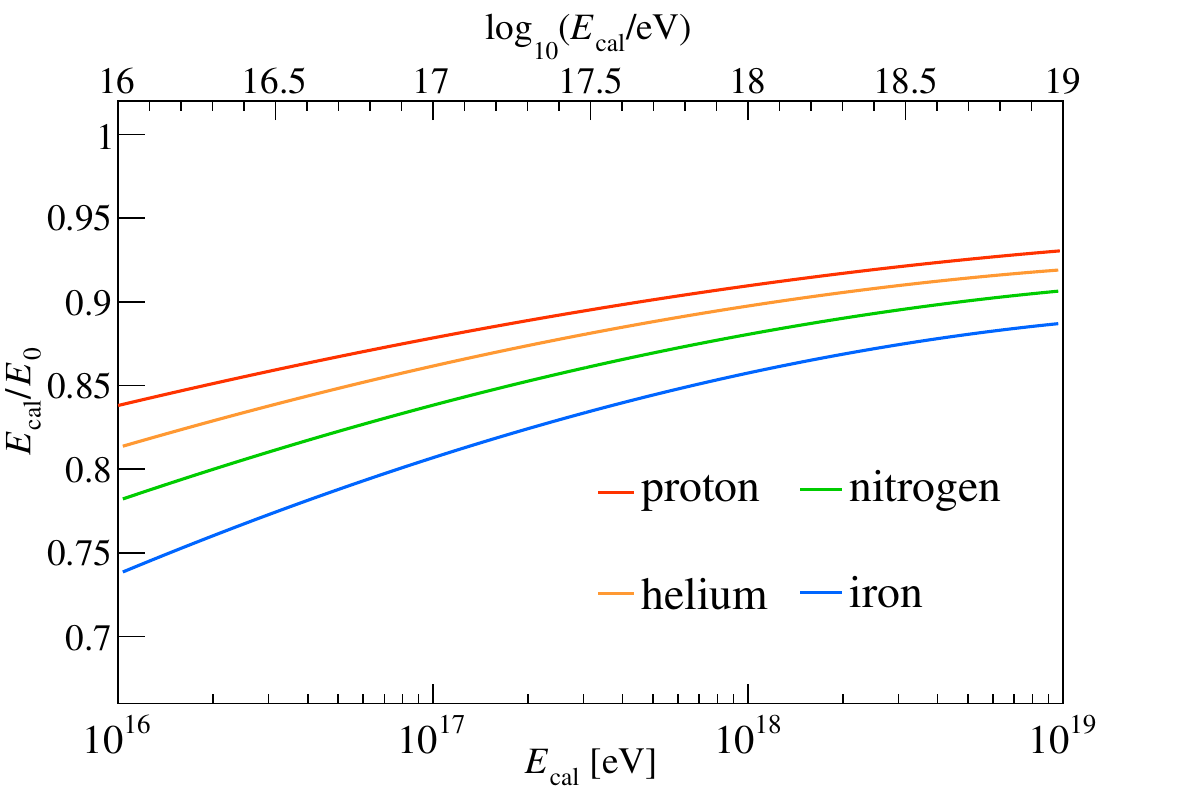}
\caption{\justifying{Primary dependence of the missing energy correction (ratio of the $E_{\rm{cal}}$ over the primary energy $E_{0}$) as a function of the $E_{\rm{cal}}$.}}
\label{fig:missingEnergyCorrection}
\vspace{-0.8cm}
\end{center}
\end{figure}
The shower longitudinal profile is then estimated along with the reconstructed shower axis by the following forward folding method.
\begin{figure*}
    \centering
    \includegraphics[width=\textwidth]{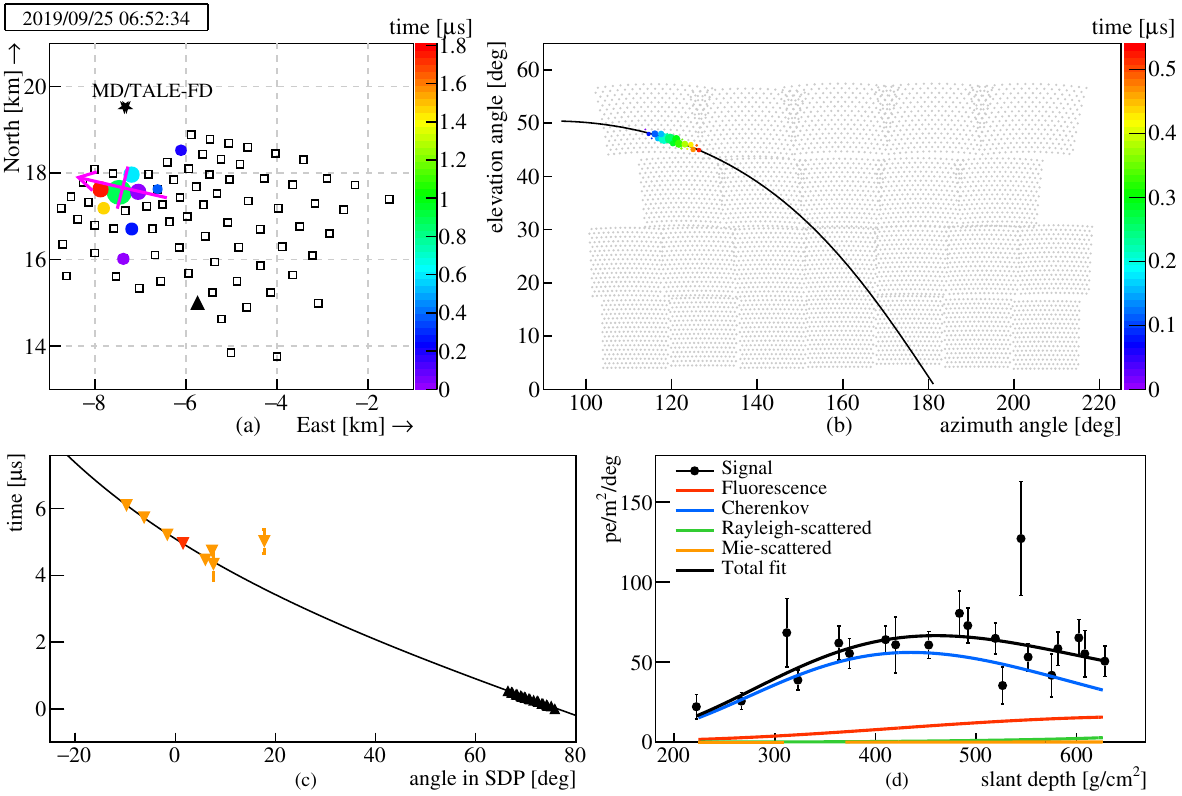}
    \caption{\justifying{Low energy Cherenkov-dominated event. 
    (a) Surface detector footprint, showing the triggered detectors (circles). The magenta arrow indicates the reconstructed azimuth angle of the shower direction projected onto the ground plane. The intersection of this arrow with the magenta line marks the reconstructed shower core position.
    (b) $\Fl$ telescope camera display, where the solid curve indicates the SDP.  
    In both panels, the color of each circle indicates the relative trigger timing, and the circle size proportional to the number of detected particles (a) or photons (b).
    (c) Time versus viewing angle fit for the selected PMTs (black points) and SDs (inverted triangles), where the solid curve represents the best-fit geometry.
    (d) Reconstructed longitudinal shower profile. Colored solid curves indicate the relative photon contributions.
    Details are described in the text.
    }}
    \label{fig:lowEnergyEvent}
\end{figure*}
The expected PMT-by-PMT signal measured by the FD is calculated by assuming the shower profile as the Gaisser-Hillas parametrization formula~\cite{bib:GH} described by Eq.\,(\ref{GHfunction}), taking into account the detector acceptance and atmospheric attenuation, 
\begin{equation}
\small
N(x) = N_{\rm{max}}\left(\frac{x - X_{0}}{X_{\rm{max}} - X_{0}}\right)^{\frac{X_{\rm{max}} - X_{0}}{\lambda}}\rm{exp} \left(\frac{\it{X}_{\rm{max}} - \it{x}}{\lambda}\right) ,
\label{GHfunction}
\end{equation}
where $N(x)$ is the number of charged particles at a given slant depth, $\it{x}$, $X_{\rm{max}}$ is the depth of shower maximum, $N_{\rm{max}}$ is the number of particles at the $X_{\rm{max}}$, $X_{0}$ is the depth of the first interaction, and $\lambda$ is an effective interaction length of the shower particles.

Once the best-fit shower profile is obtained, the calorimetric energy $E_{\rm cal}$ is calculated by integrating the longitudinal profile.  
To estimate the primary energy, a missing energy correction is then applied to account for the energy carried away mainly by neutral particles.
As shown in Fig.\,~\ref{fig:missingEnergyCorrection}, the correction varies with energy:
for proton primaries it decreases from about 15\% to 10\% between $10^{16.5}$ and $10^{18.5}$\,eV, and for iron from about 25\% to 15\%.  
For simulated events, we apply the missing energy correction corresponding to the true primary mass used in the generation.

For our observational data, an iterative approach is employed to determine the missing energy correction. Specifically, the observed $\xmax$ distribution is used to estimate the mean mass of cosmic rays at a given energy. 
Based on this estimate, the missing energy correction is estimated and applied to reconstruct the energy.
The reconstructed energy is then used to reevaluate the $\xmax$ distribution and the mean mass, which in turn provides a refined missing energy correction.
This process is repeated until the missing energy correction and the inferred mass composition converge.

\begin{figure*}
    \centering
    \includegraphics[width=\textwidth]{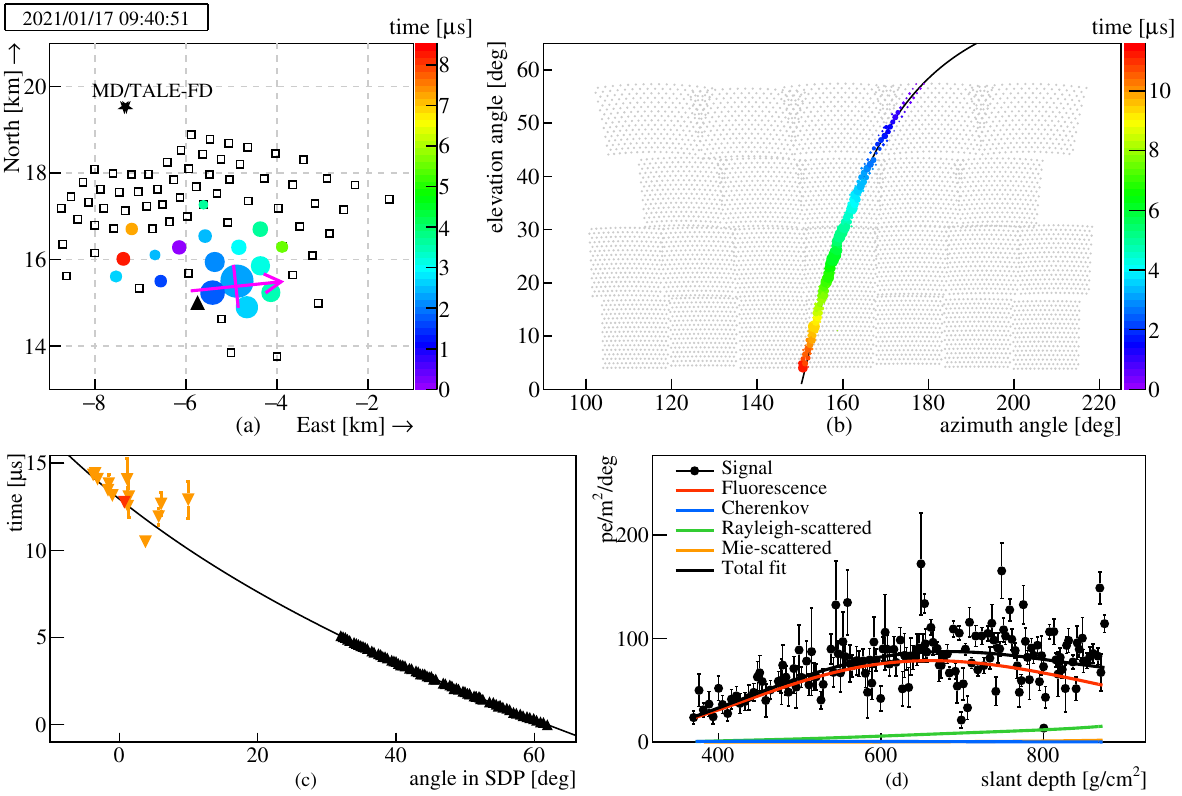}
    \caption{{Same as Fig.\,\ref{fig:lowEnergyEvent} but for a high energy fluorescence-dominated event.}}
    \label{fig:highEnergyEvent}
\end{figure*}
In the energy range below 10$^{17}$ eV, we use the TALE FD as an imaging air Cherenkov telescope to extend the energy threshold of the detector down to approximately 10$^{16}$ eV.
The $\Ch$ light produced by a shower shares a key characteristic with $\fl$ light, being directly proportional to the number of charged particles at any depth in the shower development.
This allows the observed Cherenkov signal to be used to infer shower properties such as energy and $\xmax$, similar to the use of fluorescence light.
However, unlike fluorescence light, which is emitted isotropically, Cherenkov light is strongly peaked forward along the shower direction and rapidly drops off as the viewing angle increases.

As a consequence, Cherenkov-dominated events are observed only when the shower propagates toward the telescopes, corresponding to viewing angles within a range of $\sim10^\circ$ or smaller.
In this work, such events are defined as those in which the estimated fluorescence light contributes less than 75\% of the total light signal, as determined from the shower profile reconstruction.
Because they are recorded much faster, these events exhibit shorter durations and smaller shower tracks, as shown in Fig.\,\ref{fig:lowEnergyEvent}.

In contrast, fluorescence-dominated events, defined as the contribution of estimated $\fl$ light signal over total signal greater than 75 $\%$, typically with energies above $10^{18}$ eV, are characterized by longer event durations and shower tracks (as shown in Fig.\,\ref{fig:highEnergyEvent}).
Due to these factors, reconstructing the shower geometry for lower-energy, Cherenkov-dominated events is challenging.
The short event duration and small shower track length provide only weak geometrical constraints, even when timing information from surface detectors at the ground is available.

Therefore, we processed all hybrid events using the profile-constrained geometry fit (PCGF), which simultaneously reconstructs both the shower geometry and profile.
This method, originally developed by the HiRes collaboration~\cite{bib:pcgfHiRes}, was also applied in the TALE FD monocular analysis~\cite{bib:taleSpectrum}.
In this study, we adapted the geometry reconstruction component of the PCGF to the hybrid geometry fit, further enhancing the accuracy of both the shower geometry and profile reconstruction compared to the monocular analysis.

Figures\,\ref{fig:lowEnergyEvent} and\,\ref{fig:highEnergyEvent} are examples of hybrid events observed by the TALE hybrid detector. 
Specifically, Fig.\,\ref{fig:lowEnergyEvent} shows a low-energy event dominated by Cherenkov light with a reconstructed energy of $10^{16.8}$\,eV, while Fig.\,\ref{fig:highEnergyEvent} illustrates a high-energy event dominated by fluorescence light with a reconstructed energy of $10^{18.1}$\,eV.
The triggered SDs are shown in the top-left panels of Figs.\,\ref{fig:lowEnergyEvent} and\,\ref{fig:highEnergyEvent}.
Marker size indicates the signal measured by the SD, and color indicates trigger time.
The arrow shows the reconstructed azimuthal angle of the shower direction, and the crossed point corresponds to the reconstructed shower core position.
The top-right panels of Figs.\,\ref{fig:lowEnergyEvent} and \ref{fig:highEnergyEvent} show the shower tracks of the hybrid events as seen by the fluorescence telescopes.
The marker size is proportional to the signal size, and the color indicates trigger time.
The gray dots represent each PMT direction.
The solid line is the SDP found by fitting Eq.\,(\ref{eq:chi2_sdp}).
The bottom-left panels of Figs.\,\ref{fig:lowEnergyEvent} and \ref{fig:highEnergyEvent} show the result of the hybrid geometry fit. 
The black triangles show the trigger time and viewing angle of the FD PMTs that observed the passage of the shower.
The inverted triangles indicate the corresponding signals recorded by SDs.  
Among these SD stations, the red inverted triangle marks the one used in the hybrid reconstruction, chosen as the SD yielding the smallest $\chi^2$ when multiple stations triggered.
The solid curve shows the best-fit geometry obtained from Eqs\,(\ref{eq:2}) and (\ref{eq:3}).
The bottom-right panels of Figs.\,\ref{fig:lowEnergyEvent} and\,\ref{fig:highEnergyEvent} show the reconstructed shower profiles as a function of depth for each PMT along the SDP.
Contributions from fluorescence light (red), direct Cherenkov light (blue), and scattered Cherenkov light—Rayleigh (yellow) and Mie (green)—are indicated.
The black solid line represents the sum of all components for the best-fit profile, and the black points show the observed signals.

For events like the one shown in Fig.\,\ref{fig:highEnergyEvent}, where the shower track is long and extends to lower elevations below $31^\circ$, the signals from the MD telescope covering those angles are also included in the shower profile reconstruction.
This ensures a more accurate reconstruction of the shower development.

To ensure the quality of reconstructed events and to enable reliable comparisons between the observational data and Monte Carlo simulations, we apply identical selection criteria to both observed and simulated events.
As described earlier in this Section, fluorescence-dominated (FL) and Cherenkov-dominated (CL) events exhibit distinct observational characteristics, and therefore different quality selections are applied to each event type to remove poorly reconstructed events and ensure good detector resolution.
The selection criteria used in this analysis are summarized in Table\,\ref{tb:qualityCuts}.
In addition to listing the criteria themselves, Table\,\ref{tb:qualityCuts} also presents, in the rightmost column, the selection efficiency at each step, defined as the fraction of events surviving relative to the previous selection.

\begin{figure}[t]
\begin{center}
\includegraphics[width=1.\linewidth]{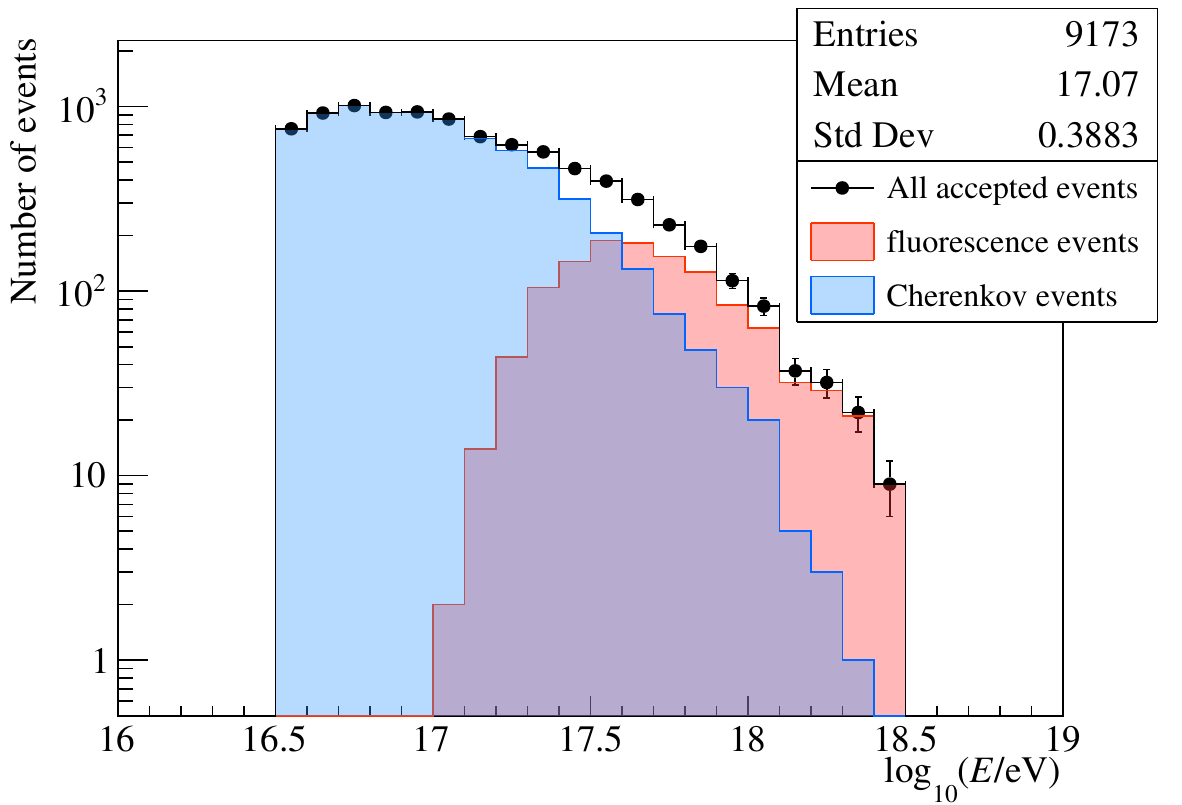}
\caption{\justifying{
Energy distribution of events obtained with the TALE hybrid observation. The distribution of all accepted events is shown by black points. Cherenkov-dominated and fluorescence-dominated events are shown in blue and red histograms, respectively.
}}
\label{fig:eventDistribution}
\end{center}
\end{figure}

For both event sets, we require the following conditions: successful reconstruction of the shower geometry and longitudinal profile, good weather conditions, no saturated signals in the shower track observed by the FD, and $\xmax$ within the field of view of FD.
We also impose a minimum total number of photoelectrons: greater than 1000 for CL events and greater than 2000 for FL events.  
For CL events only, additional requirements are applied: an event duration longer than 100\,ns, more than 10 PMTs in the shower track, and a minimum viewing angle, defined as the minimum value of the viewing angle $\alpha_i$ among the selected PMTs for each event, greater than $2.5^\circ$.
After applying these selection criteria to the observational data, a total of 9173 high-quality hybrid events remain.  
The energy distribution of accepted events is shown in Fig,\,\ref{fig:eventDistribution} by black points.
In addition to the overall energy distribution, the relative contributions of CL and FL events as a function of energy are shown in Fig,\,\ref{fig:eventDistribution}.
At lower energies, CL events constitute nearly all accepted events, because the flux of isotropically emitted fluorescence light becomes too low for effective detection below $10^{17}$\,eV.
FL events therefore appear only above this energy, where the fluorescence signal is sufficiently strong to allow reliable reconstruction.

\begin{table}[t]
\begin{center}
\caption{\justifying{Event selection criteria and selection efficiency with respect to the previous cut.}}
  \begin{tabular}{l c c c}
   \hline\hline
    Criteria & CL & FL & $\varepsilon$\,[\%]\\ \hline
    \makecell[l]{Shower geometry \\reconstruction successful} & \multicolumn{2}{c}{applied} & -\\
    \makecell[l]{Shower profile \\reconstruction successful} & \multicolumn{2}{c}{applied} & 99.9\\
    Good weather data & \multicolumn{2}{c}{applied} & 90.3\\
    No saturated PMTs in FD & \multicolumn{2}{c}{applied} & 96.7\\
    $X_{\rm{max}}$ bracketing cut &  \multicolumn{2}{c}{applied} & 75.7\\
    Event duration [ns] & $\textgreater$ 100 & -  & 99.4\\
    \# of PMTs & $\textgreater$ 10 & -  &97.2\\
    Minimum viewing angle [deg] & $\textgreater$ 2.5 & -  & 98.8\\
    \# of photoelectrons & $\textgreater$ 1000 & $\textgreater$ 2000  & 82.7 \\
    $10^{16.5}\,\textless\,\it{E}/\rm{eV}\,\textless\,10^{18.5}$ & \multicolumn{2}{c}{applied} & 82.4 \\ \hline
  \end{tabular}
\label{tb:qualityCuts}
\end{center}
\end{table}

\begin{figure*}
\begin{subfigure}{\textwidth}
\begin{minipage}[l]{0.495\hsize}
\includegraphics[width=0.95\linewidth]{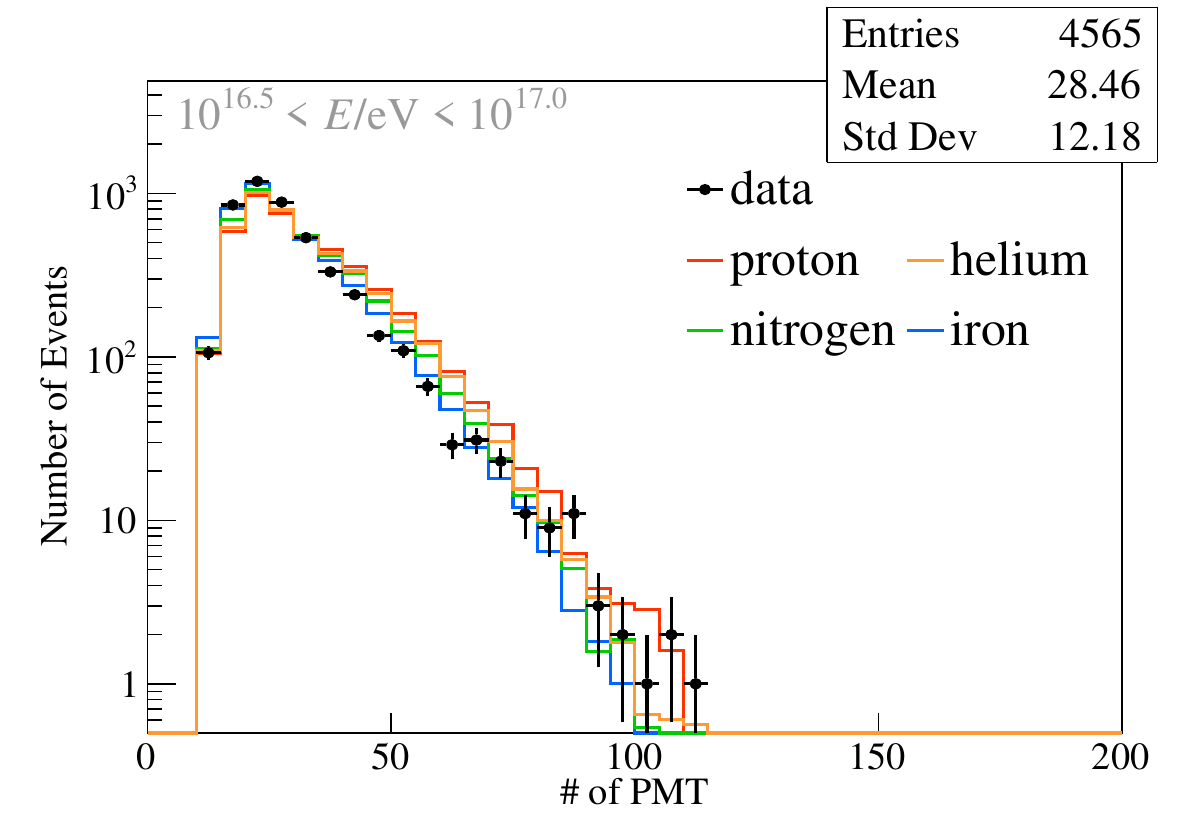}
\end{minipage}
\begin{minipage}[r]{0.495\hsize}
\includegraphics[width=0.95\linewidth]{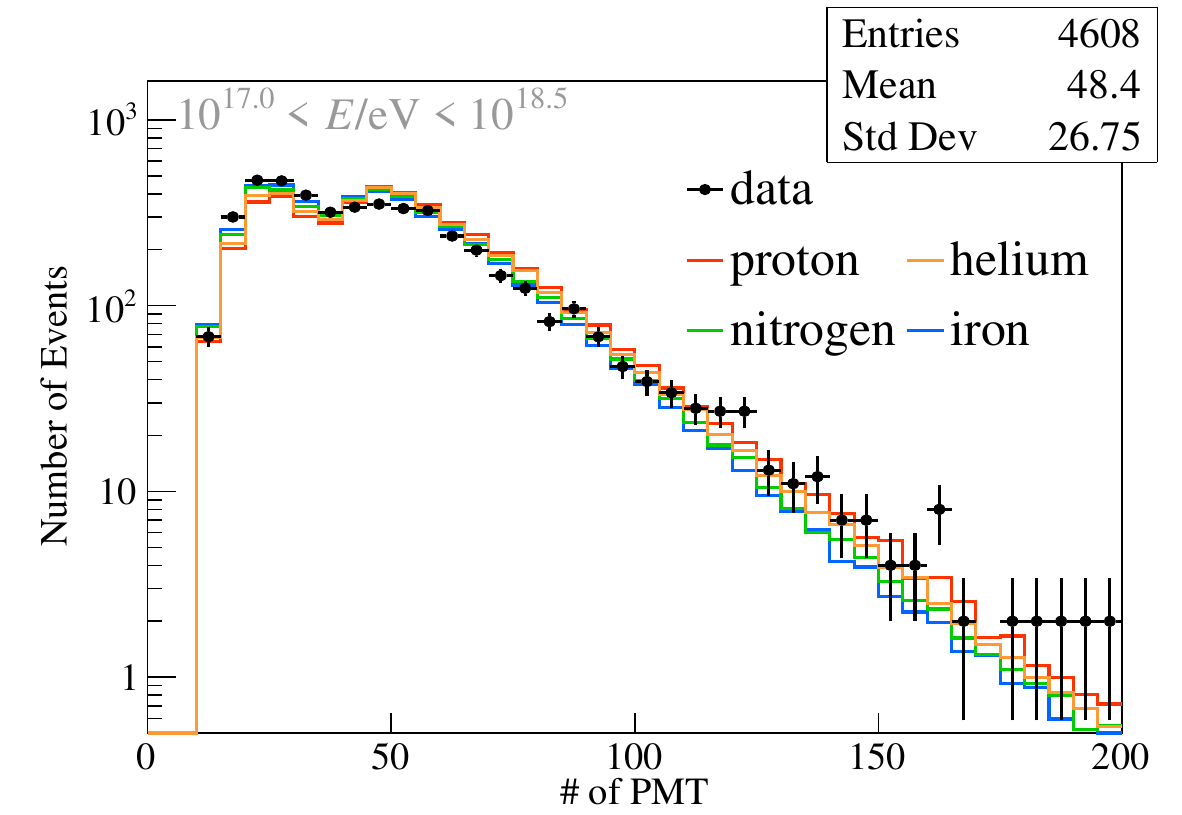}
\end{minipage}
\subcaption{Number of PMTs}
\label{fig:dataMCcomparison01}
\end{subfigure}
\vspace{+5.5mm}
\begin{subfigure}{\textwidth}
\begin{minipage}[l]{0.495\hsize}
\includegraphics[width=0.95\linewidth]{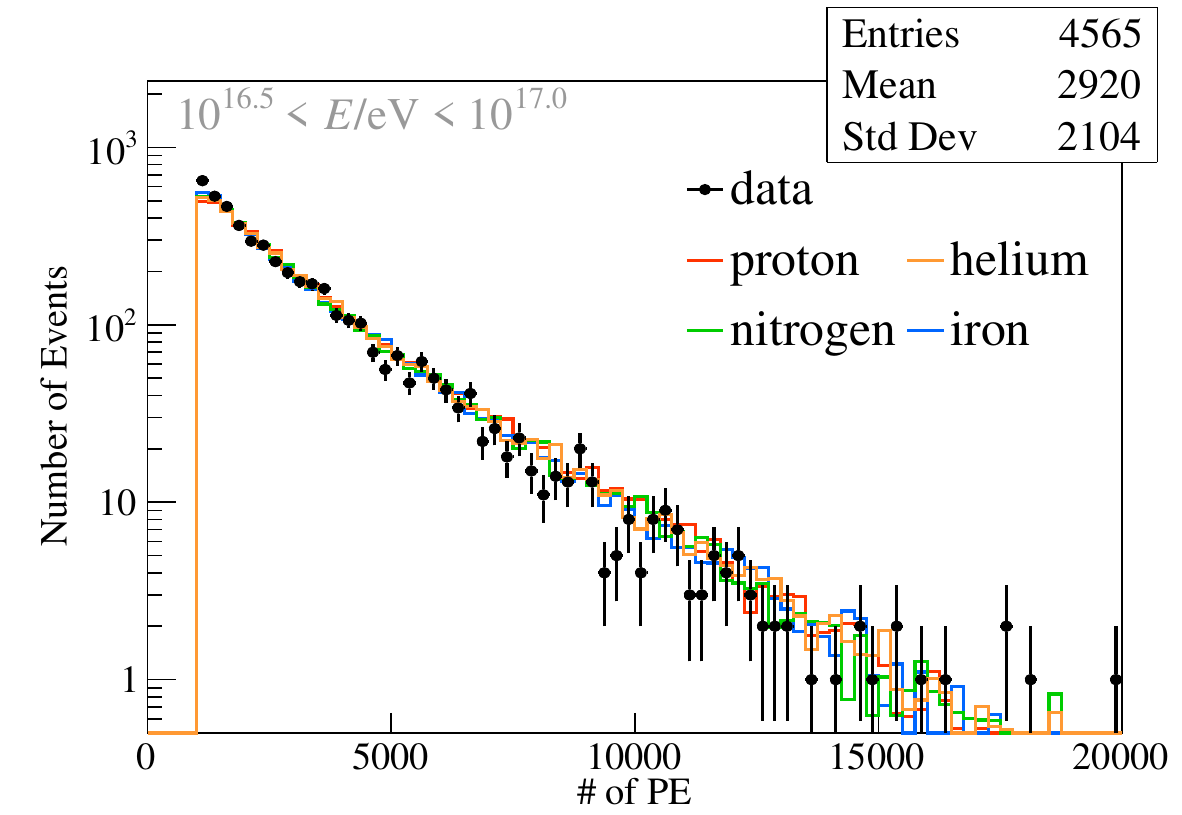}
\end{minipage}
\begin{minipage}[r]{0.495\hsize}
\includegraphics[width=0.95\linewidth]{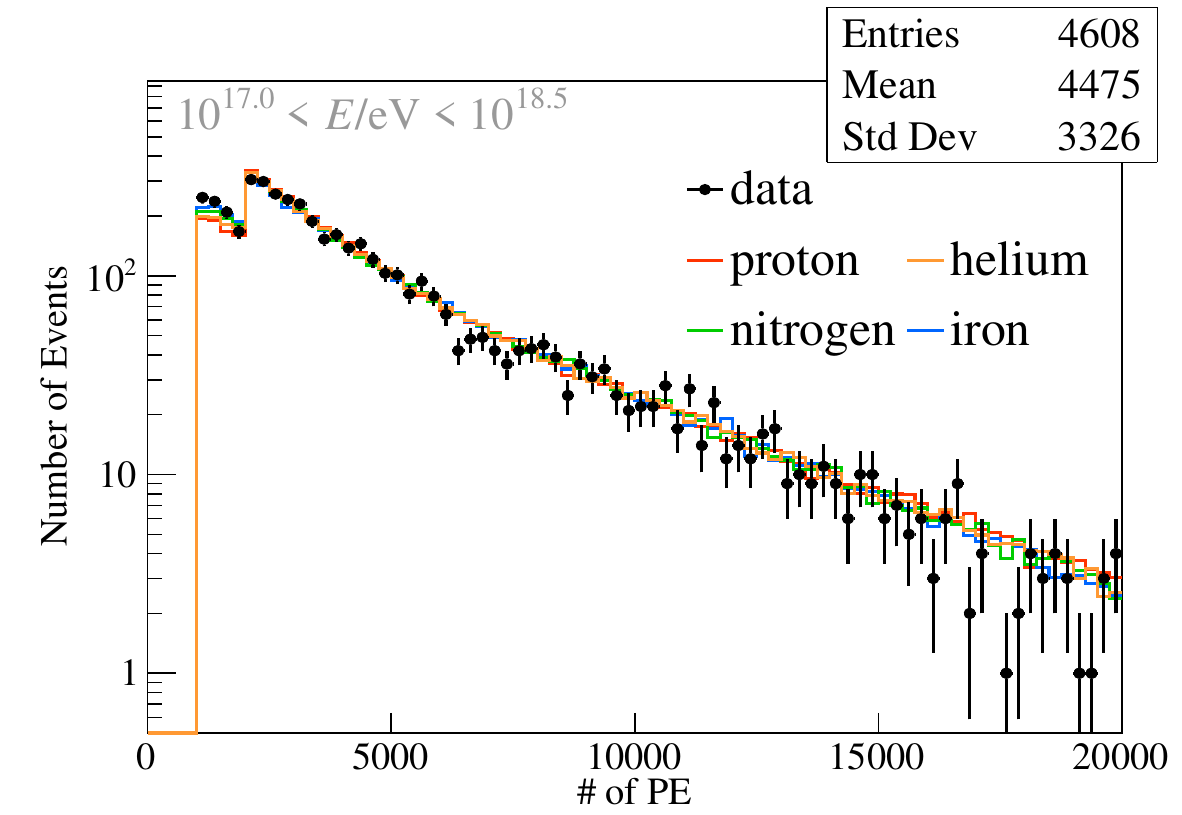}
\end{minipage}
\subcaption{Number of photoelectrons}
\label{fig:dataMCcomparison02}
\end{subfigure}
    \caption{\justifying{Data and Monte Carlo comparison plots I. Upper panels: (a) the number of PMTs. Lower panels: (b) the number of photoelectrons.
    Black points with error bars represent the observational data, while the colored histograms show the corresponding Monte Carlo simulations normalized to the same number of entries as the data.
    The datasets are binned by the reconstructed energy at $10^{17}$ eV, where the left panels correspond to the lower energy side and the right panels to the higher-energy side.
    }}
    \label{fig:dataMCcomparison001}
\end{figure*}

\section{Simulation} \label{sec:simulation}
We ran a Monte Carlo simulation to evaluate our detector resolution and response.
The TALE hybrid Monte Carlo event generation package consists of two components: the air shower generation part and the detector simulation part.
The Monte Carlo events are stored in the same format as the observational data and reconstructed in the same manner.

Air showers were first generated using the CORSIKA simulation tool~\cite{bib:CORSIKA} with the hadron interaction models FLUKA~\cite{Ferrari:2005zk} and QGSJet\,II-04~\cite{bib:qgs04} in low- and high-energy ranges, respectively.
Primary particles were simulated for four types of nuclei: proton, helium, nitrogen, and iron.
These four nuclei serve as the basis for the mass composition analysis described in Sec.\,\ref{sec:result}, and their choice is motivated as follows.  
The proton is treated as the lightest extreme, and iron represents the heaviest mass group.
Nitrogen serves as a representative of the intermediate‐mass nuclei, particularly the carbon, nitrogen, and oxygen nuclei (CNO) group.
The inclusion of helium provides an approximately uniform separation in $\mxmax$ of about $25$–$30\,\mathrm{g/cm^2}$, which is comparable to the detector resolution.
This choice also results in an approximately uniform spacing in $\ln A$ among the representative species.
Given this detector resolution in $\xmax$, introducing additional mass groups would not improve the sensitivity of the analysis and would instead destabilize the fit. 
The four-component scheme therefore offers a practical balance between physical representativeness and the robustness of the fitting procedure.

Simulated events were generated with an isotropic distribution in zenith angle from $0^{\circ}$ to $65^{\circ}$, and a uniform distribution in azimuthal angle from $0^{\circ}$ to $360^{\circ}$, respectively.
The secondary particles at the ground level obtained from these simulations were then used as input to the surface detector simulation.

\begin{figure*}
\begin{subfigure}{\textwidth}
\begin{minipage}[l]{0.495\hsize}
\includegraphics[width=0.95\linewidth]{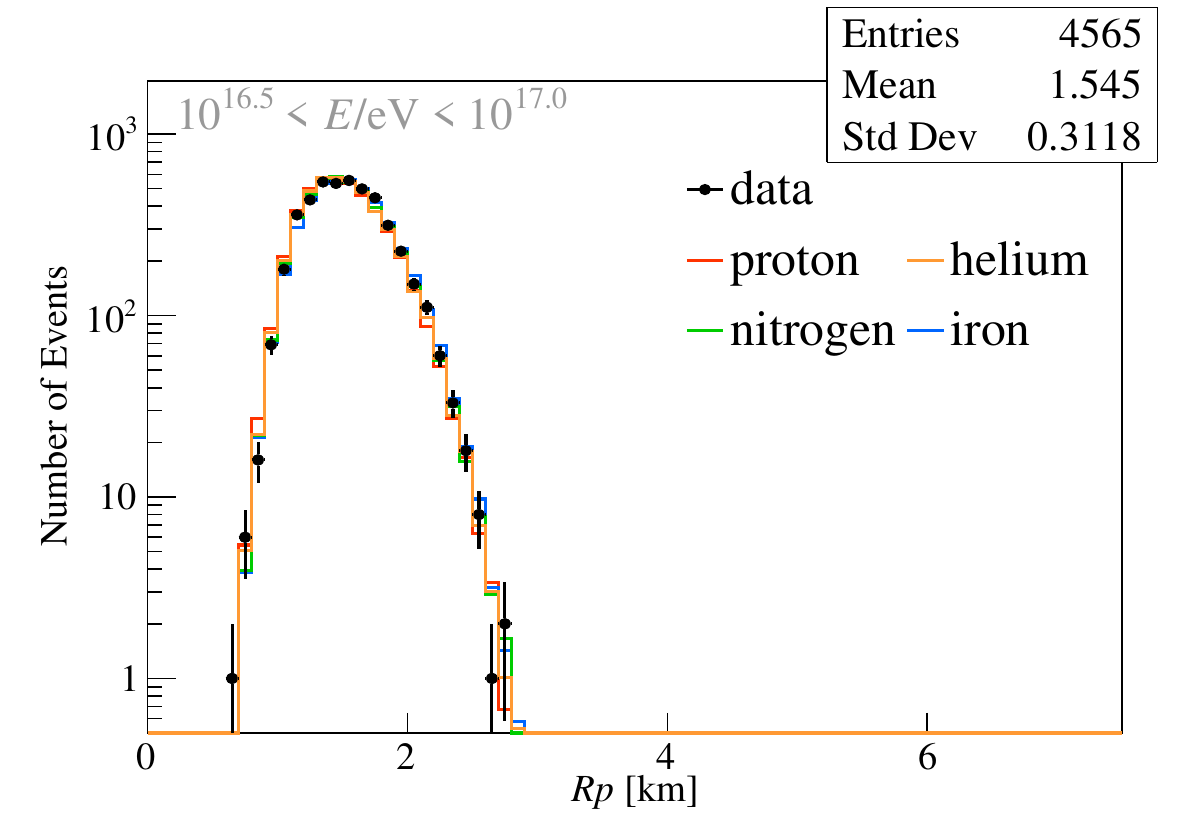}
\end{minipage}
\hfill
\begin{minipage}[r]{0.495\hsize}
\includegraphics[width=0.95\linewidth]{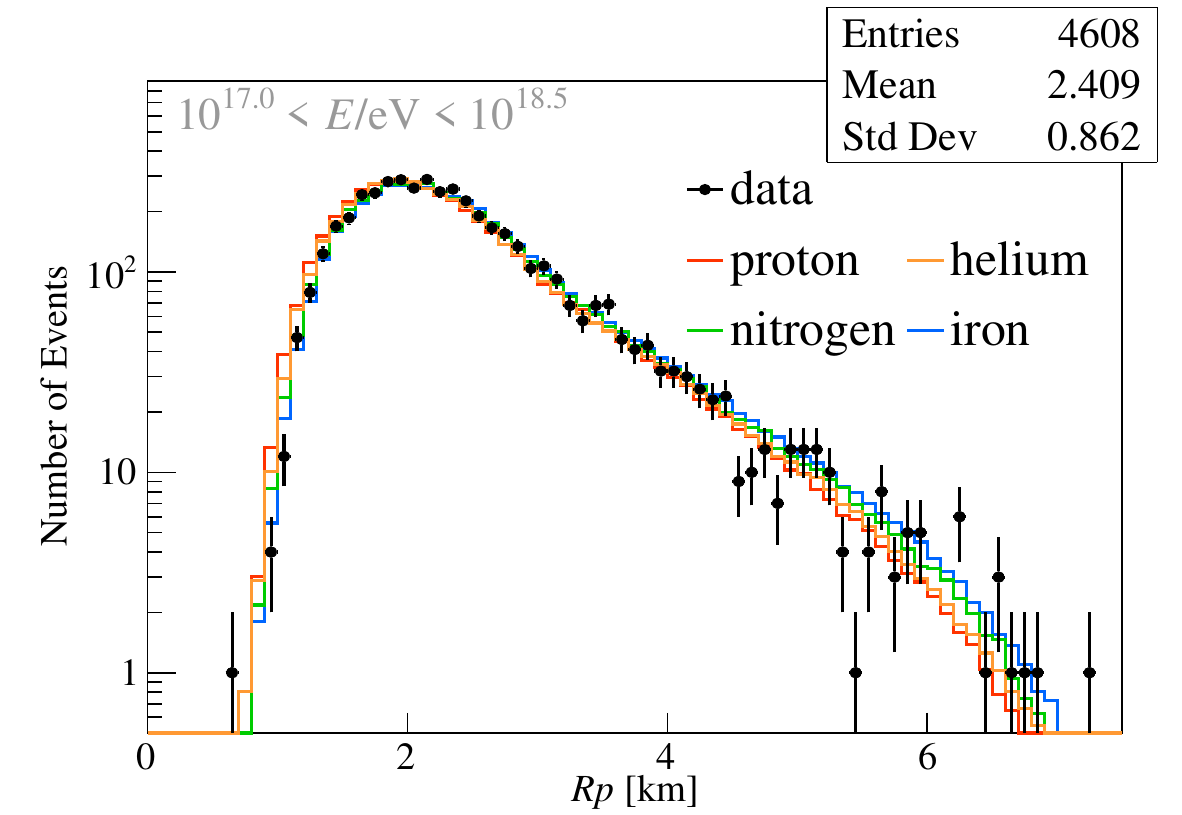}
\end{minipage}
\subcaption{Shower impact parameter, $R_{p}$}
\label{fig:dataMCcomparison03}
\end{subfigure}
\vspace{+2.5mm}
\begin{subfigure}{\textwidth}
\begin{minipage}[l]{0.495\hsize}
\includegraphics[width=0.95\linewidth]{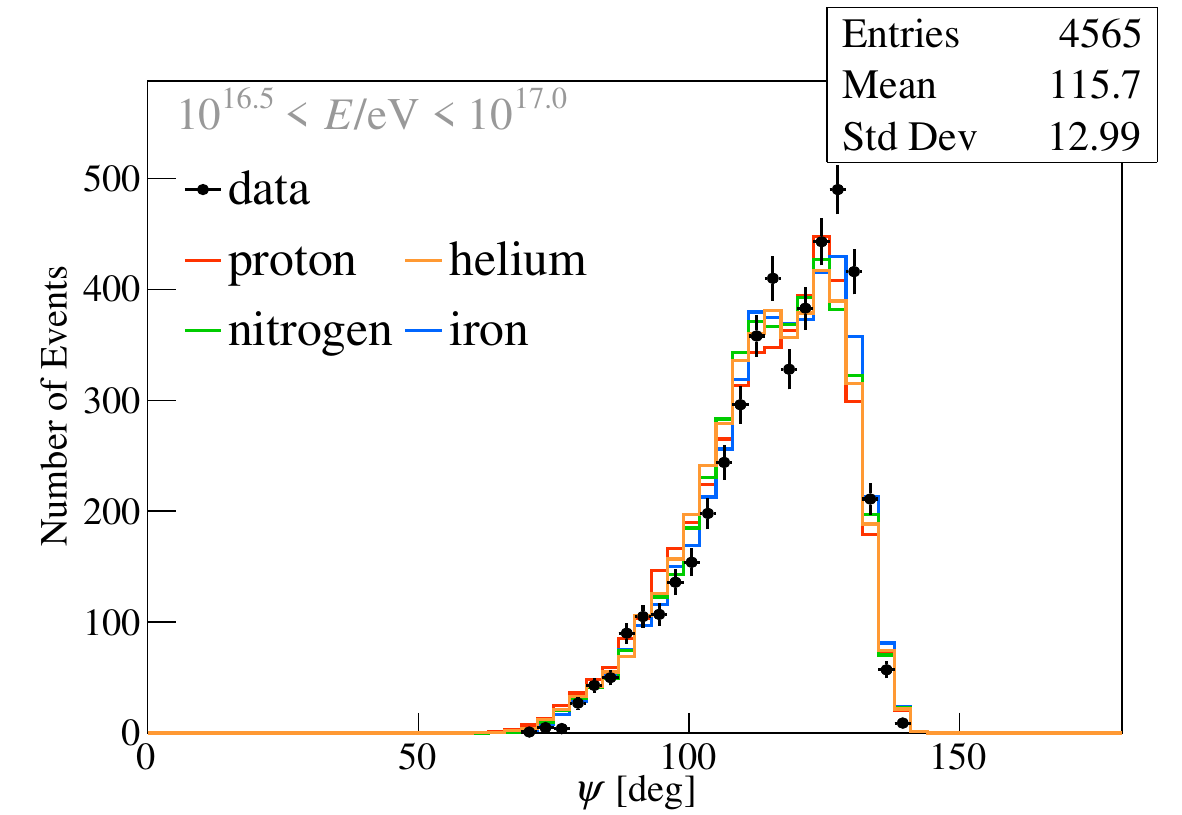}
\end{minipage}
\begin{minipage}[r]{0.495\hsize}
\includegraphics[width=0.95\linewidth]{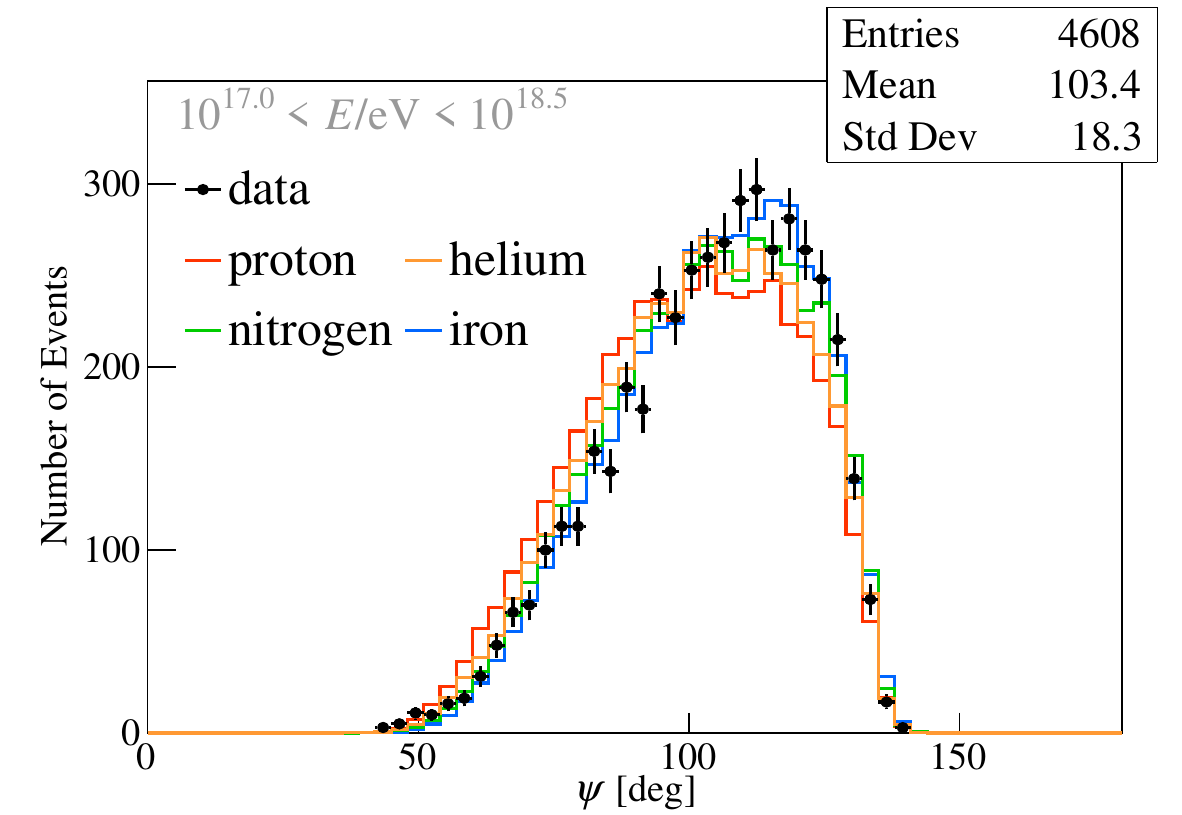}
\end{minipage}
\subcaption{$\psi$ angle in the SDP}
\label{fig:dataMCcomparison04}
\end{subfigure}
    \caption{\justifying{Data and Monte Carlo comparison plots II. Upper panels: (a) the impact parameter $R_{p}$. Lower panels: (b) the shower inclination angle in the SDP, $\psi$.}}
    \label{fig:dataMCcomparison002}
\end{figure*}

The energy deposition in each SD is calculated using the response function constructed with the \textsc{Geant4} simulation, as in the TA SD~\cite{bib:benMC}.
An event is accepted if at least one SD records a waveform within 32 $\mu$s of the event trigger during the period when the hybrid trigger is active.
Otherwise, the self-trigger condition of the SD array is applied.

The longitudinal energy deposition profiles simulated by CORSIKA are used to calculate the telescope response.
We use a fluorescence yield model that based on the absolute fluorescence yield measured by~\cite{bib:kakimoto} and the fluorescence line spectrum measured by the FLASH experiment~\cite{bib:FLASH}.
The Cherenkov light production and emission angle distribution follow the parametrization given by Nerling et al.~\cite{bib:nerling}.
The emission points of fluorescence and Cherenkov photons at a given depth are determined by taking into account the lateral spread of shower particles modeled in~\cite{bib:NKGlike}.

The attenuation and scattering of fluorescence and Cherenkov light during their propagation from the shower to the telescopes are also included in the simulation.
The atmospheric profiles, such as the pressure and atmospheric density, are provided by the GDAS database, which is updated every three hours~\cite{bib:GDAS}.
The atmospheric aerosol density used in this work is the average value of a vertical aerosol optical depth (VAOD) of 0.04.
This value corresponds to an attenuation length of 25\,km.

The telescope response is reproduced in the simulation by implementing detailed structures, including the building and supporting materials, the PMT camera response, and triggering logic.

\begin{figure*}
\begin{subfigure}{\textwidth}
\begin{minipage}[l]{0.495\hsize}
\includegraphics[width=0.95\linewidth]{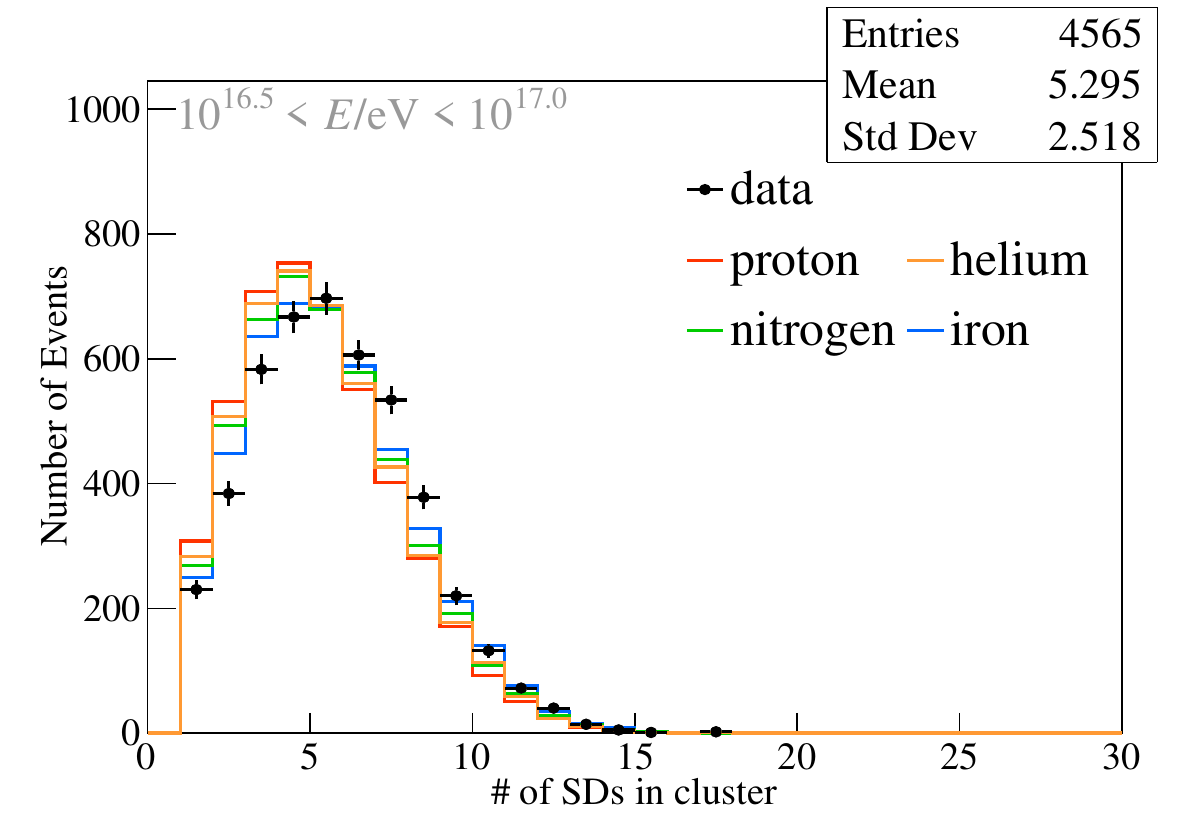}
\end{minipage}
\hfill
\begin{minipage}[r]{0.495\hsize}
\includegraphics[width=0.95\linewidth]{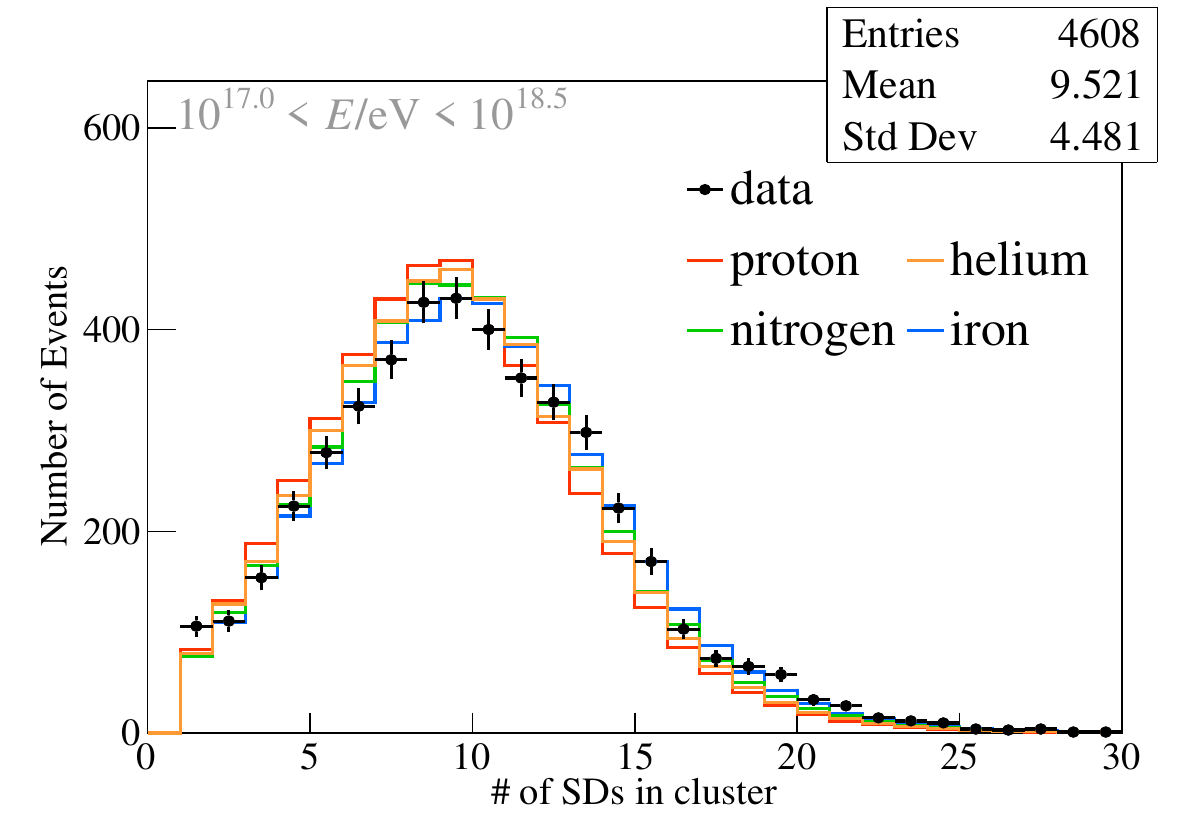}
\end{minipage}
\subcaption{Number of SDs}
\label{fig:dataMCcomparison05}
\end{subfigure}
\vspace{+2.5mm}
\begin{subfigure}{\textwidth}
\begin{minipage}[l]{0.495\hsize}
\includegraphics[width=0.95\linewidth]{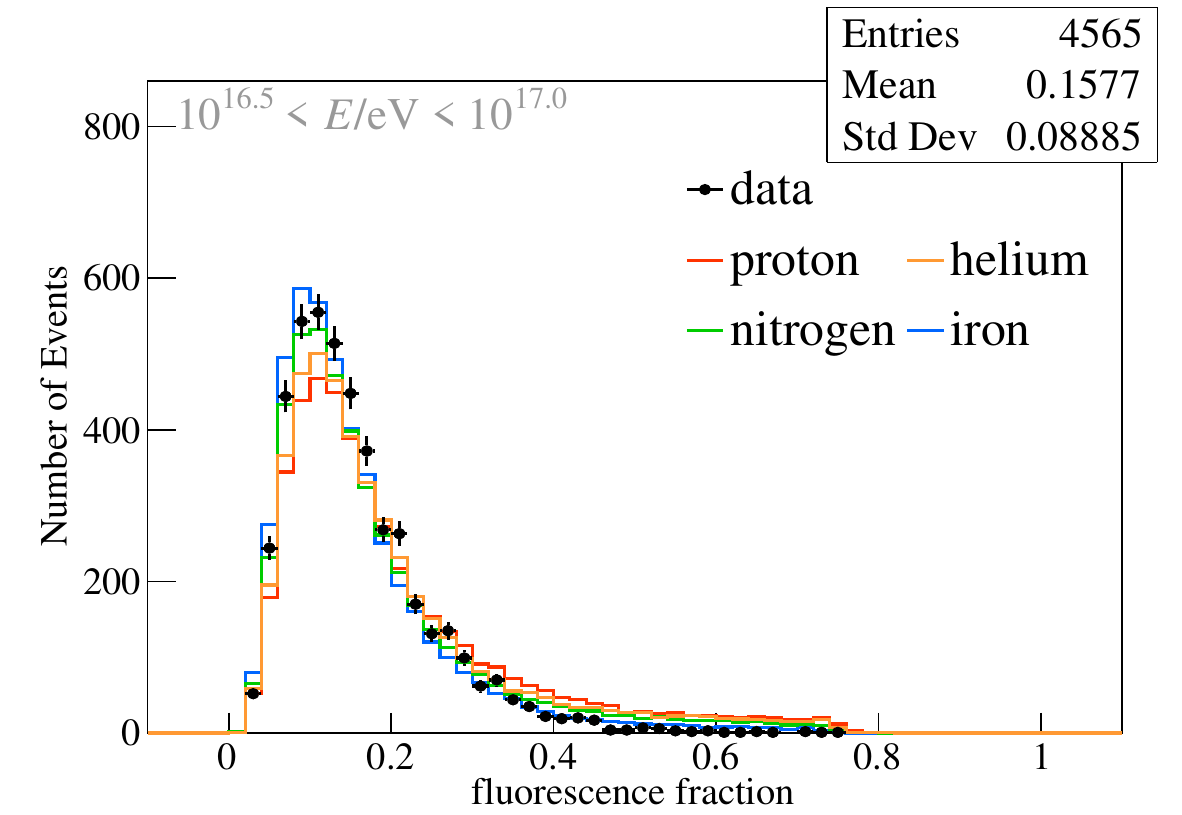}
\end{minipage}
\hfill
\begin{minipage}[r]{0.495\hsize}
\includegraphics[width=0.95\linewidth]{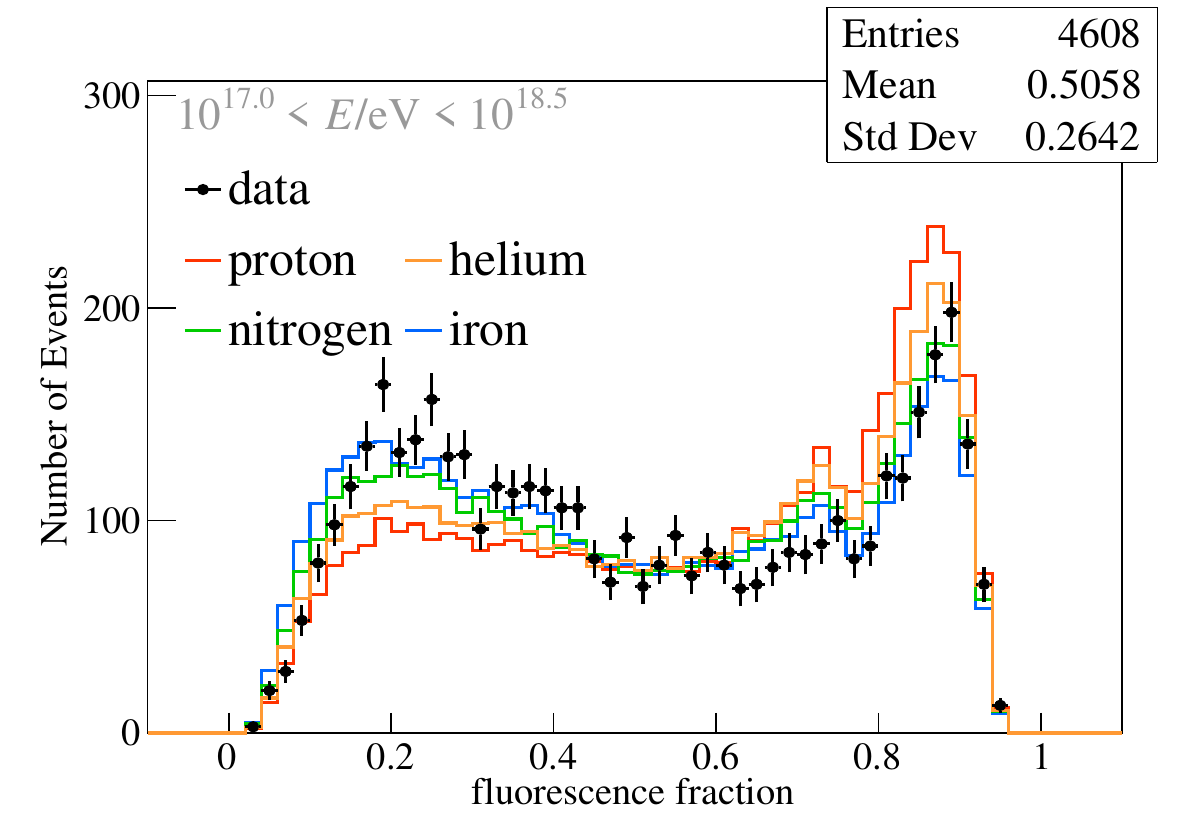}
\end{minipage}
\subcaption{$\Fl$ fraction}
\label{fig:dataMCcomparison06}
\end{subfigure}
    \caption{\justifying{Data and Monte Carlo comparison plots III. Upper panels: (a) the number of clustering SDs. Lower panels: (b) the fluorescence fraction.}
    }
    \label{fig:dataMCcomparison003}
\end{figure*}

To assess the validity of the simulation, we compare reconstructed parameters between the observational data and Monte Carlo datasets.
These comparisons for four different primaries with reconstructed energies above $10^{16.5}$\,eV are summarized in Figs.\,\ref{fig:dataMCcomparison001}, \ref{fig:dataMCcomparison002}, and \ref{fig:dataMCcomparison003}.  

The number of PMTs in the shower track (Fig.\,\ref{fig:dataMCcomparison01}), the number of photoelectrons (Fig.\,\ref{fig:dataMCcomparison02}),  the impact parameter, $R_{p}$ (Fig.\,\ref{fig:dataMCcomparison03}), the shower inclination angle in the SDP $\psi$ (Figure\,\ref{fig:dataMCcomparison04}), the number of SDs coincident in space and time (Fig.\,\ref{fig:dataMCcomparison05}), and the $\fl$ fraction (Fig.\,\ref{fig:dataMCcomparison06}) are displayed.
In the comparisons, we split the dataset with a reconstructed energy at $E$ = $10^{17}$ eV to show that the data and Monte Carlo distributions are in good agreement for both CL and FL events.
All Monte Carlo event histograms are normalized to the entries of observational data.
We find that the comparisons show overall consistency across all observables, including those from the FD, SD, and geometric parameters of the air shower.
In addition, the simulation datasets are used to determine our detector resolution.
We summarize the reconstruction biases and resolutions of the zenith angle $\theta$, azimuth angle $\phi$, the impact parameter $R_{p}$, the shower inclination angle in the SDP $\psi$, the shower maximum $\xmax$, and shower energy in Table\,\ref{table:resolution}.
We obtain the resolutions of $\sim$ 35\,m in $R_{p}$, $\sim 1^{\circ}$  in $\psi$ angle, $\sim$ 30 g/cm$^{2}$ in $\xmax$ and $\sim 10 \%$ in energy.
These values are evaluated over all simulated events above $10^{16.5}$\,eV, taking into account the expected energy spectrum, and therefore represent average resolutions over the full analysis range.
\begin{table}[t]
\caption{\justifying{Bias and resolution of the TALE hybrid measurement for events with energies above $10^{16.5}$ eV.}}
\begin{tabular}{l c c c c c c c c c c c c}
\hline \hline
  & &\multicolumn{2}{c}{proton} & & \multicolumn{2}{c}{helium} & &\multicolumn{2}{c}{nitrogen} & & \multicolumn{2}{c}{iron} \\
\cline{3-4} \cline{6-7} \cline{9-10} \cline{12-13}
   & & bias & res. & & bias & res. & & bias & res. & & bias & res. \\
\hline
$\theta$ [deg] && \,0.03 & 0.43 & & -0.03 & 0.43 & & -0.07 & 0.41 & & -0.10 & 0.41 \\
$\phi$ [deg]   && \,0.01 & 1.40 & & -0.01 & 1.21 & & -0.01 & 1.06 & & -0.02 & 0.91 \\
$R_{p}$ [m] && \,3.7  & 32.0 & & \,4.4 & 33.4 & & \,4.6 & 35.4 & & \,5.3 & 37.3 \\
$\psi$ [deg]   && \,0.34 & 0.90 & & \,0.23 & 0.89 & & \,0.18 & 0.89 & &  \,0.12 & 0.87 \\
$\xmax$ [g/cm$^{2}$] && -2.5 & 31.0 & & \,3.7 & 29.8 & & \,2.9 & 29.1 & & -1.6 & 28.6 \\
Energy [$\%$] && 1.8 & 11.8 & & 1.9 & 11.2 & & 0.6 & 10.5 & & -0.6 & 10.4 \\
\hline
\end{tabular}
\label{table:resolution}
\end{table}

\begin{figure*}
    \centering
    \begin{minipage}[l]{0.495\textwidth}    
        \centering
        \includegraphics[width=\linewidth]{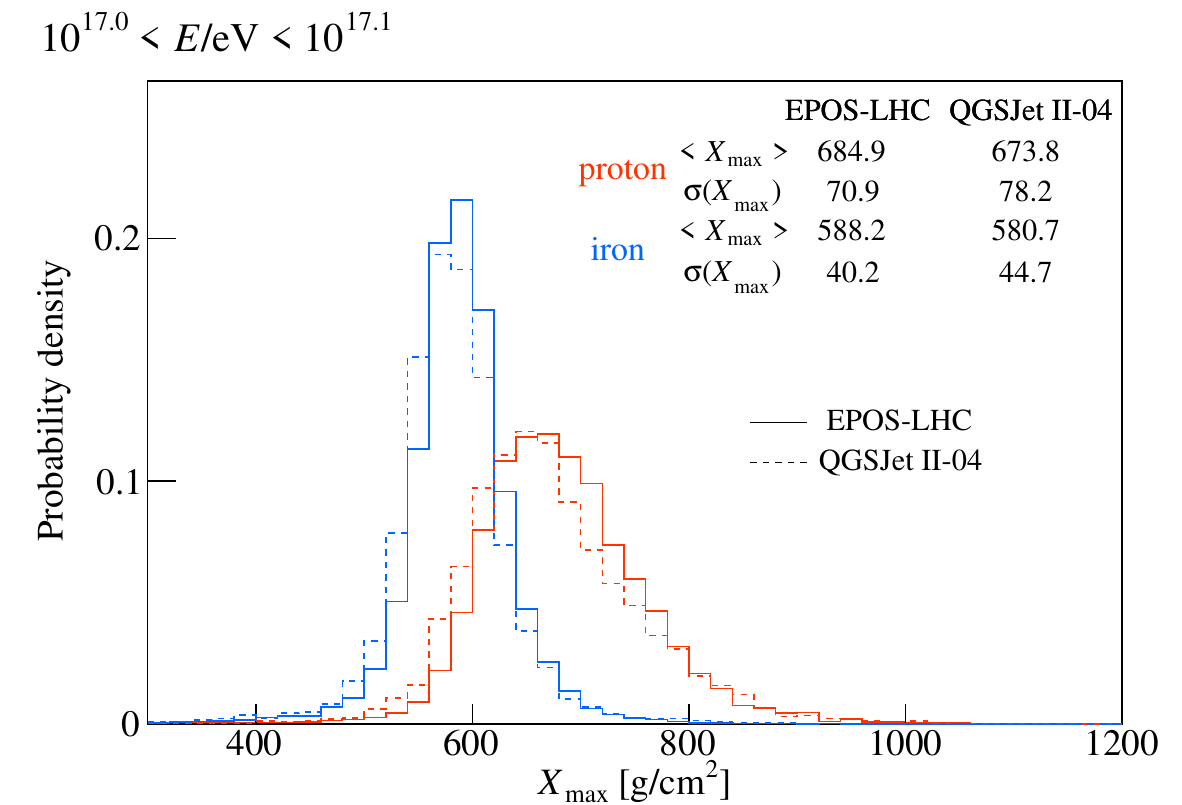}
    \end{minipage}
    \hfill
    \begin{minipage}[r]{0.495\textwidth}    
        \centering
        \includegraphics[width=\linewidth]{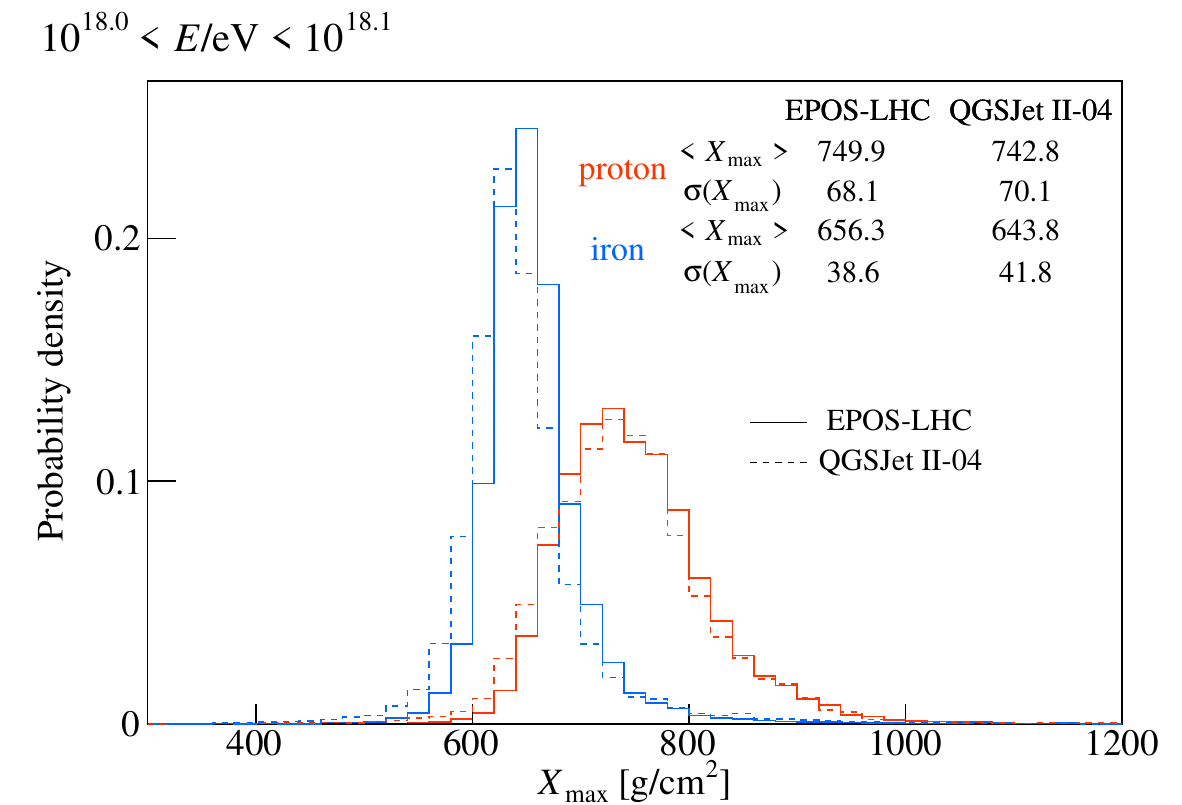}
    \end{minipage}
    \caption{\justifying{
    Comparison of the reconstructed $\xmax$ distributions obtained using two hadronic interaction models, QGSJet\,II-04 and EPOS-LHC, after full detector simulation, reconstruction, and event selection.
    Panels show proton and iron primaries in the indicated energy interval.
    The corresponding $\mxmax$ and $\sxmax$ values for each distribution are listed in the top-left corner of each panel in units of g/cm$^{2}$.
    }}
    \label{fig:interactionDependence}
\end{figure*}

\section{Systematic Uncertainties} \label{sec:systematic}
In this Section, we evaluate the systematic uncertainties associated with the measurement of $\xmax$ itself.
It is important to note that the uncertainties discussed here do not include those arising from the choice of hadronic interaction model.
The hadronic interaction model primarily affects the interpretation of mass composition rather than the $\xmax$ measurement.

The only place where the interaction model enters the reconstruction is through the missing energy correction, whose variation at the level of a few percent results in a change of no more than 2–3\,g/cm$^{2}$ in the final $\mxmax$.
This effect is small compared to the other systematic uncertainties evaluated in this Section.

By contrast, the choice of hadronic interaction model has a significant influence on the interpretation of the measured $\xmax$ distributions.
Indeed, replacing QGSJet\,II-04 with either EPOS-LHC~\cite{bib:EPOSLHC} or Sibyll\,2.3d~\cite{bib:sibyll2.3d} shifts the predicted $\mxmax$ toward deeper values by approximately +10–20\,g/cm$^{2}$~\cite{Supanitsky:2022zcw}.
This model-dependent shift remains even after applying detector response, full reconstruction, and event selection, as illustrated in Fig.\,~\ref{fig:interactionDependence}.
We also note that the $\sxmax$ predictions by different interaction models differs at a similar level.
In particular, EPOS-LHC tends to predict a narrower $\sxmax$ by about 
$5\,\mathrm{g/cm^{2}}$ compared to QGSJet\,II-04, a trend that is likewise visible in Fig.\,\ref{fig:interactionDependence}.
For these reasons, the systematic uncertainties reported in this section refer only to those affecting the $\xmax$ measurement itself, separate from the choice of hadronic interaction model.

The main sources of systematic uncertainties in the $\mxmax$ and $\sxmax$ arise from the following six sources: photonic scale, missing energy correction, detector effects, atmospheric conditions, fluorescence yield, and the modeling of Cherenkov light production.

The photonic scale uncertainty includes factors such as PMT gain, UV filter transmission, and telescope mirror reflectivity. The uncertainty in this scale has been estimated to be 10\,$\%$ by the HiRes collaboration~\cite{bib:photonicScale}.
This uncertainty directly affects the energy scale, leading to corresponding uncertainties in $\mxmax$ and $\sxmax$. 
Specifically, for $\mxmax$, the uncertainty is estimated to be $\pm$\,3.1\,g/cm$^{2}$ at $E = 10^{16.5}$\,eV, increasing to $\pm$\,10.3\,g/cm$^{2}$ at $E = 10^{18.5}$\,eV.
For $\sxmax$, the uncertainty ranges from $\pm$\,2.4\,g/cm$^{2}$ to $\pm$\,8.8\,g/cm$^{2}$ across the same energy range.

The missing energy correction introduces an additional systematic uncertainty because a single correction curve is applied to all events, whereas the true correction depends on the primary mass.
Using a correction based on the mean composition leads to a slight overcorrection for light primaries and an undercorrection for heavy primaries.
We estimate this effect by recalculating the missing energy correction separately for proton and iron, and taking the difference between them as the maximal composition–dependent shift.  
This shift is then propagated to $\mxmax$ and $\sxmax$ to obtain the resulting systematic uncertainties.
The systematic uncertainty increases from $1.6$ to $3.5\,\mathrm{g/cm^{2}}$ for $\mxmax$ and from $0.9$ to $3.2\,\mathrm{g/cm^{2}}$ for $\sxmax$ over the energy range $10^{16.5}$–$10^{18.5}$\,eV.

Uncertainties from the detector are dominated by the relative subsecond timing offset between the FD and SD systems, which is measured to be $250\,\mathrm{ns}$.
While the absolute time is synchronized to the one-second level by GPS, the subsecond timing is recorded independently by the local clocks of the two systems, leading to this offset.
The uncertainty of the subsecond synchronization is given by the SD sampling frequency of $50\,\mathrm{MHz}$, corresponding to a timing resolution of $20\,\mathrm{ns}$.
To evaluate the impact of this uncertainty, a conservative timing shift of $\pm\,25\,\mathrm{ns}$—slightly larger than the sampling resolution—was applied as a systematic test of the reconstruction.
The resulting uncertainties are $\pm$\,$3.5\,\mathrm{g/cm^2}$ for $\mxmax$ and $\pm$\,$2.5\,\mathrm{g/cm^2}$ for $\sxmax$.

Atmospheric uncertainties arise from the aerosol content and the atmospheric density profile.
Since the observed events occur at distances of up to $7\,\mathrm{km}$, as shown in Fig.\,\ref{fig:dataMCcomparison04}, the impact of the VAOD is minimal, contributing an uncertainty of $1.0\,\mathrm{g/cm^2}$ to both $\mxmax$ and $\sxmax$.
The atmospheric density profile is modeled using the GDAS in the standard analysis. To evaluate the associated uncertainty, a reanalysis was performed using radiosonde data from the NOAA National Weather Service~\cite{bib:radiosonde}, resulting in $\mxmax$ uncertainty of $\pm$\,$1.0\,\mathrm{g/cm^2}$ across the entire energy range, and an uncertainty in $\sxmax$ ranging from $\pm$\,$0.5\,\mathrm{g/cm^2}$ to $\pm$\,$2.3\,\mathrm{g/cm^2}$.

The fluorescence yield model used in this analysis is based on the absolute yield measured by Kakimoto et al.~\cite{bib:kakimoto} and the fluorescence spectrum measured by the FLASH experiment~\cite{bib:FLASH}. 
We confirmed that the effect of the difference in fluorescence spectrum is negligible.
The uncertainty in the absolute yield of $\pm\,11\%$ was accounted for through reanalysis. This results in uncertainties in $\mxmax$ ranging from $\pm$\,$5.0\,\mathrm{g/cm^2}$ to $\pm$\,$1.0\,\mathrm{g/cm^2}$.
For $\sxmax$, the uncertainty is $\pm$\,$0.5\,\mathrm{g/cm^2}$ across all energy bins.

To evaluate the impact of the Cherenkov production model, we use an alternative Cherenkov production model by Giller et al.~\cite{bib:giller} instead of the parametrization by Nerling et al.~\cite{bib:nerling}.
This resulted in uncertainties in $\mxmax$ ranging from $\pm$\,$12.9\,\mathrm{g/cm^2}$ to $\pm$\,$2.0\,\mathrm{g/cm^2}$. For $\sxmax$, the uncertainty ranges from $\pm$\,$1.5\,\mathrm{g/cm^2}$ to $\pm$\,$2.5\,\mathrm{g/cm^2}$ over the same energy range.

All individual contributions uncertainties were added in quadrature, and the total systematic on the $\mxmax$ and $\sxmax$ are summarized in Table\,\ref{table:totalSystematic}.

\begin{table}[h]
\footnotesize
\begin{center}
\caption{\justifying{Summary of systematic uncertainties in $\mxmax$ and $\sxmax$. Lines with ranges represent the values at the low and high ends of the considered energy range ($10^{16.5}$ eV to $10^{18.5}$ eV, respectively).}}
\begin{tabular}{l c c}
\hline \hline
Sources & $\mxmax$ [g/cm$^2$] & $\sxmax$ [g/cm$^2$]  \\
\hline
Photonic scale & 3.1 to 10.3 & 2.4 to 8.8 \\ 
Missing energy & 1.6 to 3.5 & 0.9 to 3.2 \\
Relative time of FD and SD & 3.5 & 2.5 \\ 
Atmosphere & 1.4 & 1.1 to 2.5 \\ 
Fluorescence yield & 5.0 to 1.0 & 0.5 \\ 
Cherenkov model & 12.9 to 2.0 & 1.5 to 2.5 \\ 
\hline
Total & 14.8 to 11.7 & 4.1 to 10.3 \\ 
\hline
\end{tabular}
\label{table:totalSystematic}
\end{center}
\end{table}

\section{Results and Discussion} 
\label{sec:result}

We present the results on the cosmic ray mass composition in the energy range from $10^{16.5}$ eV to $10^{18.5}$ eV measured with the TALE hybrid detector.
The $\xmax$ distributions for different primaries show significant differences: lighter primaries tend to produce showers that develop deeper in the atmosphere than those initiated by heavier nuclei. 

In addition, light primaries exhibit larger fluctuations in $\xmax$, resulting in broader distributions.
Therefore, both the mean and width of the $\xmax$ distributions are sensitive to the mass composition of cosmic rays.

\begin{figure}[h]
\begin{center}
\includegraphics[width=1.\linewidth]{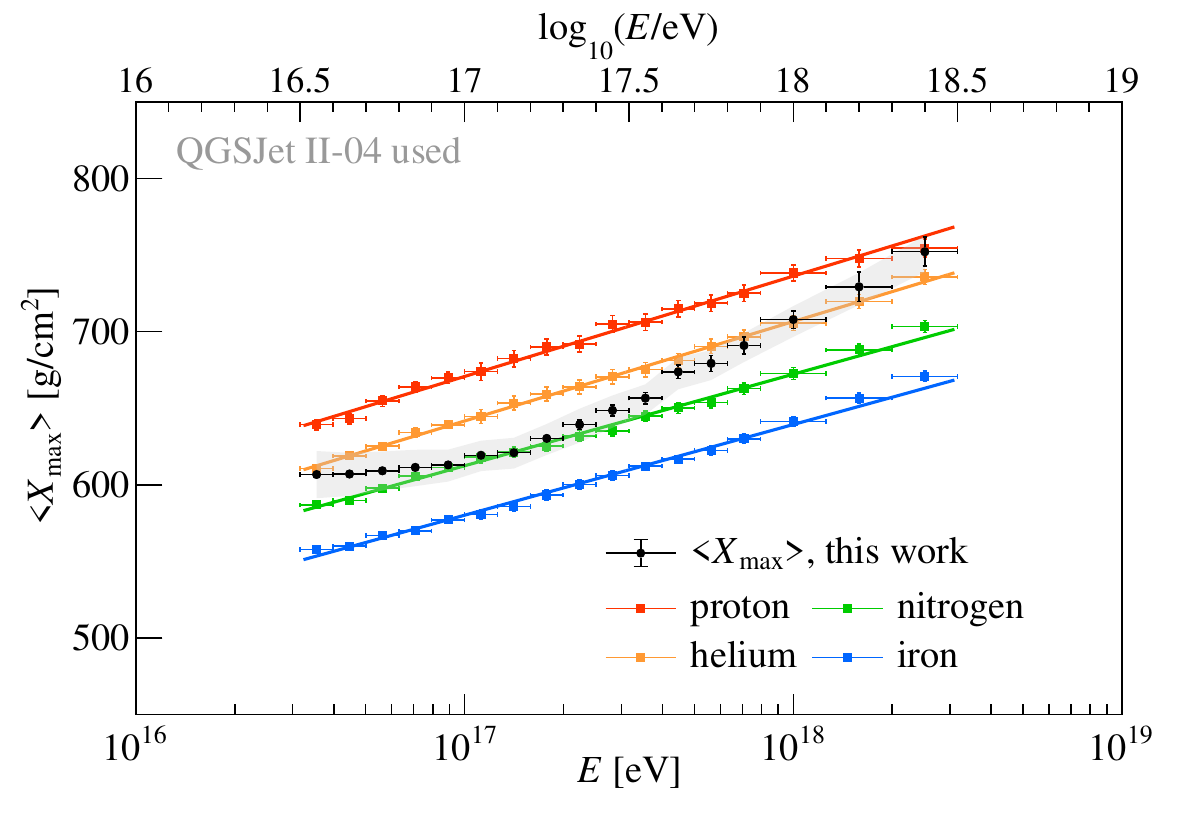}
\caption{\justifying{$\mxmax$ as a function of energy observed by the TALE hybrid mode over nearly five years of data (black points with systematic band). Colored symbols show reconstructed Monte Carlo results for four primary species generated with the QGSJet\,II-04 model, including detector effects. The solid lines represent linear fits to the $\mxmax$ distributions for each primary type.}}
\label{fig:meanXmax00}
\vspace{-5pt}
\end{center}
\end{figure}
\begin{figure}[h]
\begin{center}
\includegraphics[trim=0cm 0.025cm 0cm 0cm, clip, width=1.\linewidth]{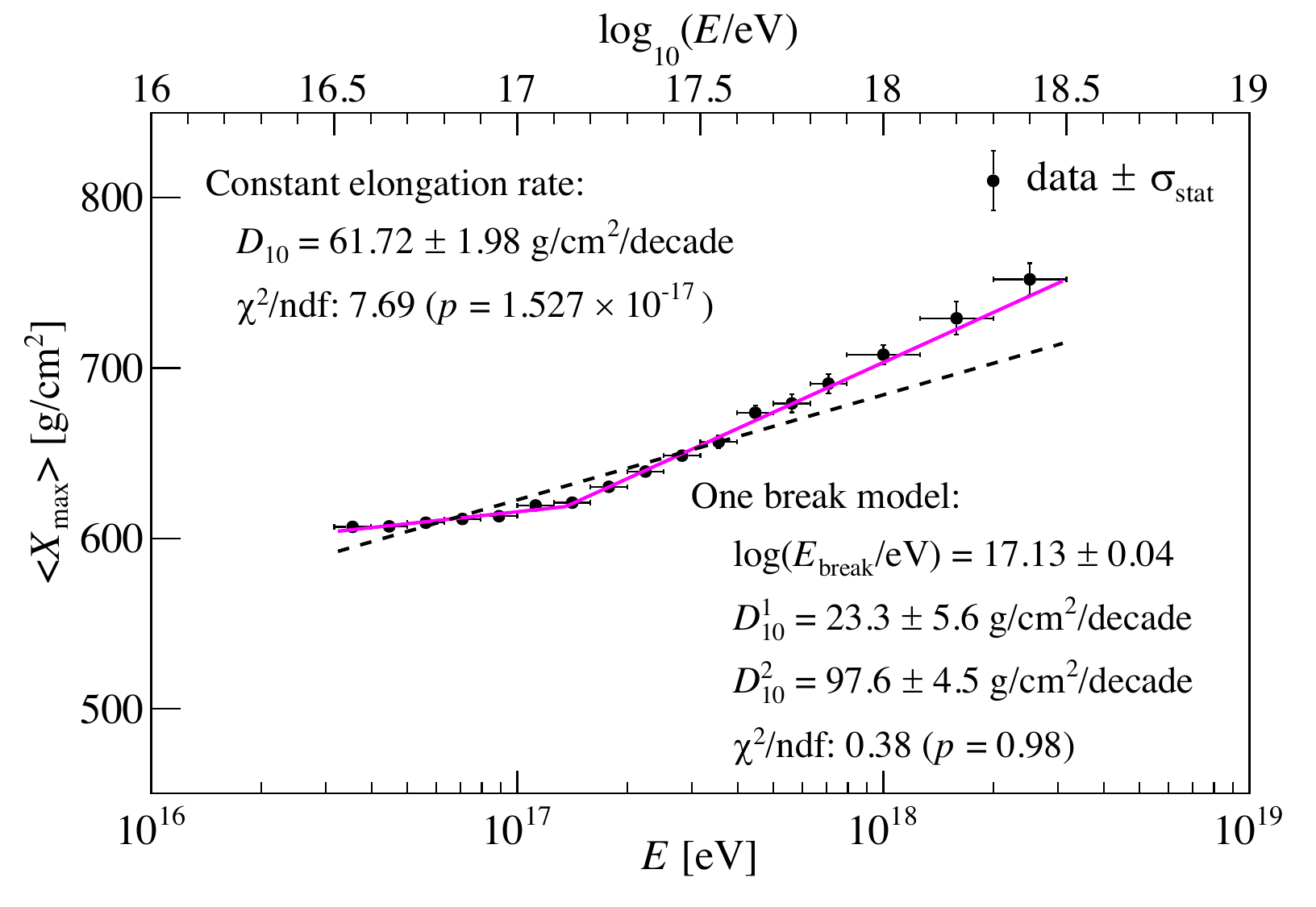}
\caption{\justifying{Evolution of $\mxmax$ as a function of energy. The best fits of a constant elongation rate and one break elongation rate are shown by a dashed line and a magenta line, respectively.}}
\label{fig:meanXmax01}
\end{center}
\vspace{-5pt}
\end{figure}

The evolution of $\mxmax$ as a function of cosmic ray energy, commonly referred to as the $elongation\,\,rate$, 
\begin{equation}
D_{10} = \frac{\rm{d}\langle \it{X}_\mathrm{max} \rangle}{\rm{dlog_{10}}(\it{E}/\rm{eV})}
\label{elongationRateDefinition}
\end{equation}
can also be examined for indications of changes in composition.
In general, the $\xmax$ depends approximately on $\ln(E/A)$, where $E$ is the primary energy and $A$ is the nuclear mass.
As a result, increasing energy leads to a deeper $\mxmax$, while heavier primaries shift $\mxmax$ to shallower depths.
Under the assumption of a pure composition, the rate of change in $\mxmax$ is therefore nearly constant and largely independent of the primary mass and the choice of the hadronic interaction model used in simulations.



Figure\,\ref{fig:meanXmax00} shows the observed $\mxmax$ as a function of energy, along with the reconstructed Monte Carlo results for four primary particle types.
The elongation rate of the simulation events is approximately 60\,–\,65\,$\rm{g/cm^{2}}$/decade and remains nearly constant regardless of the primary cosmic ray species.
In contrast, the observed elongation rate exhibits a clear break just above $10^{17}$ eV.
To quantify this evolution, we performed both a constant elongation rate fit and a broken-line fit.
In the broken-line fit, the two elongation rates and the break-point energy were treated as free parameters and determined simultaneously by minimizing the overall $\chi^{2}$.

\begin{figure}[t]
\begin{center}
\includegraphics[width=1.\linewidth]{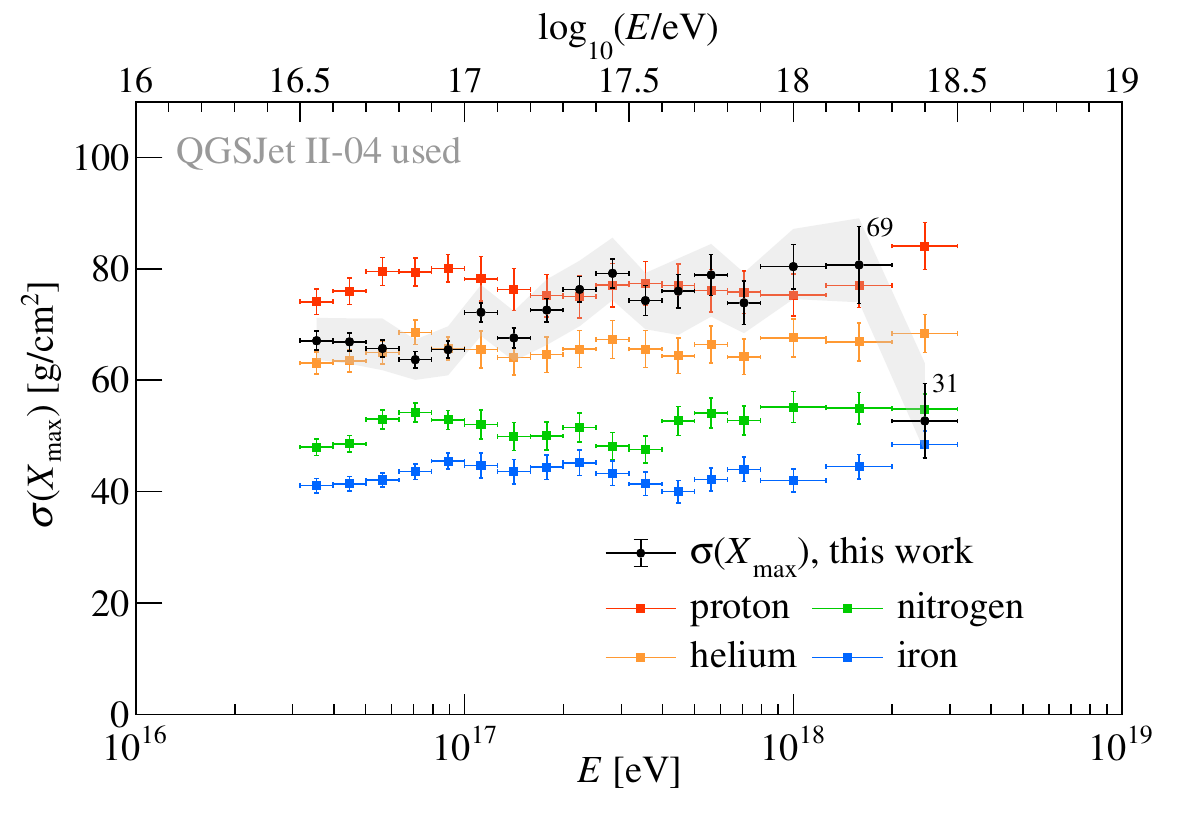}
\caption{\justifying{$\sxmax$ as a function of energy measured by the TALE hybrid mode (black points with systematic band). Colored symbols show Monte Carlo results for different primary species simulated with the QGSJet\,II-04 model, including detector effects.  
For energy bins with fewer than 100 events, the number of events is indicated next to the corresponding data point.}}
\label{fig:sigmaXmax00}
\end{center}
\end{figure}

\begin{figure}[t]
\begin{center}
\includegraphics[width=1.\linewidth]{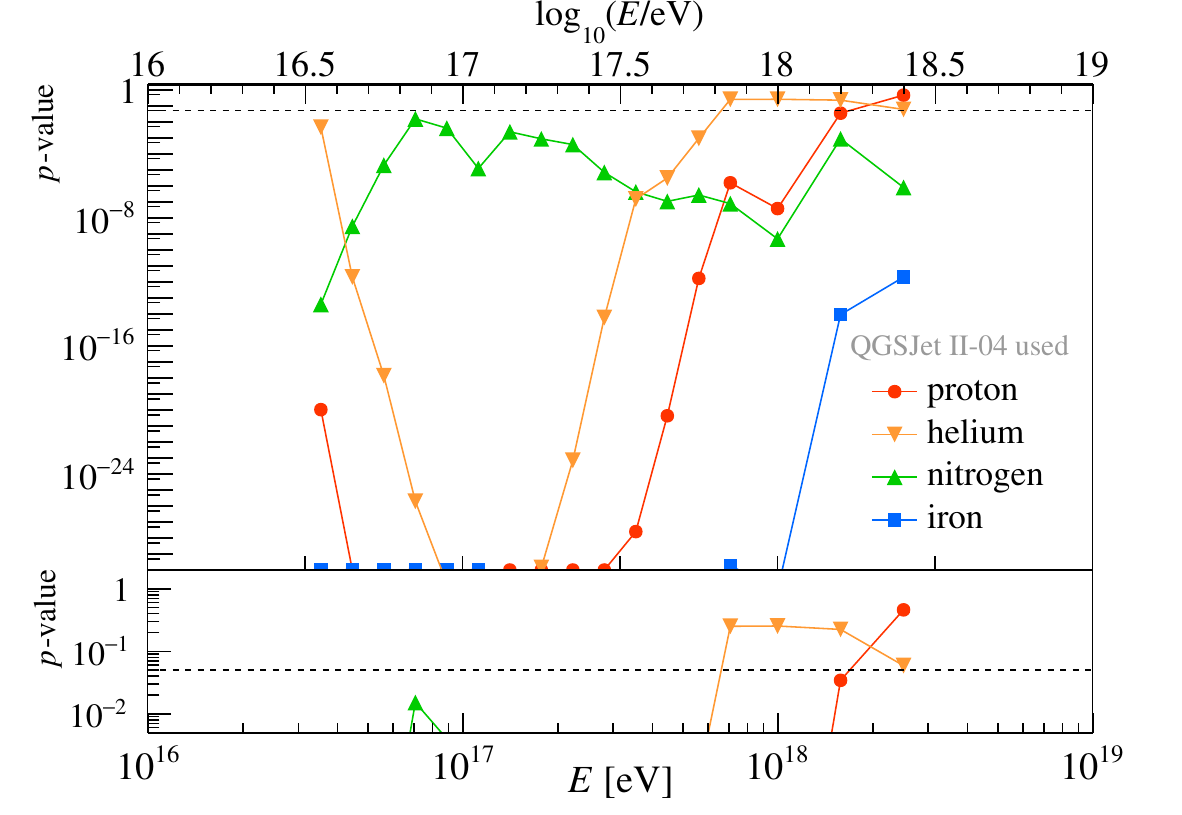}
\caption{\justifying{Kolmogorov-Smirnov test results for the $\xmax$ distributions of various primary cosmic ray species as a function of energy. The $p$ values indicate the probability that the observed distribution of $\xmax$ is consistent with the hypothesized distribution for each particle type. The bottom panel provides an enlarged view of the y-axis range shown in the top panel. The horizontal dotted line in both panels indicates $p\,=$ 0.05.}}
\label{fig:KStest}
\end{center}
\end{figure}

The result is shown in Fig.\,\ref{fig:meanXmax01}.
A single linear fit of $\mxmax$ as a function of energy does not describe our data well ($\chi^{2}$/ndf $=$ 115.7/15).
Introducing a change in the elongation rate at a break point of $E_{\rm{break}}$ results in a good $\chi^{2}$/ndf of 5.2/13, with an elongation rate of
\begin{equation}
D_{10}^{1} = 23.3 \pm 5.6(\rm{stat.})\add{^{+11.1}_{-12.2}}(\rm{sys.})\,\mathrm{g/cm^{2}}/\rm{decade}
\end{equation}
below log($E_{\rm{break}}$/eV) = 17.13 $\pm$\,0.04\,(stat.) \add{$^{+0.05}_{-0.03}$}\,(sys.) and
\begin{equation}
D_{10}^{2} = 97.6 \pm 4.5(\rm{stat.})\add{^{+2.3}_{-1.6}}(\rm{sys.})\,\mathrm{g/cm^{2}}/\rm{decade}
\end{equation}
above this energy.

It is noteworthy that the measured $\mxmax$ distribution is well described by two linear functions with different slopes.  
This behavior can be understood in the context of the composition scenario outlined in Sec.\,\ref{sec:intro}.  
In this scenario, the composition becomes gradually heavier with increasing energy up to the transition between Galactic and extra-Galactic cosmic rays.
Above this transition, the composition becomes lighter again.  
These trends naturally lead to an elongation rate that is smaller than that of any single mass component below the break and larger above it.
No additional structure is known in the energy spectrum between $10^{16.5}$ and $10^{18}$\,eV that would require a more complex description.
Therefore, the observed break in $\mxmax$ is consistent with a simple interpretation in terms of a composition change associated with the Galactic – extra-Galactic transition.

To verify that the observed break is not an artifact of combining events reconstructed from different light components, we performed an additional test only using CL events subset.
Because FL events are essentially absent below $10^{17}$\,eV, an FL event subset test is not feasible.
For the CL event subset, the elongation rate was refitted following the same procedure as in the main analysis.
The result of this test is shown in Figure\,\ref{fig:elongationFitOnlyCh} in Appendix~\ref{app:cl_only_elongation}.
The resulting fit still shows a statistically significant break, and the fitted break energy is consistent with that obtained from the full dataset.
The postbreak slope obtained from the CL-only subset is slightly smaller; however, the difference is not statistically significant and can be attributed to limited statistics at the highest energies.
This cross-check confirms that the observed change in the elongation rate is not driven by a mixture of Cherenkov- and fluorescence-dominated events.

\begin{figure*}
    \centering
    \begin{minipage}[l]{0.495\textwidth}    
        \centering
        \includegraphics[width=\linewidth]{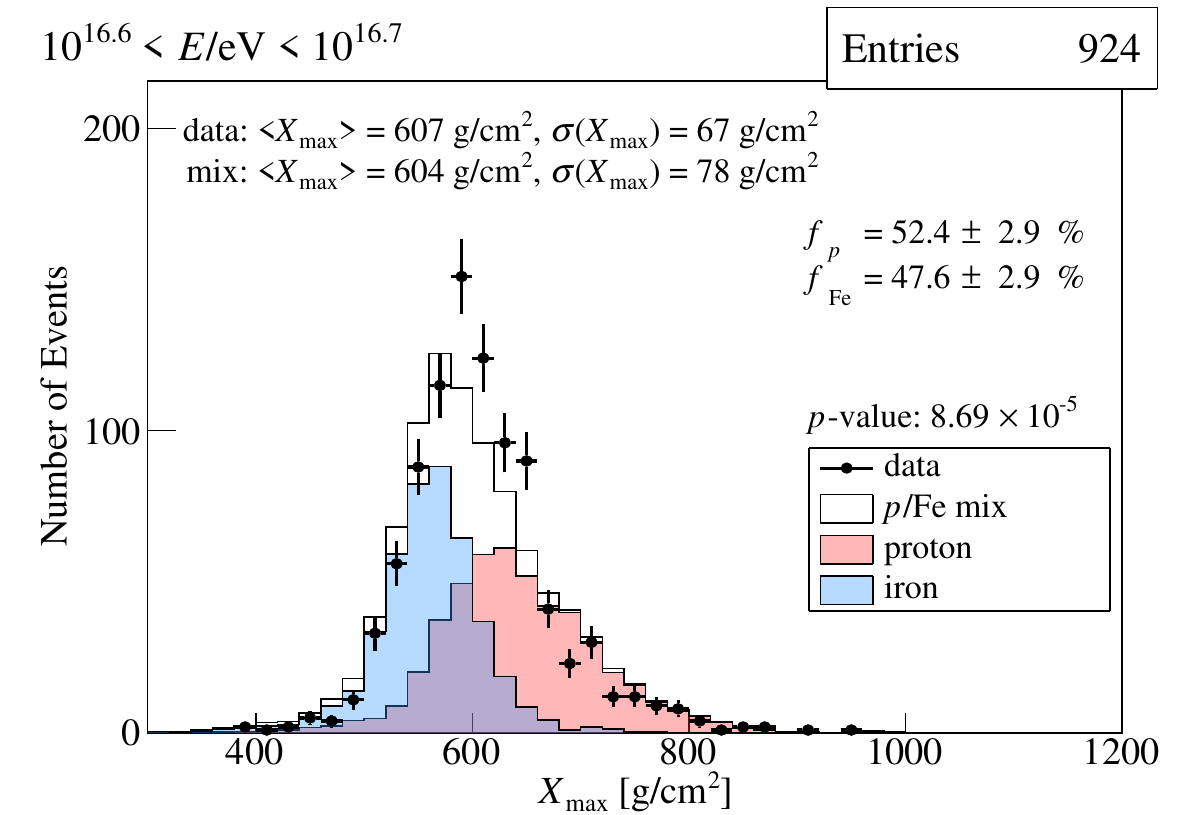}
        \vspace{+5mm}
    \end{minipage}
    \hfill
    \begin{minipage}[r]{0.495\textwidth}    
        \centering
        \includegraphics[width=\linewidth]{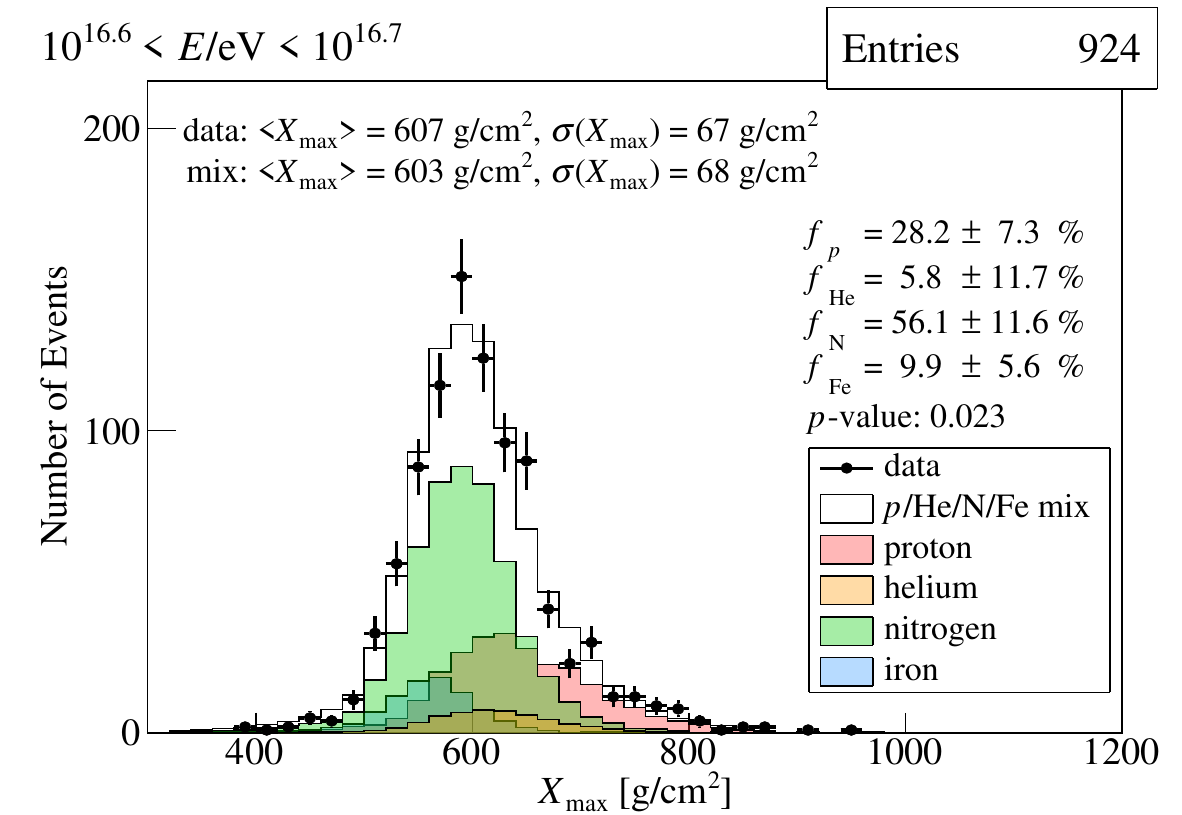}
        \vspace{+5mm}
    \end{minipage}     
    \begin{minipage}[l]{0.495\textwidth}    
        \centering
        \includegraphics[width=\linewidth]{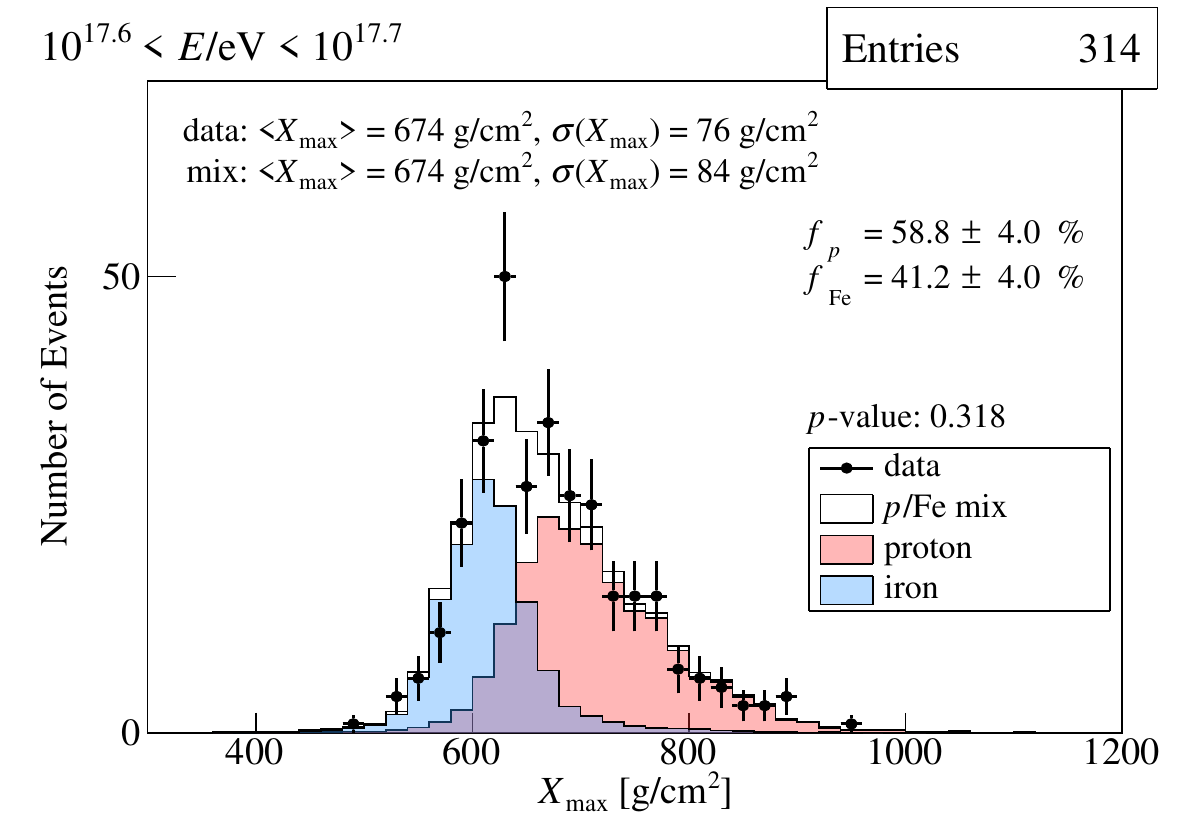}
        \subcaption{fit with \it{p}\rm{+Fe model}}
        \label{fig:xmaxFitWith2component}
    \end{minipage}    
    \hfill
    \begin{minipage}[r]{0.495\textwidth}   
        \centering
        \includegraphics[width=\linewidth]{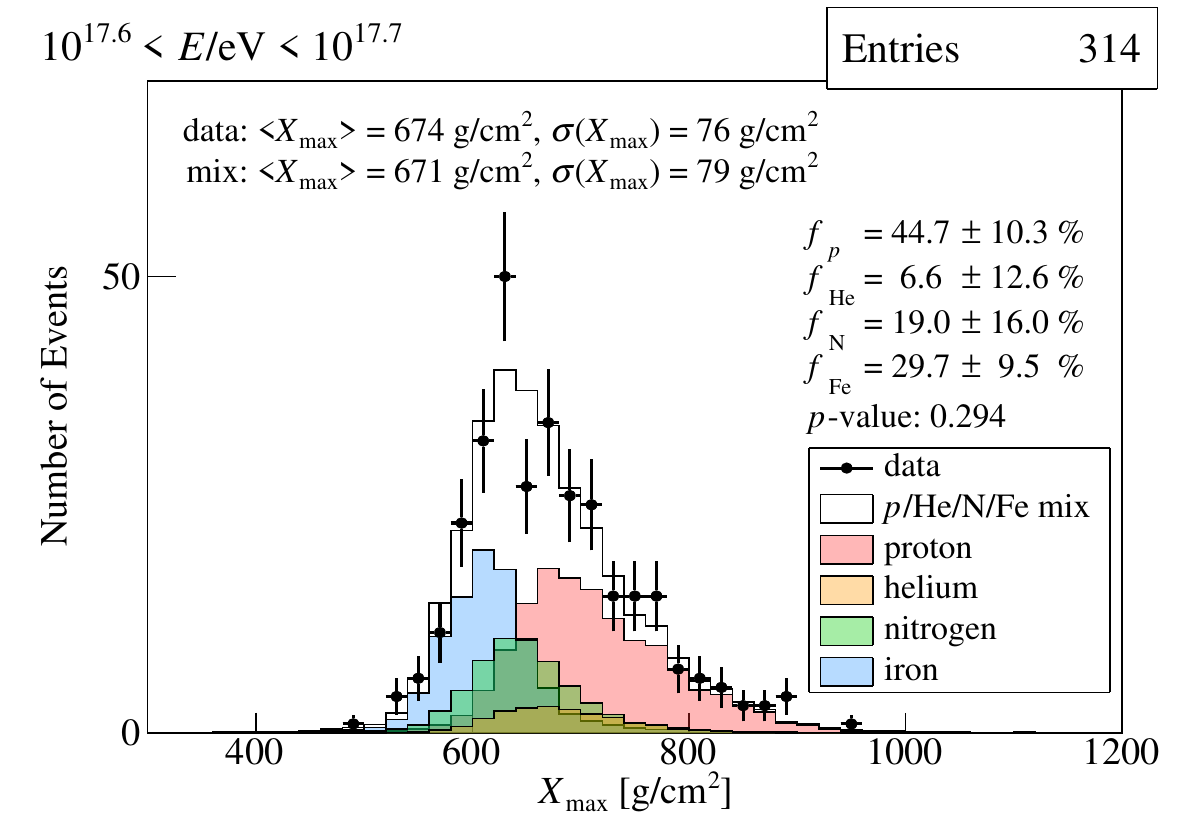}
        \subcaption{fit with \it{p}\rm{+He+N+Fe model}}
        \label{fig:xmaxFitWith4component}
    \end{minipage}    
    \caption{\justifying{Examples of $\xmax$ distribution fits for two representative energy bins: $10^{16.6} < E/\mathrm{eV} < 10^{16.7}$ (top row), and $10^{17.6} < E/\mathrm{eV} < 10^{17.7}$ (bottom row). The left panels show the fit result of the proton-iron mixture model, and the right panels include helium and nitrogen nuclei in addition. The best-fit fractions are displayed at the middle of each plot, and the best-fit $\xmax$ distribution is represented by a solid line, while the contributions from each component are represented by color-shaded distributions. The first two moments of the $\xmax$ distributions for the observational data and the best-fit mixture are displayed in the top-left corner of each plot.}}
    \label{fig:xmaxFits}
\end{figure*}

In addition, we compared the $\mxmax$ obtained from this work with the bias-corrected $\mxmax$ values derived from the TALE FD monocular observations~\cite{TelescopeArray:2020bfv}.
These values correspond to the published $\mxmax$ after correcting for the known reconstruction biases reported in that analysis.
Despite the different reconstruction methods and independent datasets, the two results are consistent within their respective systematic uncertainties.

\begin{figure*}
    \centering
    \begin{minipage}{1.0\hsize} 
    \includegraphics[width=\textwidth]{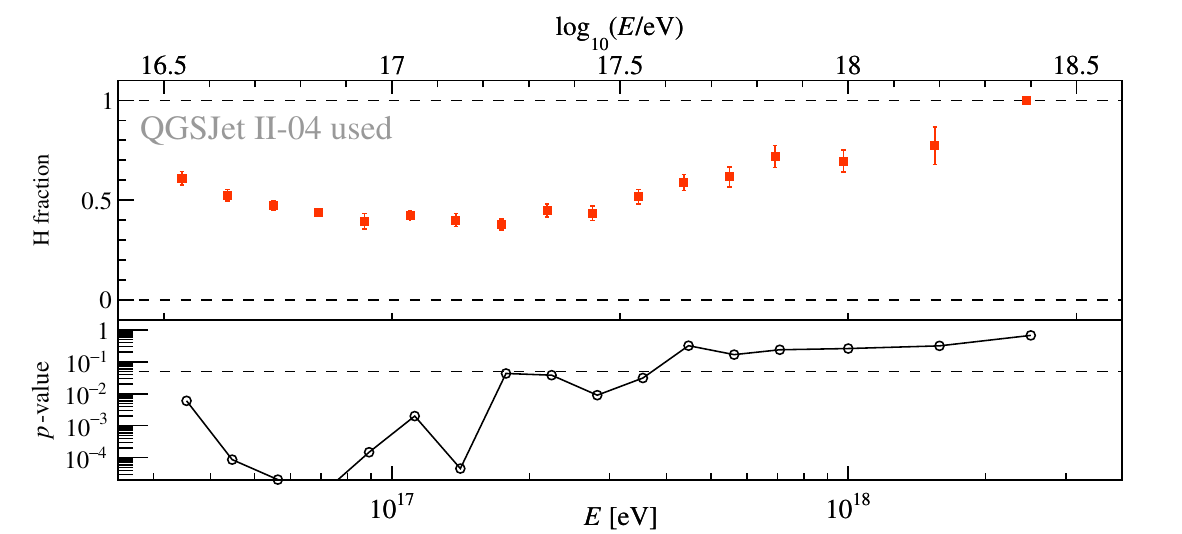}
    \subcaption{{Best-fit fractions and fit quality}}
    \label{fig:best_fit_proton_iron}
    \end{minipage}   
    \begin{minipage}[l]{0.495\hsize}    
    \begin{center}
    \includegraphics[width=1.\linewidth]{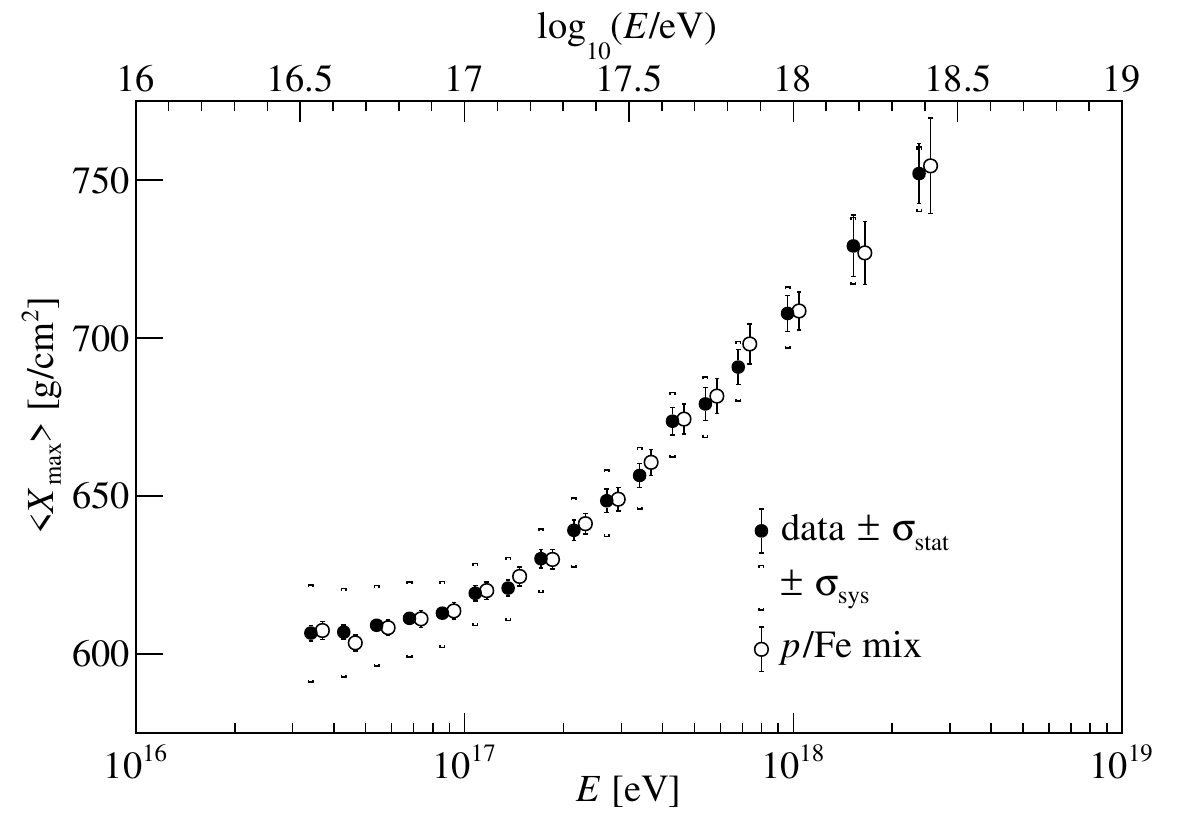}
    \subcaption{{Comparison of $\mxmax$}}
    \label{fig:meanXmax_proton_iron}
    \end{center}
    \end{minipage}    
    \begin{minipage}[r]{0.495\hsize}    
    \begin{center}
    \includegraphics[width=1.\linewidth]{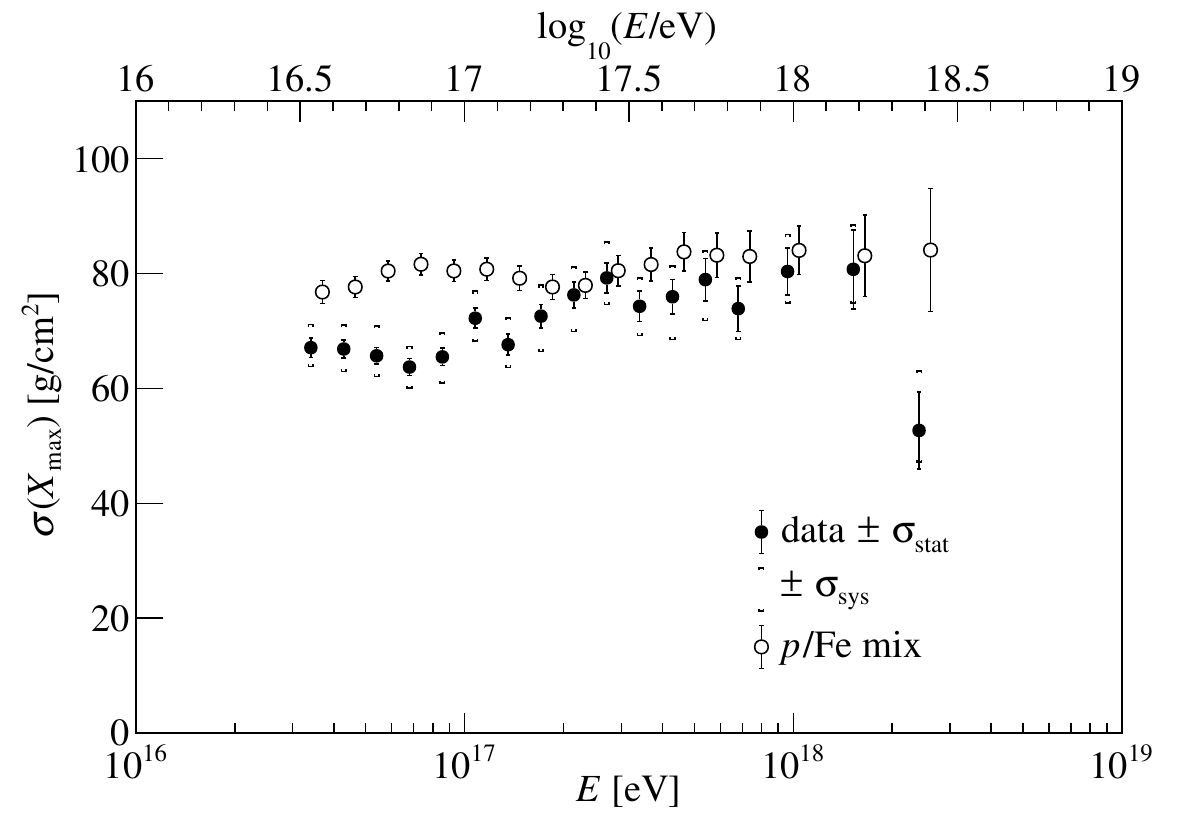}
    \subcaption{{Comparison of $\sxmax$}}
    \label{fig:sigma_proton_iron}
    \end{center}
    \end{minipage}
    \caption{\justifying{(a) Best-fit fraction after acceptance correction and $p$ values for the scenario of a mixture of proton and iron. The error bars in the upper panel represent statistical uncertainties only. The horizontal dotted line in the lower panel indicates $p\,=$ 0.05. Note that the proton fraction is labeled as ``H fraction" in the plot to avoid confusion with the $p$ value shown in the lower panel. (b) and (c) Comparison between the observed data and the $\mxmax$ and $\sxmax$ values derived from the best-fit mixture as a function of energy. For readability, the $\mxmax$ and $\sxmax$ values of the best-fit mixture are slightly offset in energy.}}
    \label{fig:2componentFit}
\end{figure*}

Following the discussion on the $\mxmax$, we now examine the standard deviation of the $\xmax$ distribution ($\sxmax$), which provides additional information on the cosmic ray mass composition.
A comparison of $\sxmax$ with Monte Carlo predictions for four primaries, including the detector and reconstruction effects, is shown in Fig.\,\ref{fig:sigmaXmax00}.
The observed $\mxmax$ is close to that of nitrogen primaries around $10^{17}$ eV, as shown in Fig.\,\ref{fig:meanXmax00}. 
However, the corresponding $\sxmax$ is broader than that predicted for nitrogen, as shown in Fig.\,\ref{fig:sigmaXmax00}.
It should be noted that the last energy bin contains only 31 events, which makes a precise evaluation challenging.

We performed a Kolmogorov-Smirnov (KS) test to evaluate the compatibility between the observed $\xmax$ distributions and the Monte Carlo predicted $\xmax$ distributions assuming a single primary composition.
The result of the probabilities calculated by the KS tests for each energy bin for four primaries is shown in Fig.\,\ref{fig:KStest}, indicating that the observed $\xmax$ distributions at energies less than $10^{17.8}$ eV cannot be explained by a single primary composition.
This suggests that the observed $\sxmax$ can naturally be interpreted as resulting from a mixture of different cosmic ray components.

\begin{figure*}
    \centering
    \begin{minipage}{1.0\hsize} 
    \includegraphics[width=\textwidth]{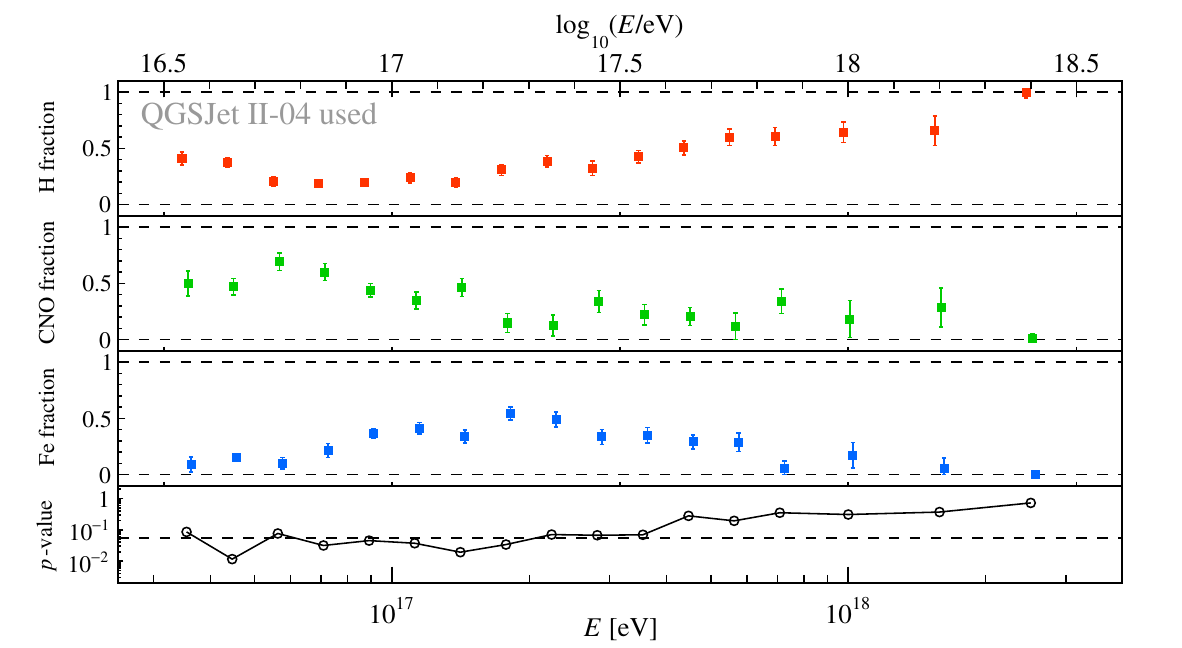}
    \subcaption{{Best-fit fractions and fit quality}}
    \label{fig:best_fit_proton_nitrogen_iron}
    \end{minipage}   
    \begin{minipage}[l]{0.495\hsize}    
    \begin{center}
    \includegraphics[width=1.\linewidth]{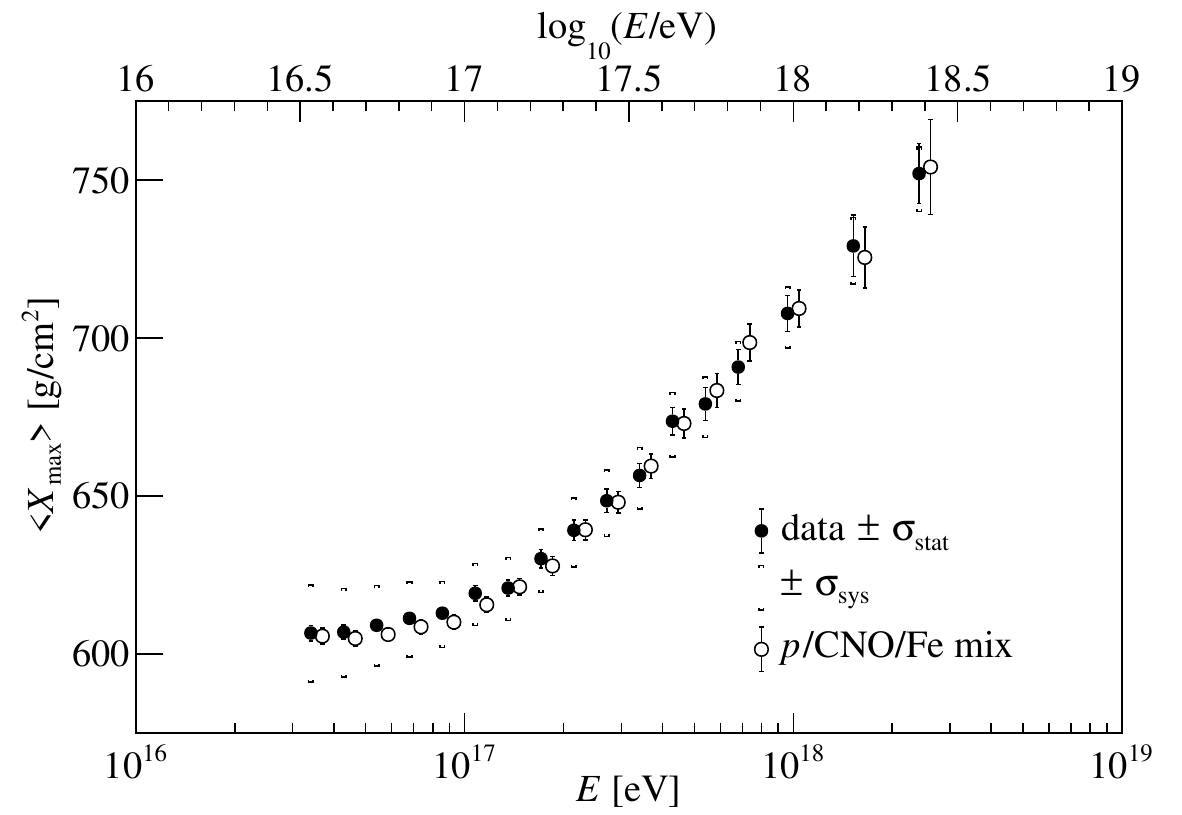}
    \subcaption{{Comparison of $\mxmax$ as a function of energy}}
    \label{fig:meanXmax_proton_nitrogen_iron}
    \end{center}
    \end{minipage}    
    \begin{minipage}[r]{0.495\hsize}    
    \begin{center}
    \includegraphics[width=1.\linewidth]{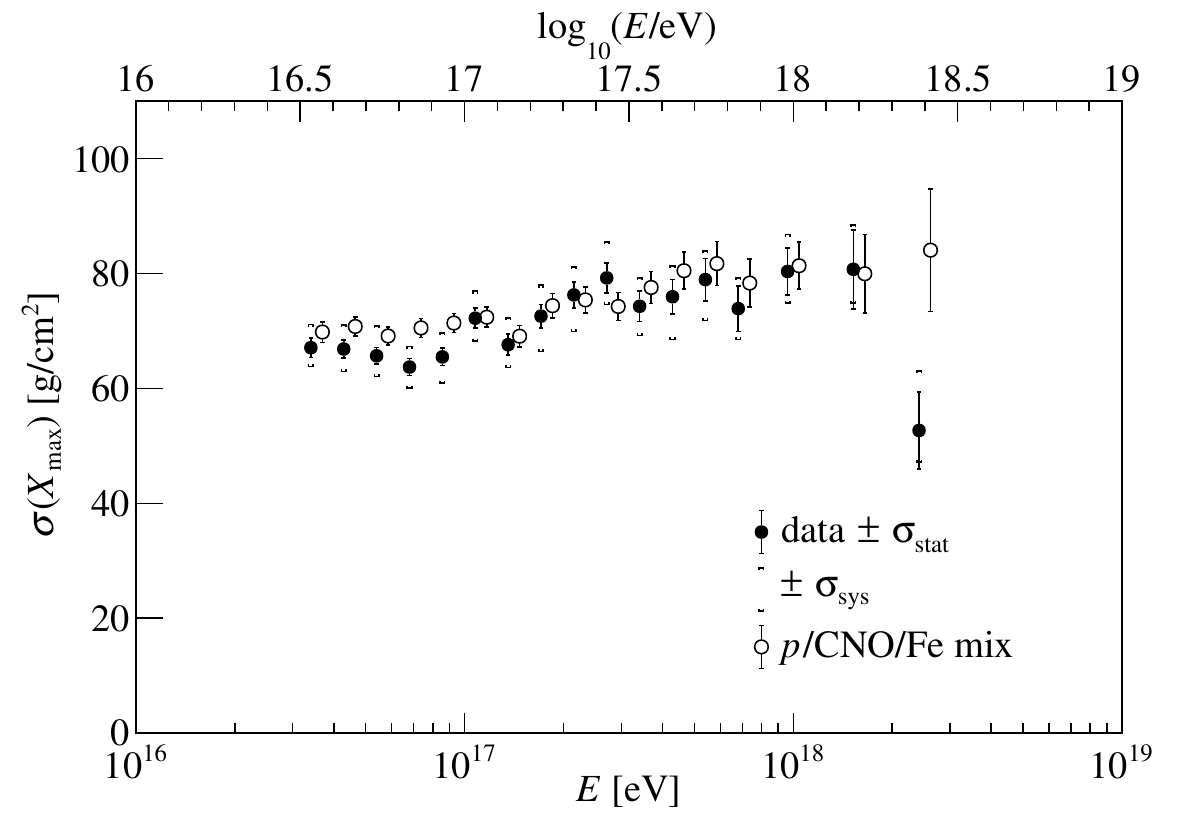}
    \subcaption{{Comparison of $\sxmax$ as a function of energy}}
    \label{fig:sigma_proton_nitrogen_iron}
    \end{center}
    \end{minipage}
    \caption{\justifying{Best-fit fraction and quality, comparison of $\mxmax$ and $\sxmax$ as a function of energy for the scenario of a mixture of proton, nitrogen, and iron.}}      \label{fig:3componentFit}
\end{figure*}
\begin{figure*}
    \centering
    \begin{minipage}{1.0\hsize} 
    \includegraphics[width=\textwidth]{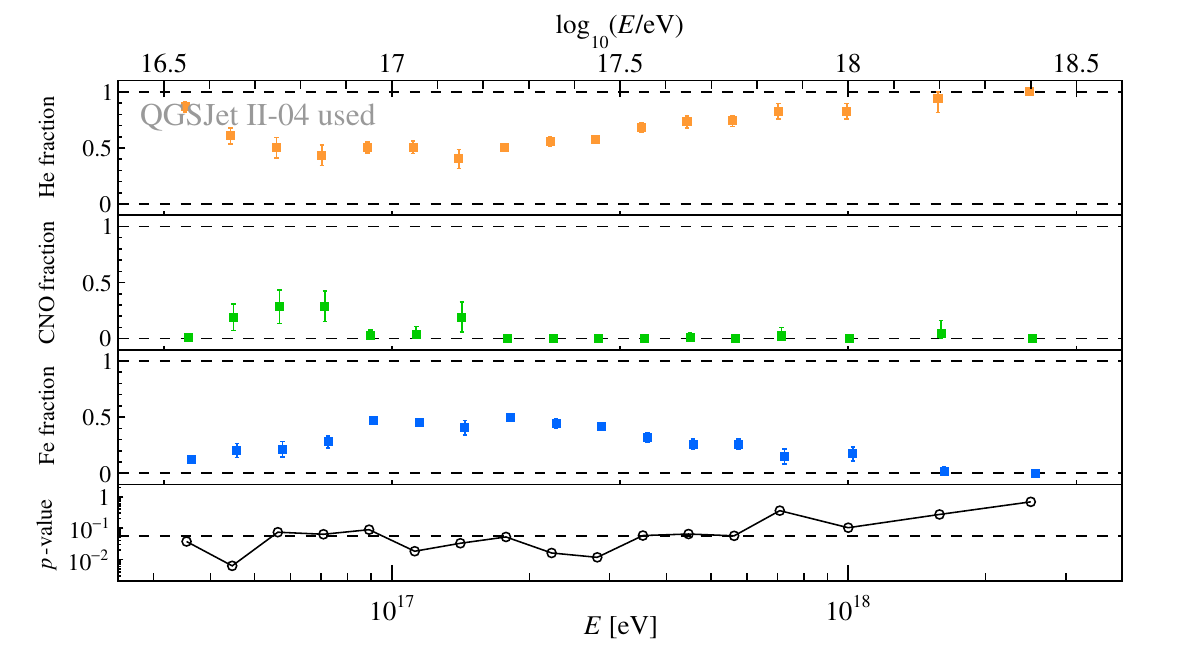}
    \subcaption{{Best-fit fractions and fit quality}}
    \label{fig:best_fit_helium_nitrogen_iron}
    \end{minipage}   
    \begin{minipage}[l]{0.495\hsize}    
    \begin{center}
    \includegraphics[width=1.\linewidth]{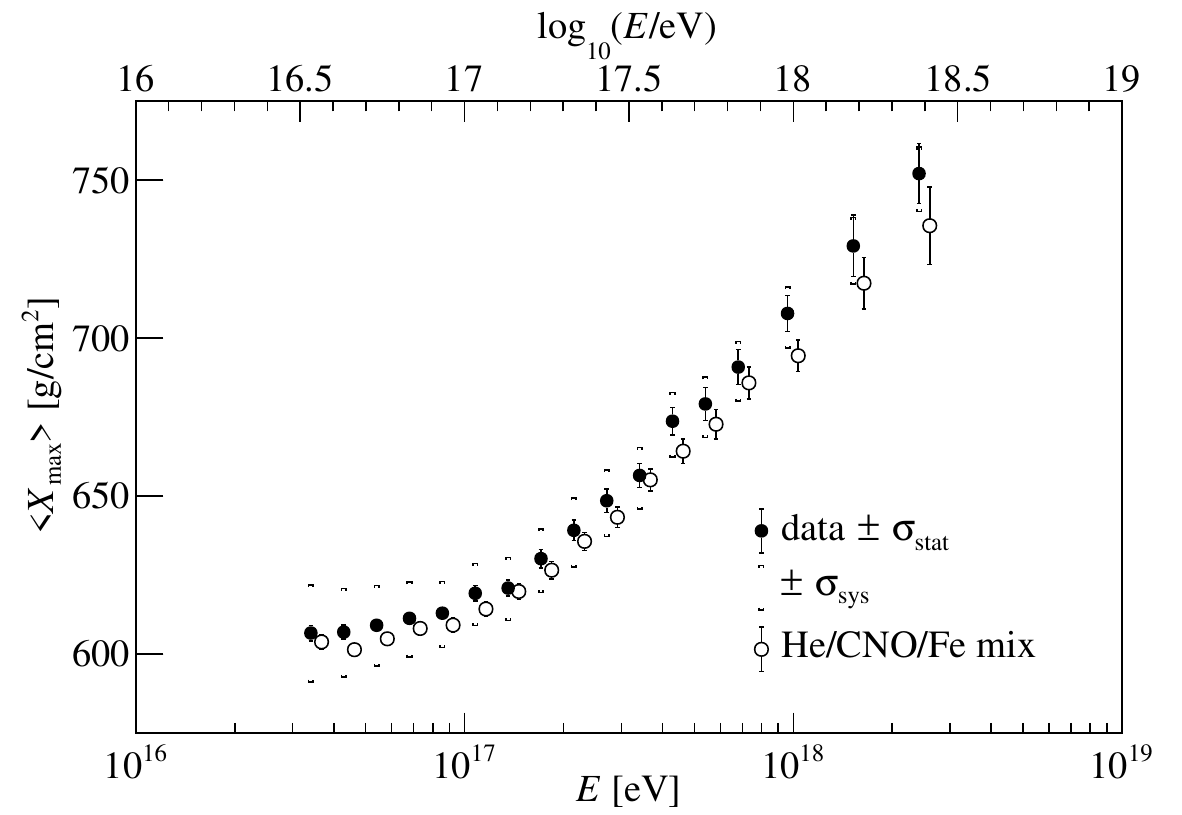}
    \subcaption{{Comparison of $\mxmax$ as a function of energy}}
    \label{fig:meanXmax_helium_nitrogen_iron}
    \end{center}
    \end{minipage}    
    \begin{minipage}[r]{0.495\hsize}    
    \begin{center}
    \includegraphics[width=1.\linewidth]{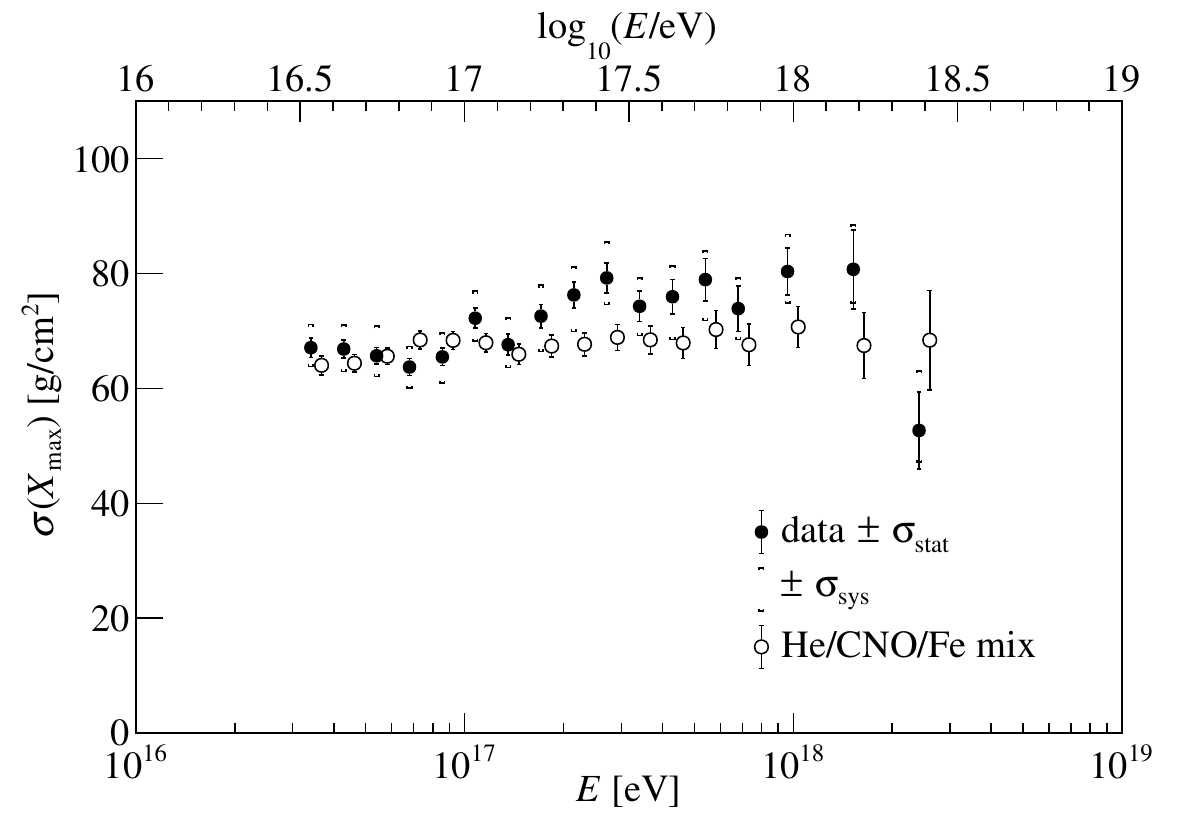}
    \subcaption{{Comparison of $\sxmax$ as a function of energy}}
    \label{fig:sigma_helium_nitrogen_iron}
    \end{center}
    \end{minipage}
    \caption{\justifying{Best-fit fraction and quality, comparison of $\mxmax$ and $\sxmax$ as a function of energy for the scenario of a mixture of helium, nitrogen, and iron.}} 
    \label{fig:3componentFitWithHeNFe}
\end{figure*}
\begin{figure*}
    \centering
    \begin{minipage}{1.0\hsize} 
    \includegraphics[width=\textwidth]{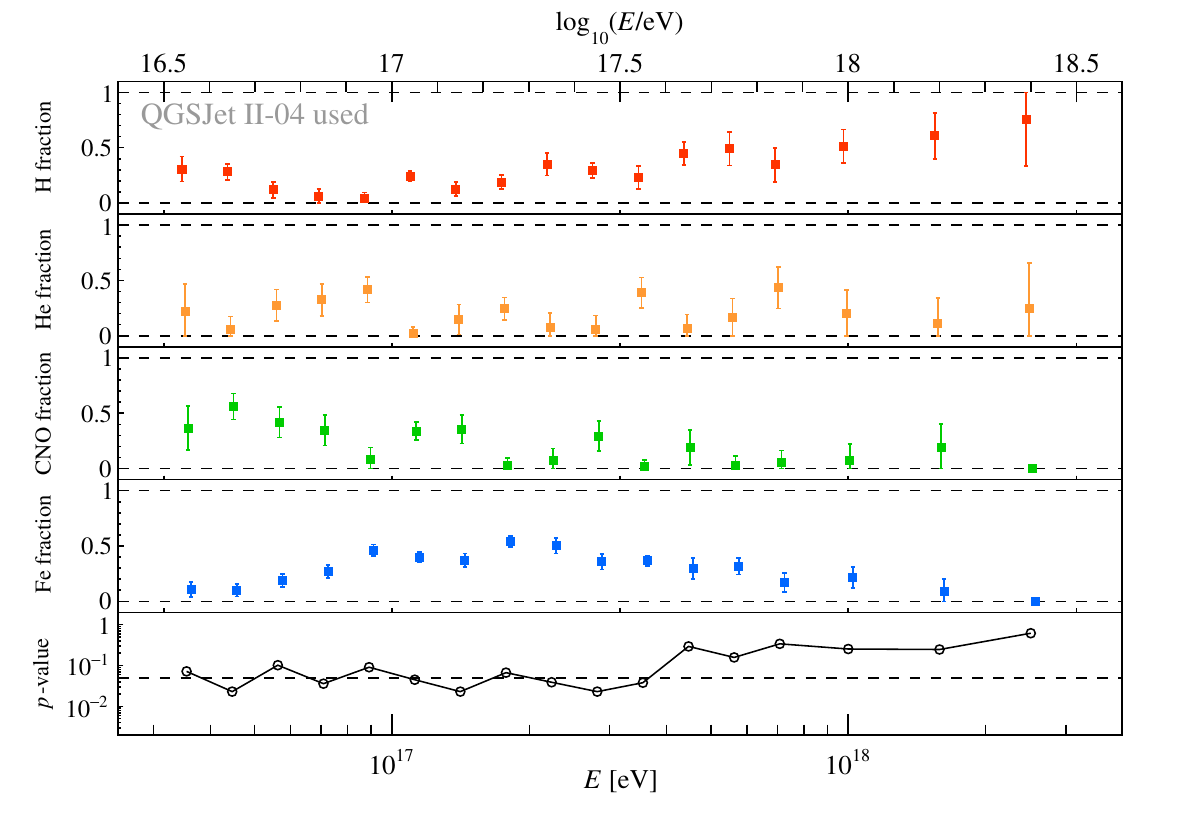}
    \subcaption{{Best-fit fractions and fit quality}}
    \label{fig:best_fit_proton_helium_nitrogen_iron}
    \end{minipage}   
    \begin{minipage}[l]{0.495\hsize}    
    \begin{center}
    \includegraphics[width=1.\linewidth]{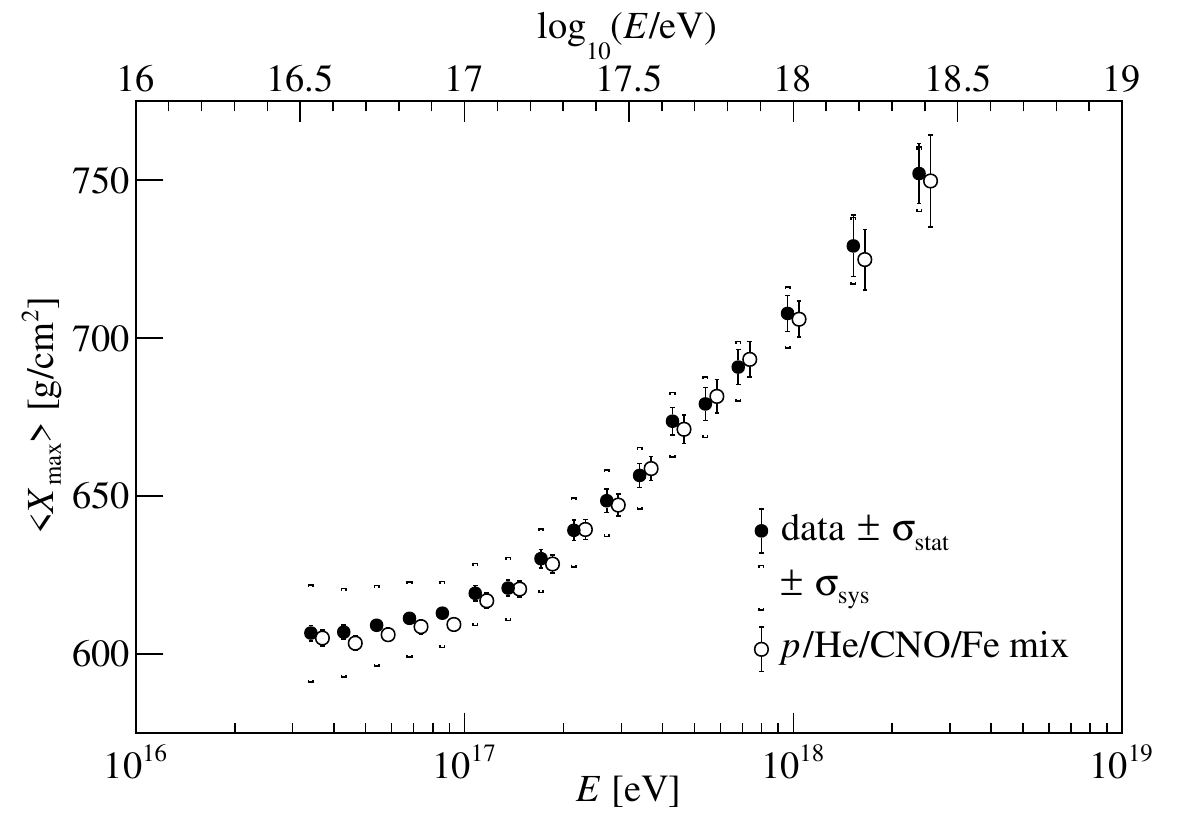}
    \subcaption{{Comparison of $\mxmax$ as a function of energy}}
    \label{fig:meanXmax_proton_helium_nitrogen_iron}
    \end{center}
    \end{minipage}    
    \begin{minipage}[r]{0.495\hsize}    
    \begin{center}
    \includegraphics[width=1.\linewidth]{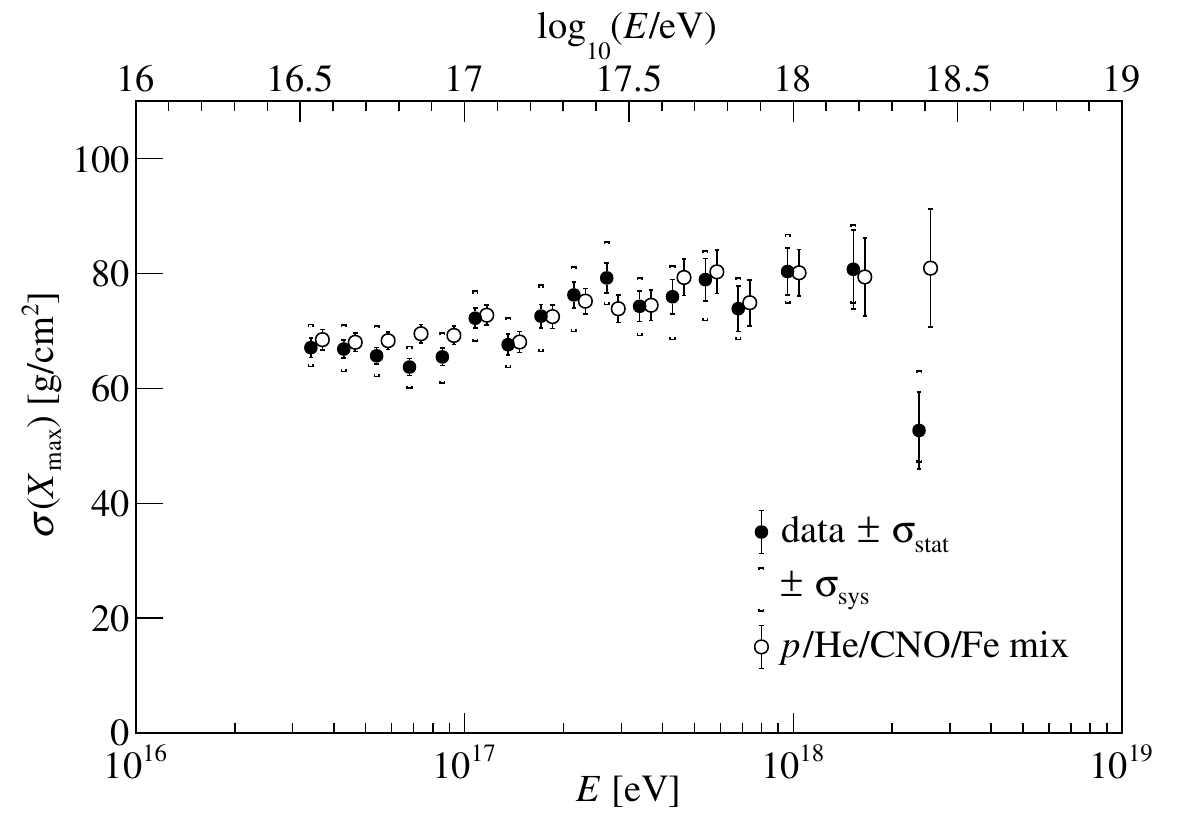}
    \subcaption{{Comparison of $\sxmax$ as a function of energy}}
    \label{fig:sigma_proton_helium_nitrogen_iron}
    \end{center}
    \end{minipage}
    \caption{\justifying{Best-fit fraction and quality, comparison of $\mxmax$ and $\sxmax$ as a function of energy for the scenario of a mixture of proton, helium, nitrogen, and iron.}}
    \label{fig:4componentFit}
\end{figure*}
\begin{figure*}
\begin{center}
\includegraphics[width=1.\linewidth]{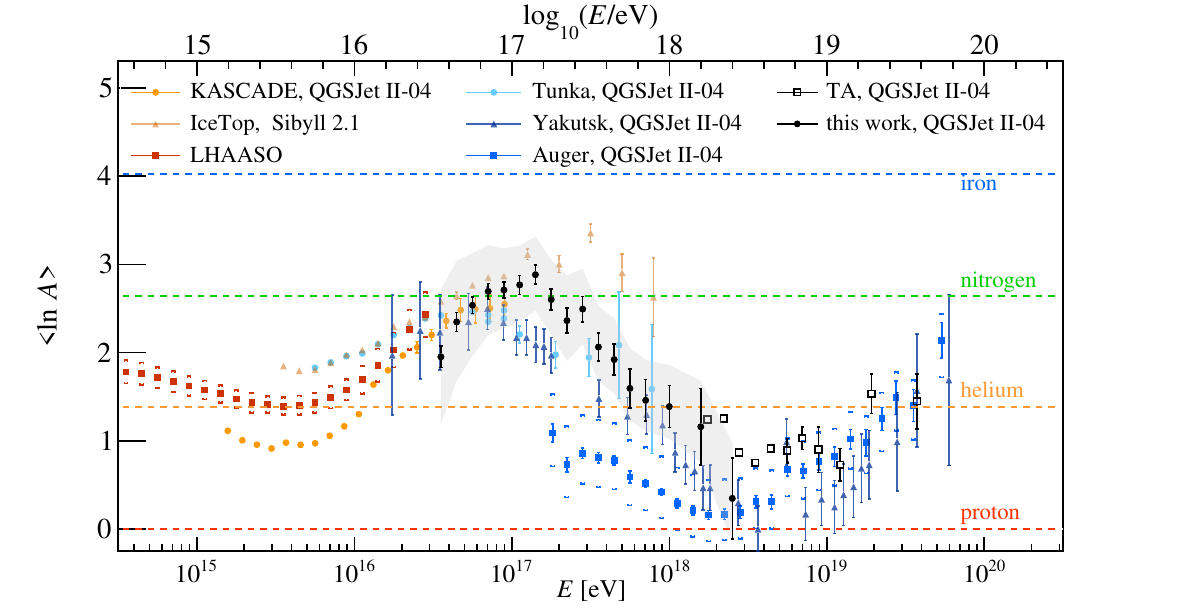}
\caption{\justifying{Mean logarithmic mass number as a function of energy. The black circles represent the result of this work, with the systematic uncertainty shown as a shaded band. For comparison, results from KASCADE~\cite{Kuznetsov:2023pvo}, IceTop~\cite{bib:iceTopResults}, LHAASO~\cite{bib:lhaasoCRSpectrumAndLnA}, Tunka~\cite{bib:tunkaComposition}, Yakutsk~\cite{bib:Knurenko:2019oil}, Auger~\cite{bib:AugerXmax}, and 8.5 yrs TA BRM/LR hybrid~\cite{bib:tahybrid} are shown. The interaction models used in each measurement are indicated next to the experiment names in the figure. For LHAASO data, the systematic uncertainties are displayed as brackets in the figure, including the dependence on the interaction model.}}
\label{fig:lnA}
\end{center}
\end{figure*}

To further investigate this possibility, we performed similar types of tests to assess the viability of various mixtures of cosmic ray elements.
More complete information can be obtained from the full $\xmax$ distributions.
We therefore perform quantitative comparisons between the observed and simulated $\xmax$ distributions.
For this purpose, we binned events by reconstructed energy and constructed the $\xmax$ distribution in each energy bin.
We used a bin width of 0.1 in log$_{10}(E/\mathrm{eV})$ below $10^{17.9}$\,eV and 0.2 up to $10^{18.5}$\,eV, ensuring $\gtrsim 50$ events per bin.
To achieve this, we compared the observed $\xmax$ distributions with those predicted by Monte Carlo simulation, including the detector and reconstruction effects, focusing on different combinations of primary cosmic ray compositions.
To determine the combination of cosmic ray compositions that best reproduces the observed Xmax distributions, we utilized the \texttt{TFractionFitter} tool in ROOT~\cite{bib:ROOT, bib:TFractionFitter1, bib:TFractionFitter2}.
Examples of fits are displayed in Fig.\,\ref{fig:xmaxFits}.

For the calculation of the final results for each primary fraction, we applied a first-order correction to account for the detector acceptance bias, ensuring that the results are as accurate and unbiased as possible.
This correction allows us to infer the primary fractions at the top of the atmosphere by compensating for the differences in detection efficiencies among different cosmic ray primaries.
At an energy of $10^{16.5}$\,eV, the detection efficiencies relative to protons are approximately $R_{\mathrm{Fe}/p} \sim 0.8, \quad R_{\mathrm{N}/p} \sim 0.9, \quad R_{\mathrm{He}/p} \sim 0.95$.
For energies above \( 10^{17.5} \,\mathrm{eV} \), these ratios become flat and independent of the primary.
Details of this first-order correction, including the derivation of the detector acceptance bias correction and its application, are provided in Appendix\,\ref{app:acceptanceCorrection}.

As a first step, we adopted a simple composition model, combining light cosmic rays (protons) and heavy cosmic rays (iron).
The best-fit fractions for this model, along with the corresponding $\mxmax$ and $\sxmax$ values, are illustrated in Fig.\,\ref{fig:2componentFit}, demonstrating the failure of this simple composition model.
Although the model successfully reproduces the $\mxmax$ values, it fails to describe the $\sxmax$ values in the energy range below $10^{17.5}$\,eV, as shown in Fig.\,\ref{fig:sigma_proton_iron}. 
This discrepancy indicates that the simplistic two-component model is insufficient to reproduce the full complexity of the cosmic ray composition at these energies.

Next, we explored a more complex scenario by introducing an intermediate-mass component, corresponding to the CNO group, in addition to protons and iron.
This three-component model significantly improved the fit quality, yielding better agreement between the observed and predicted $\xmax$ distributions in the energy range of $E < 10^{17.5}$\,eV.
In this scenario, we observed a CNO group peak at around 50 PeV and an iron peak at approximately 150 PeV, which may indicate the presence of a Peters cycle, as illustrated in Figs.\,\ref{fig:best_fit_proton_nitrogen_iron}, \ref{fig:meanXmax_proton_nitrogen_iron}, and \ref{fig:sigma_proton_nitrogen_iron}.
These figures demonstrate a substantially improved alignment between the observed data and the Monte Carlo simulations, indicating that intermediate mass components constitute a significant fraction of the cosmic ray composition at these energies.

We further extended our analysis by testing a combination of helium, CNO group, and iron, motivated by direct experimental measurements~\cite{bib:caletProton, bib:caletHelium, bib:dampeProton, bib:dampeHelium} suggesting that helium may dominate over protons near the cosmic ray spectrum knee when extrapolating to higher energies.
Despite this hypothesis, our analysis shows that this combination cannot reproduce the $\sxmax$ values at higher energies, as depicted in Fig.\,\ref{fig:sigma_helium_nitrogen_iron}.
Furthermore, when comparing the proton/CNO group/iron and helium/CNO group/iron scenarios, the latter consistently yields lower $p$-values across most of the energy range.
The corresponding best-fit fractions and $\mxmax$ comparisons, presented in Figs.\,\ref{fig:best_fit_helium_nitrogen_iron} and \ref{fig:meanXmax_helium_nitrogen_iron}, indicate that the features observed at higher energies cannot be fully explained by a model without protons.

In addition, we performed a fit using four components: proton, helium, CNO group, and iron.
Due to the detector resolution of $\xmax$ and the significant overlap in the $\xmax$ distributions of protons and helium, the fractions of these two components exhibit some correlation.
However, the critical finding is that, even with the inclusion of helium component in the mixture of proton, CNO group, and iron assumption, the results clearly indicate the dominant presence of CNO group below $E\,\textless\,10^{17}$\,eV.
Additionally, around $E\,\sim\,10^{17.5}$\,eV, iron remains the most significant contributor.
These trends are consistent and robust, as shown in Fig.\,\ref{fig:4componentFit}.

We additionally note that the relatively small $p$ values observed at energies below $10^{17.5}$\,eV in Fig.\,\ref{fig:4componentFit} do not appear to originate from the choice of mass components used in the fit.
As shown in Figs.\,\ref{fig:meanXmax_proton_helium_nitrogen_iron} and 
\ref{fig:sigma_proton_helium_nitrogen_iron}, the fit results are consistent across almost all energy bins within the systematic uncertainties of $\mxmax$ and $\sxmax$.
Rather, a similar behavior has been reported by the Pierre Auger Collaboration and may reflect limitations of current hadronic interaction models in this energy range.  
In the Auger $\xmax$ study~\cite{PierreAuger:2014gko}, the fit quality in the lower-energy bins becomes systematically small when using QGSJet\,II-04 or Sibyll, whereas EPOS-LHC provides acceptable $p$ values over the full range.
This suggests that differences among interaction models, particularly in their predicted combinations of $\mxmax$ and $\sxmax$, may contribute to the reduced $p$-values observed at the lower energies.

For comparison with other experiments, we evaluated the mean logarithmic mass number, $\mlna$, defined as
\begin{equation}
\mlna = \Sigma_{ip} \left( f_{ip} \times \rm{ln}\it{A_{ip}} \right),
\end{equation}
where $ip$ stands for one of {p, He, N, or Fe}, and $f_{ip}$ is the best-fit fraction after applying the acceptance correction derived from the four-component fit.

Figure\,\ref{fig:lnA} summarizes our result together with measurements from other experiments, categorized according to their detection techniques.
The orange symbols represent experiments that infer the composition from measurements of the electromagnetic and muonic components of extensive air showers using surface and underground detector arrays (e.g., KASCADE, LHAASO, and IceTop). 
The blue and black symbols correspond to experiments that detect Cherenkov or fluorescence light emitted by the air shower (e.g., Tunka, Yakutsk, Auger, TA, and TALE). 

For LHAASO and Auger, the corresponding systematic uncertainty ranges are indicated by brackets.
Note that the interaction models adopted in the $\mlna$ derivation differ between experiments; for instance, IceTop and LHAASO use models distinct from the one employed in this work.
The KASCADE measurement shown in the figure is based on public KASCADE data, recently reanalyzed using machine learning techniques with convolutional neural networks~\cite{Kuznetsov:2023pvo}.
This result is consistent with our measurement within systematic uncertainties.
Our result is also consistent with recent measurements reported by LHAASO~\cite{bib:lhaasoCRSpectrumAndLnA} within the systematic uncertainties.
Notably, the evolution of the mean logarithmic mass number at energies below $10^{17}$\,eV shows a similar trend observed by KASCADE and LHAASO.
For energies below $10^{17.2}$\,eV, our results also agree with those from IceTop~\cite{bib:iceTopResults} within systematic uncertainties.
Above this energy, discrepancies become apparent, which may be attributed to differences in detection techniques and the influence of the so-called ``muon puzzle''~\cite{bib:muonPuzzle}.

On the other hand, our result is consistent with those from Tunka~\cite{bib:tunkaComposition} and Yakutsk~\cite{bib:Knurenko:2019oil}, which employ nonimaging Cherenkov detection techniques. 
At the higher energy end, this work smoothly connects with our previous measurement by using the TA hybrid mode~\cite{bib:tahybrid}.  
In contrast, a significant discrepancy remains when compared to the result reported by Auger~\cite{bib:AugerXmax}, even after accounting for the systematic uncertainties of both experiments.
The numerical values of the measured $\mxmax$, $\sxmax$, and $\mlna$ for each energy bin, including both statistical and systematic uncertainties, are summarized in Appendix\,\ref{app:dataTables}.

Overall, the observed break in the elongation rate, the consistently broad $\xmax$ distributions across all energy ranges, the results of the KS test assuming single components, and the fraction fits of $\xmax$ distributions suggest the following interpretation: Below $10^{17}$\,eV, the mass of the dominant cosmic ray nuclei gradually transitions from intermediate-mass nuclei to heavy nuclei, resulting in an increase in the mean mass.
At around $10^{17}$\,eV, this trend reverses, as indicated by the characteristic change in the elongation rate, which suggests an increasing dominance of proton cosmic rays.
Furthermore, at all energy ranges, a mixture of different primary nuclei is required, naturally explaining the reason for the consistently broad $\xmax$ distribution.

In this study, we used the QGSJet\,II-04 hadronic interaction model for all data interpretation.
It should be noted that this model predicts $\mxmax$ values that are 10-20 g/cm$^2$ shallower than those of other post-LHC models, such as EPOS-LHC and Sibyll\,2.3d.
In other words, this work corresponds to the lightest estimate of the mass composition.
Looking forward, hadronic interaction models are expected to be updated with the input from the proton–oxygen collision run at the LHC~\cite{bib:futureLHC}, which was successfully carried out in 2025.
These experiments will closely replicate the conditions of cosmic ray interactions in the upper atmosphere, potentially leading to substantial updates in interaction models and enabling more precise interpretations of cosmic ray observations.

\section{Conclusion}
\label{sec:conclusion}

TALE has successfully measured the cosmic ray mass composition in the energy range from $10^{16.5}$ eV to $10^{18.5}$ eV using almost five years of data collected in hybrid mode.
The analysis presented in this work is based on detailed comparisons of observed $\xmax$ distributions with Monte Carlo simulations. These simulations were generated using CORSIKA with the QGSJet\,II-04 hadronic interaction model, considering four primary species: proton, helium, nitrogen, and iron. 
The reliability of the Monte Carlo simulations was confirmed through comparisons of various reconstructed observables between the observational data and the simulations, as discussed in Sec. \ref{sec:simulation}.

On the basis of these results, our analysis incorporates both the elongation rate and the observed $\xmax$ distributions.
Up to approximately $10^{17}$\,eV, the composition gradually evolves from lighter to heavier nuclei, with iron becoming dominant and reaching roughly 50\%.
This behavior is consistent with an interpretation in which the knee corresponds to the acceleration limit of Galactic protons, while heavier nuclei reach their respective limits at charge-scaled higher energies, ending with iron around the second knee.
Beyond this energy, the composition shifts back toward being proton dominated, indicating the onset of a significant contribution from extra-Galactic components.
This transition in composition is reflected by a break in the elongation rate around $10^{17.1}$\,eV, near the energy where a change of slope in the spectrum is experimentally observed.

In future work, we plan to verify our results using different hadronic interaction models to assess the robustness of our findings.
Additionally, we intend to perform similar analyses using $\xmax$ measurements obtained in the newly established hybrid mode.  
This mode utilizes data from 50 SDs, deployed with 100\,m spacing in 2023, in combination with the TALE FD.
We expect this new setup to enhance the precision of cosmic ray composition studies, particularly at lower energies below $10^{16.5}$\,eV.
By covering the knee in the cosmic ray spectrum, this configuration will provide further insights into the characteristics of cosmic ray, potentially revealing new details about the transition from Galactic to extra-Galactic sources.

\input{TAacknowledgements-20250704}


%

\appendix
\setcounter{figure}{0}
\renewcommand{\thefigure}{A.\arabic{figure}}
\section{Additional Test of the Elongation Rate Fit}
\label{app:cl_only_elongation}
\begin{figure}[h]
\begin{center}
\includegraphics[trim=0cm 0.025cm 0cm 0cm, clip, width=1.\linewidth]{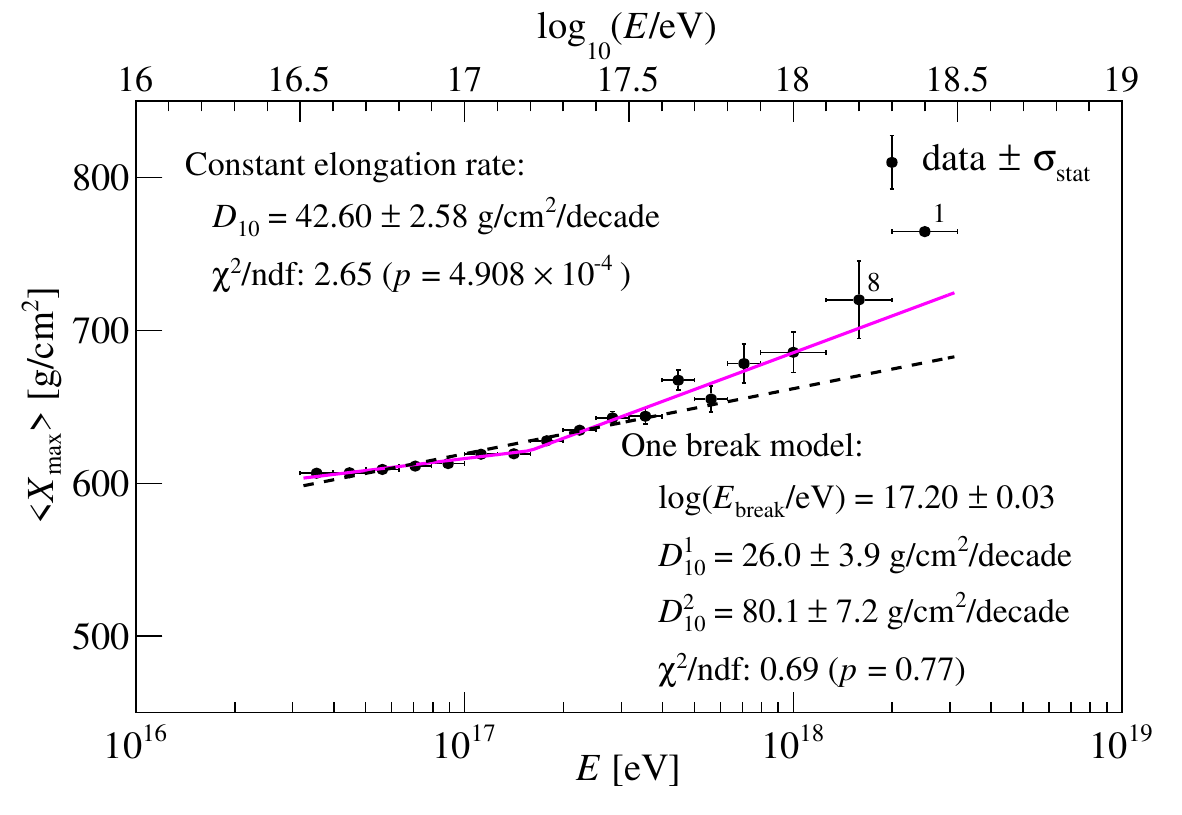}
\caption{\justifying{Evolution of $\mxmax$ as a function of energy only using Cherenkov-dominated event dataset. The best fits of a constant elongation rate and one break elongation rate are shown by a dashed line and a magenta line, respectively. For energy bins with fewer than ten events, the number of events is indicated next to the corresponding data point.}}
\label{fig:elongationFitOnlyCh}
\end{center}
\end{figure}

\add{
As an additional cross-check of the elongation rate study, we performed a fit only using Cherenkov-dominated events.
As discussed in Sec.\,\ref{sec:result}, fluorescence-dominated events are essentially absent below $10^{17}$\,eV, making a fluorescence-dominated only test infeasible.
The fit result is shown in Fig.\,\ref{fig:elongationFitOnlyCh}.
The Cherenkov-only sample still shows a statistically significant break in the elongation rate, and the fitted break energy is consistent with that obtained from the full dataset.
}

\section{Acceptance Correction}
\label{app:acceptanceCorrection}
Let \( f_i \) denote the fraction of each primary cosmic ray species \( i \) (proton, helium, nitrogen, and iron) at the top of the atmosphere. Assuming that only these four types of primaries exist, their fractions must satisfy the condition,
\vspace{-3mm}
\[
f_p + f_\mathrm{He} + f_\mathrm{N} + f_\mathrm{Fe} = 1.
\]
On the other hand, the number of cosmic ray events observed by the detector reflects differences in acceptance for each primary species, which vary due to the difference of shower development properties of nuclei. 
Lighter nuclei, such as protons, tend to have deeper $\xmax$ values compared to heavier nuclei as shown in Fig.\,\ref{fig:meanXmax00}, making them more likely to be observed.
The relative detection efficiencies for iron, nitrogen, and helium compared to protons were evaluated using Monte Carlo events.
As described in Sec.\,\ref{sec:result}, at an energy of \( 10^{16.5} \,\mathrm{eV} \), the ratios of detection efficiencies relative to protons are approximately,
\vspace{-3mm}
\[
R_{\mathrm{Fe}/p} \sim 0.8, \quad R_{\mathrm{N}/p} \sim 0.9, \quad R_{\mathrm{He}/p} \sim 0.95.
\]
For energies above \( 10^{17.5} \,\mathrm{eV} \), these ratios become flat and independent of the primary.
Considering these differences, the observed number of cosmic ray events at a certain energy bin, \( N_\mathrm{CR}^\mathrm{obs}(E) \), can be expressed as,
\vspace{-3mm}
\begin{multline*}
N_\mathrm{CR}^\mathrm{obs}(E) = N_\mathrm{CR}^\mathrm{top}(E) \Big\{ f_p(E) 
+ f_\mathrm{He}(E) \cdot R_{\mathrm{He}/p}(E) \\
+ f_\mathrm{N}(E) \cdot R_{\mathrm{N}/p}(E) 
+ f_\mathrm{Fe}(E) \cdot R_{\mathrm{Fe}/p}(E) \Big\}.
\end{multline*}
\allowdisplaybreaks
Here, $N_\mathrm{CR}^\mathrm{top}(E)$ denotes the number of cosmic ray events arriving at the top of the atmosphere in the given energy bin. 
This quantity reflects the true flux of cosmic rays before accounting for detector acceptance effects.
From this, the observed fraction of each primary is given by,
\begin{align*}
f_{p}^\mathrm{obs}(E) &= \frac{f_{p}(E) \cdot N_{\mathrm{CR}}^\mathrm{top}}{N_\mathrm{CR}^\mathrm{obs}}, \\
f_{\rm{He}}^\mathrm{obs}(E) &= \frac{f_{\rm{He}}(E) \cdot N_{\mathrm{CR}}^\mathrm{top} \cdot R_{\mathrm{He}/p}(E)}{N_\mathrm{CR}^\mathrm{obs}}, \\
f_{\rm{N}}^\mathrm{obs}(E) &= \frac{f_{\rm{N}}(E) \cdot N_{\mathrm{CR}}^\mathrm{top} \cdot R_{\mathrm{N}/p}(E)}{N_\mathrm{CR}^\mathrm{obs}}, \\
f_{\rm{Fe}}^\mathrm{obs}(E) &= \frac{f_{\rm{Fe}}(E) \cdot N_{\mathrm{CR}}^\mathrm{top} \cdot R_{\mathrm{Fe}/p}(E)}{N_\mathrm{CR}^\mathrm{obs}}.
\end{align*}
The observed fractions $f^{\rm{obs}}(E)$ for each primary are the fractions obtained directly from the fitting procedure using the \texttt{TFractionFitter} tool.
From these relationships, the fractions \( f_\mathrm{He}(E) \), \( f_\mathrm{N}(E) \), and \( f_\mathrm{Fe}(E) \) can be expressed in terms of \( f_p(E) \) as:
\begin{align*}
f_\mathrm{He}(E) &= f_p(E) \cdot \frac{f_\mathrm{He}^\mathrm{obs}(E)}{f_p^\mathrm{obs}(E)} \cdot \frac{1}{R_{\mathrm{He}/p}(E)}, \\
f_\mathrm{N}(E) &= f_p(E) \cdot \frac{f_\mathrm{N}^\mathrm{obs}(E)}{f_p^\mathrm{obs}(E)} \cdot \frac{1}{R_{\mathrm{N}/p}(E)}, \\
f_\mathrm{Fe}(E) &= f_p(E) \cdot \frac{f_\mathrm{Fe}^\mathrm{obs}(E)}{f_p^\mathrm{obs}(E)} \cdot \frac{1}{R_{\mathrm{Fe}/p}(E)}.
\end{align*}
The condition that the sum of the fractions must equal 1 leads to the following equation for \( f_p(E) \):
\begin{multline*}
f_p(E) \cdot \Bigg\{ 1 
+ \frac{f_\mathrm{He}^\mathrm{obs}(E)}{f_p^\mathrm{obs}(E)} \cdot \frac{1}{R_{\mathrm{He}/p}(E)} \\
+ \frac{f_\mathrm{N}^\mathrm{obs}(E)}{f_p^\mathrm{obs}(E)} \cdot \frac{1}{R_{\mathrm{N}/p}(E)} \\
+ \frac{f_\mathrm{Fe}^\mathrm{obs}(E)}{f_p^\mathrm{obs}(E)} \cdot \frac{1}{R_{\mathrm{Fe}/p}(E)} \Bigg\} = 1.
\end{multline*}
Solving for \( f_p(E) \), we obtain:
\begin{multline*}
f_p(E) = 1 \scalebox{2}{/} \Bigg\{ 1 
+ \frac{f_\mathrm{He}^\mathrm{obs}(E)}{f_p^\mathrm{obs}(E)} \cdot \frac{1}{R_{\mathrm{He}/p}(E)} \\
+ \frac{f_\mathrm{N}^\mathrm{obs}(E)}{f_p^\mathrm{obs}(E)} \cdot \frac{1}{R_{\mathrm{N}/p}(E)} \\
+ \frac{f_\mathrm{Fe}^\mathrm{obs}(E)}{f_p^\mathrm{obs}(E)} \cdot \frac{1}{R_{\mathrm{Fe}/p}(E)} \Bigg\}.
\end{multline*}
Once \( f_p(E) \) is determined, the fractions \( f_\mathrm{He}(E) \), \( f_\mathrm{N}(E) \), and \( f_\mathrm{Fe}(E) \) can be calculated directly using the earlier expressions.

\renewcommand{\thetable}{A.\arabic{table}}
\setcounter{table}{0} 
\section{Data Table}
\label{app:dataTables}
The measured $\mxmax$, $\sxmax$ and $\mlna$ in this work are summarized in Table\,\ref{table:XmaxData}.
\begin{table*}[ht]
\begin{center}
\caption{\justifying{First two moments of the $\xmax$ distributions and $\mlna$. Energies in $\log_{10}(\it{E}/\rm{eV})$\,,\,$\mxmax$ and $\sxmax$ given in [g/cm$^{2}$], and $\mlna$ are listed followed by their statistical and systematic uncertainties. The number of selected events in each energy bin is given in the second column.}}
\centering
\renewcommand{\arraystretch}{1.35}
\begin{tabular*}{0.8\textwidth}{@{\extracolsep{\fill}} c r c c c @{}}
\hline \hline
\makecell{Energy bin \\ in $\log_{10}(\it{E}/\rm{eV})$} & $N$ & $\mxmax$ [g/cm$^2$] & $\sxmax$ [g/cm$^2$]  & $\mlna$\\
\hline
$ 16.5 - 16.6 $ & $  759 $ & $  606.7\,\pm\,   2.4\,^{+  15.3}_{ -15.5} $ & $   67.1\,\pm\,   1.7\,^{+   4.1}_{-   3.4} $ 	& $   1.95 \pm   0.12 ^{+  0.76}_{-  0.77} $ \\
$ 16.6 - 16.7 $ & $  924 $ & $  607.0\,\pm\,   2.2\,^{+  13.9}_{ -14.3} $ & $   66.9\,\pm\,   1.6\,^{+   4.2}_{-   3.9} $ 	& $   2.35 \pm   0.11 ^{+  0.67}_{-  0.69} $ \\
$ 16.7 - 16.8 $ & $ 1014 $ & $  609.1\,\pm\,   2.1\,^{+  12.6}_{ -12.9} $ & $   65.7\,\pm\,   1.5\,^{+   5.4}_{-   3.9} $ 	& $   2.54 \pm   0.09 ^{+  0.58}_{-  0.59} $ \\
$ 16.8 - 16.9 $ & $  931 $ & $  611.3\,\pm\,   2.1\,^{+  11.5}_{ -12.3} $ & $   63.7\,\pm\,   1.5\,^{+   3.6}_{-   3.6} $ 	& $   2.69 \pm   0.09 ^{+  0.49}_{-  0.52} $ \\
$ 16.9 - 17.0 $ & $  937 $ & $  612.9\,\pm\,   2.1\,^{+  10.0}_{ -10.9} $ & $   65.5\,\pm\,   1.5\,^{+   4.2}_{-   4.6} $ 	& $   2.71 \pm   0.09 ^{+  0.43}_{-  0.47} $ \\
$ 17.0 - 17.1 $ & $  857 $ & $  619.2\,\pm\,   2.5\,^{+   9.6}_{ -10.3} $ & $   72.2\,\pm\,   1.7\,^{+   4.9}_{-   4.2} $ 	& $   2.77 \pm   0.11 ^{+  0.41}_{-  0.44} $ \\
$ 17.1 - 17.2 $ & $  688 $ & $  620.9\,\pm\,   2.6\,^{+   9.8}_{ -10.3} $ & $   67.6\,\pm\,   1.8\,^{+   4.8}_{-   4.2} $ 	& $   2.88 \pm   0.11 ^{+  0.41}_{-  0.43} $ \\
$ 17.2 - 17.3 $ & $  622 $ & $  630.2\,\pm\,   2.9\,^{+   9.5}_{ -10.7} $ & $   72.6\,\pm\,   2.1\,^{+   5.5}_{-   6.3} $ 	& $   2.60 \pm   0.12 ^{+  0.39}_{-  0.45} $ \\
$ 17.3 - 17.4 $ & $  569 $ & $  639.2\,\pm\,   3.2\,^{+  10.4}_{ -11.7} $ & $   76.3\,\pm\,   2.3\,^{+   5.2}_{-   6.6} $ 	& $   2.36 \pm   0.14 ^{+  0.46}_{-  0.51} $ \\
$ 17.4 - 17.5 $ & $  462 $ & $  648.5\,\pm\,   3.7\,^{+   9.8}_{ -11.3} $ & $   79.2\,\pm\,   2.6\,^{+   6.4}_{-   4.8} $ 	& $   2.49 \pm   0.14 ^{+  0.40}_{-  0.46} $ \\
$ 17.5 - 17.6 $ & $  395 $ & $  656.5\,\pm\,   3.7\,^{+   9.0}_{ -10.8} $ & $   74.3\,\pm\,   2.6\,^{+   5.0}_{-   5.1} $ 	& $   2.06 \pm   0.16 ^{+  0.39}_{-  0.46} $ \\
$ 17.6 - 17.7 $ & $  314 $ & $  673.7\,\pm\,   4.3\,^{+   9.3}_{ -11.4} $ & $   76.0\,\pm\,   3.0\,^{+   5.9}_{-   7.9} $ 	& $   1.92 \pm   0.18 ^{+  0.38}_{-  0.47} $ \\
$ 17.7 - 17.8 $ & $  229 $ & $  679.2\,\pm\,   5.2\,^{+   8.7}_{ -10.9} $ & $   78.9\,\pm\,   3.7\,^{+   5.6}_{-   7.5} $ 	& $   1.59 \pm   0.22 ^{+  0.37}_{-  0.46} $ \\
$ 17.8 - 17.9 $ & $  175 $ & $  690.9\,\pm\,   5.6\,^{+   8.2}_{ -10.8} $ & $   73.9\,\pm\,   4.0\,^{+   5.4}_{-   5.5} $ 	& $   1.46 \pm   0.24 ^{+  0.35}_{-  0.46} $ \\
$ 17.9 - 18.1 $ & $  197 $ & $  707.8\,\pm\,   5.7\,^{+   8.8}_{ -11.3} $ & $   80.4\,\pm\,   4.1\,^{+   6.8}_{-   5.8} $ 	& $   1.39 \pm   0.24 ^{+  0.37}_{-  0.47} $ \\
$ 18.1 - 18.3 $ & $   69 $ & $  729.2\,\pm\,   9.7\,^{+   9.1}_{ -11.9} $ & $   80.7\,\pm\,   6.9\,^{+   8.4}_{-   6.7} $ 	& $   1.16 \pm   0.43 ^{+  0.40}_{-  0.52} $ \\
$ 18.3 - 18.5 $ & $   31 $ & $  752.1\,\pm\,   9.5\,^{+  10.5}_{ -13.2} $ & $   52.7\,\pm\,   6.7\,^{+  10.4}_{-   5.5} $ 	& $   0.35 \pm   0.46 ^{+  0.50}_{-  0.63} $ \\
\hline
\end{tabular*}
\label{table:XmaxData}
\end{center}
\end{table*}


\bibliographystyle{apsrev4-1}
\bibliography{talehybrid.bib}

\end{document}

%% file: TA-author-20250704-revtex.tex
\author{R.U.~Abbasi}
\affiliation{Department of Physics, Loyola University-Chicago, Chicago, Illinois 60660, USA}

\author{T.~Abu-Zayyad}
\affiliation{High Energy Astrophysics Institute and Department of Physics and Astronomy, University of Utah, Salt Lake City, Utah 84112-0830, USA}
\affiliation{Department of Physics, Loyola University-Chicago, Chicago, Illinois 60660, USA}

\author{M.~Allen}
\affiliation{High Energy Astrophysics Institute and Department of Physics and Astronomy, University of Utah, Salt Lake City, Utah 84112-0830, USA}

\author{J.W.~Belz}
\affiliation{High Energy Astrophysics Institute and Department of Physics and Astronomy, University of Utah, Salt Lake City, Utah 84112-0830, USA}

\author{D.R.~Bergman}
\affiliation{High Energy Astrophysics Institute and Department of Physics and Astronomy, University of Utah, Salt Lake City, Utah 84112-0830, USA}

\author{F.~Bradfield}
\affiliation{Graduate School of Science, Osaka Metropolitan University, Sugimoto, Sumiyoshi, Osaka 558-8585, Japan}

\author{I.~Buckland}
\affiliation{High Energy Astrophysics Institute and Department of Physics and Astronomy, University of Utah, Salt Lake City, Utah 84112-0830, USA}

\author{W.~Campbell}
\affiliation{High Energy Astrophysics Institute and Department of Physics and Astronomy, University of Utah, Salt Lake City, Utah 84112-0830, USA}

\author{B.G.~Cheon}
\affiliation{Department of Physics and The Research Institute of Natural Science, Hanyang University, Seongdong-gu, Seoul 426-791, Korea}

\author{K.~Endo}
\affiliation{Graduate School of Science, Osaka Metropolitan University, Sugimoto, Sumiyoshi, Osaka 558-8585, Japan}

% ...

\collaboration{The Telescope Array Collaboration}
\noaffiliation

%% file: TAacknowledgements-20250704.tex
\begin{acknowledgements}

The Telescope Array experiment is supported by the Japan Society for
the Promotion of Science(JSPS) through
Grants-in-Aid
for Priority Area
431,
for Specially Promoted Research
JP21000002,
for Scientific  Research (S)
JP19104006,
for Specially Promoted Research
JP15H05693,
for Scientific  Research (S)
JP19H05607,
for Scientific  Research (S)
JP15H05741,
for Science Research (A)
JP18H03705,
for Young Scientists (A)
JPH26707011,
for Transformative Research Areas (A)
JP25H01294,
for International Collaborative Research
24KK0064,
and for Fostering Joint International Research (B)
JP19KK0074,
by the joint research program of the Institute for Cosmic Ray Research (ICRR), The University of Tokyo;
by the Pioneering Program of RIKEN for the Evolution of Matter in the Universe (r-EMU);
by the U.S. National Science Foundation awards
PHY-1806797, PHY-2012934, PHY-2112904, PHY-2209583, PHY-2209584, and PHY-2310163, as well as AGS-1613260, AGS-1844306, and AGS-2112709;
by the National Research Foundation of Korea
(2017K1A4A3015188, 2020R1A2C1008230, and RS-2025-00556637) ;
by the Ministry of Science and Higher Education of the Russian Federation under the contract 075-15-2024-541, IISN project No. 4.4501.18, by the Belgian Science Policy under IUAP VII/37 (ULB), by National Science Centre in Poland grant 2020/37/B/ST9/01821, by the European Union and Czech Ministry of Education, Youth and Sports through the FORTE project No. CZ.02.01.01/00/22\_008/0004632, and by the Simons Foundation (MP-SCMPS-00001470, NG). This work was partially supported by the grants of the joint research program of the Institute for Space-Earth Environmental Research, Nagoya University and Inter-University Research Program of the Institute for Cosmic Ray Research of University of Tokyo. The foundations of Dr. Ezekiel R. and Edna Wattis Dumke, Willard L. Eccles, and George S. and Dolores Dor\'e Eccles all helped with generous donations. The State of Utah supported the project through its Economic Development Board, and the University of Utah through the Office of the Vice President for Research. The experimental site became available through the cooperation of the Utah School and Institutional Trust Lands Administration (SITLA), U.S. Bureau of Land Management (BLM), and the U.S. Air Force. We appreciate the assistance of the State of Utah and Fillmore offices of the BLM in crafting the Plan of Development for the site.  We thank Patrick A.~Shea who assisted the collaboration with much valuable advice and provided support for the collaboration’s efforts. The people and the officials of Millard County, Utah have been a source of steadfast and warm support for our work which we greatly appreciate. We are indebted to the Millard County Road Department for their efforts to maintain and clear the roads which get us to our sites. We gratefully acknowledge the contribution from the technical staffs of our home institutions. An allocation of computing resources from the Center for High Performance Computing at the University of Utah as well as the Academia Sinica Grid Computing Center (ASGC) is gratefully acknowledged.
\end{acknowledgements}